\documentclass[twocolumn]{aastex63}

\usepackage{amsmath}
\usepackage{natbib}
\usepackage{multirow} 
\usepackage{anyfontsize}
\usepackage{indentfirst}
\usepackage{comment}
\usepackage{booktabs}
\usepackage{bigints}
\usepackage{mathrsfs,amsmath} 
\usepackage{makecell}
\usepackage{graphicx}

\usepackage[normalem]{ulem}
\useunder{\uline}{\ul}{}
\usepackage{hyperref}
\usepackage{lineno}

\hypersetup{linkcolor=magenta,citecolor=cyan,filecolor=yellow,urlcolor=blue}

\received{ddmmyyyy}
\revised{\today}
\accepted{ddmmyyyy}
\published{ddmmyyyy}

\shorttitle{Precursors versus Main Bursts}
\shortauthors{Li et al.}

\begin{document}

\title{Pulse-resolved Classification and Characteristics of Long-duration GRBs with \emph{Swift}/BAT Data.I.\\ Precursors versus Main Bursts}

\author[0000-0002-1343-3089]{Liang Li}
\affiliation{Institute of Fundamental Physics and Quantum Technology, Ningbo University, Ningbo, Zhejiang 315211, People's Republic of China}
\affiliation{School of Physical Science and Technology, Ningbo University, Ningbo, Zhejiang 315211, People's Republic of China}

\author{Yu Wang}
\affiliation{ICRANet, Piazza della Repubblica 10, I-65122 Pescara, Italy}
\affiliation{ICRA and Dipartimento di Fisica, Universit\`a  di Roma ``La Sapienza'', Piazzale Aldo Moro 5, I-00185 Roma, Italy}
\affiliation{INAF -- Osservatorio Astronomico d'Abruzzo, Via M. Maggini snc, I-64100, Teramo, Italy}

\author{Jin-Jun~Geng}
\affiliation{Purple Mountain Observatory, Chinese Academy of Sciences, Nanjing 210023, P. R. China}

\author{Yong-Feng Huang}
\affiliation{School of Astronomy and Space Science, Nanjing University, Nanjing 210023, China}
\affiliation{Key Laboratory of Modern Astronomy and Astrophysics (Nanjing University),
    Ministry of Education, Nanjing 210023, China}

\author{Rong-Gen~Cai}
\affiliation{Institute of Fundamental Physics and Quantum Technology, Ningbo University, Ningbo, Zhejiang 315211, People's Republic of China}

\correspondingauthor{Liang-Li, and Rong-Gen~Cai}
\email{liliang@nbu.edu.cn; cairg@itp.ac.cn}

\begin{abstract}

We present a systematic pulse-by-pulse analysis of 22 long-duration GRBs observed by \emph{Swift}, each exhibiting a well-separated precursor before the main burst. We compare duration, spectral hardness ratio, minimum variability timescale (MVT), and spectral lag between these components. Both precursors and main bursts have durations and hardness broadly consistent with Type II GRBs. However, precursors show longer MVTs (by factors of 3–10) and diverse lags with near-zero median values, while main bursts display variable MVTs and positive lags. These differences suggest precursors may originate from distinct dissipation conditions, possibly due to cocoon shock breakout or early magnetically dominated outflows. Despite temporal differences, both episodes are consistent with a single collapsar origin, providing no evidence for dual-progenitor events. Our findings support pulse-resolved classification and show that precursors offer critical insights into jet formation and pre-burst activity.

\end{abstract}

\keywords{Gamma-ray bursts (629); Astronomy data analysis (1858); Time domain astronomy (2109)}

\section{Introduction} \label{sec:intro}

Gamma-ray bursts (GRBs) are the most luminous explosive events in the Universe. Traditionally, GRBs are empirically classified into two groups based on their $T_{90}$ durations\footnote{The time interval during which 90\% of the burst's fluence is detected.}: long-duration GRBs (lGRBs, $T_{90} > 2$ s) and short-duration GRBs (sGRBs, $T_{90} < 2$ s) \citep{Kouveliotou1993}. This bimodal duration distribution corresponds to different progenitor types. Long GRBs are typically associated with the core-collapse of massive stars (the collapsar model; Type II progenitors) whereas short GRBs are thought to originate from the mergers of compact binary objects (e.g., binary neutron star or neutron star--black hole systems, Type I) \citep{Woosley2006, Zhang2009b, Berger2014}. These two populations differ not only in host environments and supernova associations, but also in prompt-emission properties, including light curve structure and morphology, spectral hardness, spectral lag, and minimum variability timescale (MVT) \citep[e.g.,][]{Norris2000, Norris2005, Yi2006, Horvath2010, Golkhou2014, Golkhou2015}.

Over the decades, many prompt-emission characteristics have been found to differ statistically between the short (Type I) and long (Type II) GRB populations \citep{Berger2014, Zhang2011Review}. Relative to long GRBs, short GRBs tend to exhibit promptly peaked, rapidly varying light curves, often composed of one or a few sharp, narrow pulses with durations of 0.1--2 s. Long GRBs, in contrast, show more extended, complex light curves lasting tens to hundreds of seconds, often with multiple overlapping pulses \citep[e.g.,][]{Norris2005, Li2019a, Li2021a, Li2021b,Li2024}. Spectrally, short GRBs are typically harder (i.e., emission peaks at higher photon energies) than long GRBs \citep{Dezalay1992, Dezalay1995}, as evidenced by their higher hardness ratios, though the precise values are instrument-dependent \citep[e.g.,][]{Horvath2010, Goldstein2012GBMSpec, QinY2013}. Short GRBs also characteristically show negligible or even slightly negative spectral lags ($\tau$), meaning there is little to no time delay between high-energy and lower-energy photons. Conversely, long GRBs often exhibit measurable positive spectral lags, with lower-energy emission arriving slightly later than higher-energy emission \citep{Norris2000, Norris2006, Yi2006, Bernardini2015}. Another distinguishing property is the minimum variability timescale (MVT). Short GRBs typically vary on much shorter timescales (tens of milliseconds) compared to long GRBs (hundreds of milliseconds) \citep{Golkhou2014, Golkhou2015}. Furthermore, the two classes occupy distinct regions in empirical correlations relating spectral peak energy to energetics or luminosity \citep[e.g.,][and references therein]{Amati2002, Amati2006, Yonetoku2004, Yonetoku2010, Liang2005, Xu2012, ZhangFuWen2012, Heussaff2013, Dainotti2018, Li2023c}. Most notable are the $E_{\rm p,z}$--$E_{\gamma,\rm iso}$ (Amati) relation \citep{Amati2002} and the lag--luminosity relation \citep{Norris2000}, where $E_{\rm p,z}\equiv (1+z) E_{\rm p}$ is the cosmological rest-frame $\nu F_\nu$ peak energy, $E_{\gamma,\rm iso}$ is the isotropic-equivalent bolometric emission energy, and $L_{\rm p,iso}$ is the isotropic-equivalent bolometric peak luminosity. 

Despite the broad utility of the $T_{90}=2$~s threshold in distinguishing short and long GRBs, it remains an imperfect divider with obvious limitations. The $T_{90}$ distributions overlap near the boundary, allowing for the existence of both short-duration collapsars and long-duration merger events. This ambiguity was already apparent in BATSE data \citep{Kouveliotou1993, Bromberg2013, Lien2016} and is exacerbated by instrument-dependent duration measurements \citep{Lien2016, Sakamoto2011BATcat}. Several GRBs critically challenge the standard classification. GRB~200826A ($T_{90} \sim 1$~s) exhibits collapsar-like features \citep{Ahumada2021NatAs, ZhangBB2021NatAs}, while GRB~060614 and GRB~211211A (both $T_{90} > 2$~s) show merger signatures \citep{Gehrels2006Nature, Fynbo2006Nature, GalYam2006Nature, Troja2022Nature, Rastinejad2022Nature, Yang2022Nature}. Such events highlight the need for caution when using duration alone as a diagnostic. The situation is further complicated by multi-episode GRBs, which include either an early-time precursor emission \citep[hereafter PE;][]{Lazzati2005, WangXY2007, LiLX2007, Burlon2008, LiXJ2021} or a later-time soft extended emission tail \citep[hereafter EE;][]{Norris2006, Kaneko2015}. These components can significantly alter the total $T_{90}$. For instance, soft EE tails in short GRBs can extend $T_{90}$ beyond 10~s, while precursors in long GRBs can increase the measured duration via quiescent gaps. Thus, duration is an unreliable classifier for bursts with complex temporal structures.

Multi-episode prompt emission has emerged as a key feature in understanding GRB progenitors \citep{Li2026}. Short GRBs often consist of a main emission (ME) spike within 2~s, occasionally followed by a soft, low-fluence EE tail lasting up to tens, or even hundreds, of seconds \citep[e.g.,][]{Norris2006, Kaneko2015, Hu2014, LiXJ2021}. Long GRBs frequently exhibit weak PE that precedes the main burst by several to tens, or even hundreds, of seconds. These precursors are typically low in peak flux and may differ spectrally from the main burst \citep{Koshut1995, Lazzati2005, Burlon2008, Coppin2020}. In some events (e.g., GRB~041219A and GRB~160625B), distinct emission episodes are separated by quiescent intervals of tens to hundreds of seconds \citep{McBreen2006, Zhang2018NA}. The diversity of light curve morphology suggests that the central engine can produce multiple temporally separated outflows \citep{NakarPiran2002}. These complex cases emphasize the critical importance of analyzing each emission episode individually to fully understand the underlying physics \citep{Zhang2018NA}.

In light of the complexities above, it is evident that pulse-by-pulse classification within individual GRBs, especially those with quiescent intervals separating sub-bursts, is crucial for correctly inferring their physical origin. Rather than classifying an entire GRB by a single global $T_{90}$ or other bulk properties, we can gain deeper insight by analyzing each distinct emission episode independently. Previous studies of multi-episode (or “two-step”) GRBs have mostly focused on light curve morphologies and spectral differences (e.g. noting that an extended emission is softer than the main burst spike, or that a precursor has a thermal component). However, a systematic classification and characteristics of sub-bursts within the same GRB using the established observational criteria (duration, hardness, lag, variability, etc.) has been lacking. Key questions remain unexplored. In a long GRB with a precursor, does the precursor pulse meet the criteria of a Type-II (collapsar-origin) burst just like its main burst emission, or could it resemble a Type-I (merger-origin) burst? Likewise, in a short GRB with extended emission, do the short burst main spike and the extended emission tail share the same classification (both Type I), or does the EE tail more closely resemble a Type II long-burst component? In multi-pulse long GRBs with quiescent gaps, are all pulses homogeneous in their properties, or do they differ, potentially indicating different emission regions or physical processes within the same event? Addressing these questions can shed light on whether a single GRB event with multiple episodes is driven by one central engine mechanism manifesting in different forms, or if we might occasionally be witnessing two different phenomena coincident in one event.

In this paper, we present a comprehensive, systematic study of GRB prompt emission episodes at the sub-burst level. Using a large sample of GRBs detected by the \emph{Swift} satellite (with known redshifts), we identify events that have clearly separable prompt emission episodes (those with significant quiescent intervals between distinct pulses or sub-bursts). We then examine each episode using the standard classification observables, including the duration $T_{90}$, the hardness ratio, the spectral lag, and the minimum variability timescale. By comparing these parameters for sub-bursts within the same GRB, we evaluate whether they would be classified the same or differently under Type I/II schemes. Our focus is on identifying commonalities or differences that might indicate the same or distinct physical origins for the sub-bursts\footnote{We note that the \emph{Swift}-BAT's limited 15–150~keV bandpass leads to poorly constrained spectral peak energies ($E_{\rm p}$) for most bursts. Due to this limitation, we do not emphasize $E_{\rm p}$-related classification measures, such as the Amati or Yonetoku relations. Instead, our analysis focuses on several key observables that \emph{Swift} can measure robustly: $T_{90}$ (and other duration-related quantities), hardness ratio, spectral lag, and minimum variability timescale.}.

We will report our results in a series of papers. In this first paper, we examine long-duration GRBs with early-time weak precursors. These are events in which a relatively weak pulse precedes the main burst emission, separated by a distinct quiescent gap with no emission. Our goal is to determine whether such precursors are simply smaller manifestations of the collapsar engine (and thus Type II-like), or if they show evidence of a different origin. This effort aims to achieve a “clean” pulse-based GRB classification and characterization. We have compiled an “ideal” sample of 22 well-defined precursor events (see Section \ref{sec:SS}) from the \emph{Swift} dataset with measured redshift. For each event, we analyze and compare the properties of the precursor versus the main burst. The remainder of this paper is organized as follows. In Section~2, we describe our sample selection, data reduction, and analysis methodology, including the definitions of the observed parameters and statistical methods used. Section~3 presents the results of our comparative analysis between the precursor and main emission episodes, covering their durations, hardness ratios, minimum variability timescales, and spectral lags. In Section~4, we discuss the implications of these results for GRB classification and progenitor physics, including a case-by-case discussion of particularly interesting events. Section~5 summarizes our conclusions. Throughout the paper, the standard $\Lambda$-CDM cosmology with the parameters $H_{0}= 67.4$ ${\rm km s^{-1}}$ ${\rm Mpc^{-1}}$, $\Omega_{M}=0.315$, and $\Omega_{\Lambda}=0.685$ are adopted \citep{PlanckCollaboration2018}.

\section{Methodology} \label{sec:methodology}

\subsection{Sample Selection}\label{sec:SS}

To investigate the Type I/II classification properties of distinct prompt-emission episodes in gamma-ray bursts, we constructed a sample of long-duration GRBs that exhibit early-time precursors and well-separated sub-bursts. Our starting point was all \emph{Swift}-BAT GRBs with measured redshifts, which comprise over 200 events in total. We restrict our analysis to GRBs with measured redshifts because several key observables in our classification scheme require rest-frame corrections. Specifically, the rest-frame duration $T_{90}/(1+z)$, and quiescent gap duration $t_{\rm gap}$/(1+z) are essential for deriving physical parameters (e.g., emission radius; see Section \ref{sec:discussion}) and for uniform comparison across bursts at different cosmological distances, following standard practice in GRB classification studies \citep{Golkhou2015,Norris2000}. Observationally, bursts with well-separated double or multiple peaks are the most crucial subclasses for this task. Unlike continuously emitting multi-peak GRBs, episodes separated by quiescent intervals allow us to probe whether physically distinct emission phases (e.g., the precursor vs. the main burst) within a single burst share the same classification properties. We therefore visually inspected the 15–150 keV mask-weighted light curves of the parent sample to identify events that exhibit multiple prompt-emission episodes separated by quiescent intervals with count rates consistent with background fluctuations.

Before describing our selection criteria, we clarify our terminology. Following standard usage in GRB time-series analysis \citep{Norris2005}, a ``pulse" denotes a single-peaked emission structure with coherent rise and decay properties. An ``emission episode" or ``sub-burst`` (terms we use interchangeably) refers to a temporally distinct phase of prompt emission, which may contain one or multiple pulses. In our analysis, $G_1$ (precursor) and $G_2$ (main emission) denote the two sub-bursts separated by quiescent intervals. We define a GRB as having well-separated sub-bursts if they satisfy two criteria: (i) a temporal gap of several seconds or more separating distinct emission episodes, and (ii) count rates during this gap that drop to levels statistically consistent with background fluctuations (typically $<$ 0.01 counts s$^{-1}$ det$^{-1}$ in the 15-150 keV band), ensuring minimal temporal overlap between episodes. Based on this criterion, we compiled a subset of \textit{cleanly separated} GRBs with two or more temporally distinct episodes. For this work, we restrict our analysis to the \emph{quiescent} two-episode subsample with exactly two well-separated prompt emission episodes ($G_1+G_2$). This selection provides a clean, pulse-resolved (sub-burst) comparison within the same burst. These two-episode GRBs typically fall into three widely discussed morphological categories in the literature: (i) an early \emph{precursor emission} followed by a \emph{main burst}, (ii) a \emph{main burst} followed by \emph{an later-time softer extended emission}, and (iii) two well-separated \emph{main burst-like} episodes.
Four morphological groups can be therefore identified within this quiescent populations:
\begin{itemize}
\item Group I: A weak precursor emission ($G_1$) followed by a brighter main burst ($G_2$).
\item Group II: A main spike ($G_1$) followed by an softer extended emission ($G_2$), as commonly seen in short GRBs with later-time extended emission tails.
\item Group III: Two comparable-intensity pulses ($G_1$ and $G_2$) with a quiescent gap, effectively representing two main burst-like episodes.
\item Group IV: More complex cases with three or more distinct episodes (e.g., precursor+main+extended emissions).
\end{itemize}

For the scope of this paper, we specifically focus on the first subsample, which consists of long-duration GRBs where an early-time precursor ($G_1$) is followed by a main burst episode ($G_2$). Our selection requires two strict quantitative criteria for this subsample: (i) The peak flux of the main burst event in the \emph{Swift}-BAT 15–150 keV band is significantly higher than that of the precursor event, with $F_{\rm p}(G_1)/F_{\rm p}(G_2)<0.5$, and (ii) the flux falls below the background level before the start of the main burst, a clearly \emph{quiescent} interval (with count rates consistent with background levels) of at least several seconds separating $G_1$ and $G_2$ within the same bandpass, which ensure uniform comparison across the two pulses. These quantitative thresholds, combined with initial visual screening, yielded 22 GRBs that unambiguously satisfy both criteria.\footnote{We exclude more complex events with three or more distinct sub-bursts. These typically mix precursor, main, and extended emission components, thus complicating the pulse-resolved comparisons we seek.} We have assembled a ``ideal" sample of 22 such events based on the criteria and these  events span a broad range of redshifts ($0.1 \lesssim z \lesssim 2.8$) and light-curve morphologies. Their shared, well-defined precursor–main structure enables a clean, sub-burst-level statistical comparison within the individual bursts.

All \emph{Swift}-BAT data for the selected GRBs were retrieved from the UK \emph{Swift} Science Data Centre (UKSSDC) repository\footnote{\url{https://www.swift.ac.uk}}. Standard data reduction utilized HEASoft v6.30 with the latest calibration files \citep{Evans2007, Evans2009}. BAT light curves were extracted in multiple energy bands and at various time resolutions (from 100~$\mu$~s up to 1.024~s bins) to facilitate different analyses (e.g. fine temporal structure for MVT analysis, and broader overviews for $T_{90}$). All quantitative results presented in this study are derived solely from \emph{Swift}-BAT observations.

\subsection{Bivariate Normal Distribution} \label{sec:BND}

To visually represent the clustering properties of Type I (short) and Type II (long) GRBs in our classification diagrams, such as hardness ratio versus duration or spectral lag versus luminosity, we overlay confidence ellipses. These contours represent the $1\sigma$ and $2\sigma$ levels of a bivariate normal distribution fitted to established reference GRB samples, and they are used primarily for illustrative purposes.

We briefly outline the construction of these ellipses for completeness. Let $x$ and $y$ be two observables assumed to follow a bivariate normal distribution. The distribution is characterized by means $\mu_x$ and $\mu_y$, standard deviations $\sigma_x$ and $\sigma_y$, and a correlation coefficient $\rho$. The joint probability density function $f(x,y)$ is given by:
\begin{equation}
f(x,y) = \frac{1}{2\pi \sigma_x \sigma_y \sqrt{1-\rho^2}} \exp\left[-\frac{1}{2(1-\rho^2)}\left(\frac{(x-\mu_x)^2}{\sigma_x^2} + \frac{(y-\mu_y)^2}{\sigma_y^2} - \frac{2\rho (x-\mu_x)(y-\mu_y)}{\sigma_x \sigma_y}\right)\right].
\end{equation}

Contours of constant probability density correspond to ellipses in the $x$–$y$ plane, defined by a constant Mahalanobis distance ($\Delta$) from the mean vector $\mu$. These contours satisfy the relation:
\begin{equation}
(\mathbf{X}-\boldsymbol{\mu})^T \Sigma^{-1} (\mathbf{X}-\boldsymbol{\mu}) = \Delta^2,
\end{equation}
where $\mathbf{X} = (x, y)^T$, the mean vector is $\boldsymbol{\mu} = (\mu_x, \mu_y)^T$, and $\Sigma$ is the covariance matrix. The specific values $\Delta = 1$ and $\Delta =2$ correspond to the $1\sigma$ and $2\sigma$ confidence contours in two dimensions, which enclose approximately 39.3\% and 86.5\% of the total probability, respectively.

In practice, we fit these bivariate Gaussians in the $\log$-transformed observable space (e.g., $\log T_{90}$) using reference populations of BATSE or \emph{Swift} GRBs with well-established classifications from the literature. The resulting ellipses provide a useful statistical visualization of the central tendencies and dispersions of the two classes. However, we note that the real distributions of GRB parameters often exhibit non-Gaussian tails and inter-class overlaps. Thus, the ellipses are not intended to be strict classification boundaries but rather effective visual guides for interpreting the relative positions of our sub-burst data points.

\subsection{Temporal Properties of Sub-Bursts} \label{sec:sub}

To extract the timing characteristics of individual prompt-emission episodes, we decomposed the emission of each GRB in our target sample into a sequence of two sub-bursts: the precursor ($G_1$) and the main burst ($G_2$). We then measured key temporal properties for each episode, including the start and end times ($t_1$, $t_2$), duration measurements ($W$, $T_{90}$), peak time ($t_{\rm p}$), peak flux ($F_{\rm p}$), and the quiescent interval ($\Delta t_{\rm gap}$) separating them.

Background-subtracted light curves were generated in the 15–150 keV BAT energy band. The time binning was chosen based on signal strength, typically using 256~ms or 1024~ms resolution. For each emission episode, the peak time $t_{\rm p}$ was identified as the time bin registering the maximum count rate. From $t_{\rm p}$, we searched backward and forward to define the pulse boundaries ($t_1$, $t_2$). These boundaries were set at the first time bins on either side of the peak where the count rate returns to a statistically consistent background level. Specifically, a bin was considered background-dominated if its count rate fell below a conservative threshold (typically $<0.01$~counts\,s$^{-1}$\,det$^{-1}$) and exhibited minimal variation relative to adjacent bins. This robust method ensures the inclusion of the full emission wings while minimizing contamination from statistical noise or inter-pulse background fluctuations.

The total pulse width was defined as $W = t_2 - t_1$. To obtain a more robust duration, we also calculated the $T_{\rm pulse}$ duration by integrating the photon flux between $t_1$ and $t_2$ to obtain the total net counts for the pulse. The times $t_{05}$ and $t_{95}$, corresponding to the accumulation of 5\% and 95\% of the total net counts, were then used to define the $T_{90} = t_{95} - t_{05}$ duration. Uncertainties in $T_{90}$ were estimated via standard error propagation from the photon counting statistics and background fluctuations, yielding typical uncertainties of $\sim$0.1–0.5~s, depending on the chosen binning and signal quality.

We compute the quiescent interval separating $G_1$ and $G_1$ as:
\begin{equation}
\Delta t_{\rm gap} = t_1^{(G_2)} - t_2^{(G_1)},
\end{equation}
which is the temporal separation between the end of the precursor and the start of the main burst. The associated uncertainty was derived through the propagation of the time resolution and the errors on the determined pulse boundaries ($t_1$ and $t_2$). Although $\Delta t_{\rm gap}$ could be computed for multiple pulse pairs in more complex events, our analysis is strictly confined to the precursor-main pair ($G_1$ and $G_2$).

This entire procedure was implemented using an automated pulse-searching algorithm developed in Python, with all results validated through visual inspection. The algorithm consistently records the peak flux $F_{\rm p}$ and its $1\sigma$ uncertainty, derived from Poisson statistics for each pulse. 

The temporal properties derived here establish the foundational framework for calculating subsequent prompt-emission observables, including the hardness ratio, minimum variability timescale, and spectral lag, all of which are computed over the pulse interval $[t_1, t_2]$ for each sub-burst and are detailed in the following subsections.

\section{Results}\label{sec:results}

\begin{figure*}
\includegraphics[width=1.0\textwidth]{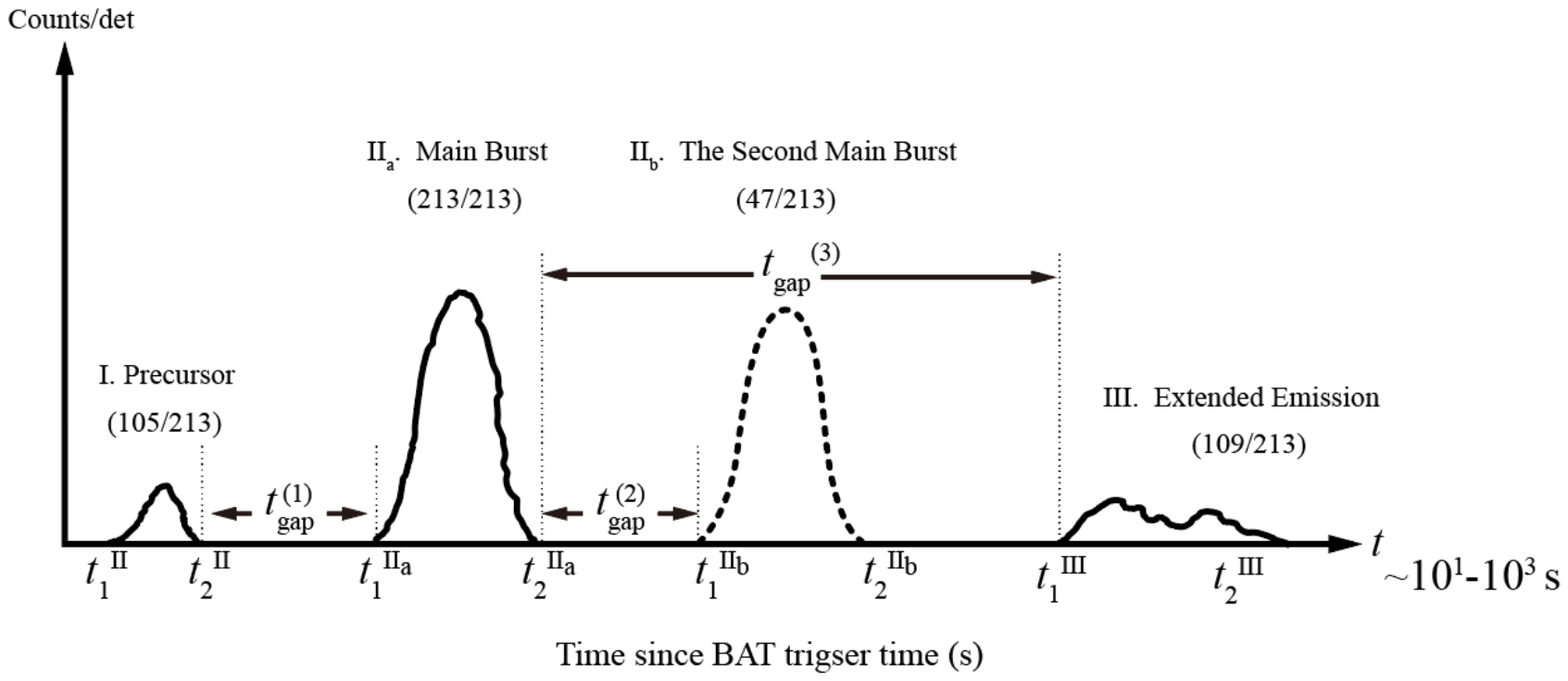}
\caption{Synthetic schematic light curve illustrating the prompt-emission temporal morphology of \emph{Swift}-BAT GRBs with measured redshift. Four components are identified. I. \emph{Precursor}, a weaker early-time emission episode. IIa. \emph{Main burst}, the first main burst emission episode. IIb. \emph{The second main burst}, a subsequent main burst emission episode with an intensity comparable to IIa, separated from IIa by a clear quiescent gap. III. \emph{Extended emission}, a late and weaker extended emission tail. Quiescent gaps are denoted as $ t_{\mathrm{gap}}^{(1)}$ between the precursor and the main burst, $t_{\mathrm{gap}}^{(2)}$ between the first and the second main bursts, and $t_{\mathrm{gap}}^{(3)}$ between the main burst and the extended emission. The time axis is the observer-frame time in seconds since the BAT trigger.}
\label{fig:schematic}
\end{figure*}

To establish a clear time-domain framework for subsequent statistical analysis, we decompose the prompt-emission light curves of all \emph{Swift}-BAT GRBs with known redshift into four morphological components, as illustrated in Figure~\ref{fig:schematic}. {\bf Component~I} is an early, faint precursor, usually consisting of a single pulse, although in a few cases it shows two or more pulses of comparable intensity. It typically occurs within a few seconds after the trigger, while in some weaker precursors, it may precede the trigger time by several to tens of seconds. 
{\bf Component~IIa} represents the first main burst, typically composed of multi-peak overlapping pulses. In some cases, it appears as a single-pulse event or as several separated pulses with short quiescent gaps. {\bf Component~IIb} is a second main burst with an intensity comparable to IIa, separated from it by a quiescent interval. A small fraction of events exhibit a third or more such main episodes. {\bf Component~III} corresponds to the extended emission, a weaker later-time emission tail that may appear as a stretched decay or as two or more short, low-intensity pulses. We define three quiescent intervals. $t_{\mathrm{gap}}^{(1)}$ between the precursor and IIa, $t_{\mathrm{gap}}^{(2)}$ between IIa and IIb, and $t_{\mathrm{gap}}^{(3)}$ between IIa and III. Observationally, $t_{\mathrm{gap}}^{(1)}$ typically ranges from $\sim$ 0 to a few tens seconds, with rare cases $\gtrsim 10^{2}$ s, $t_{\mathrm{gap}}^{(2)}$ is typically ranges from $\sim$ 0 to a few seconds, and $t_{\mathrm{gap}}^{(3)}$ covers a similar range, occasionally extending to $\gtrsim 10^{2}$ seconds. Not all components are necessarily present in a single burst. This framework provides a consistent temporal reference for comparing sub-bursts properties such as hardness ratio, spectral lag, and minimum variability timescale in the analyses that follow.

In this section, we present a comparative analysis of the prompt emission properties for the precursors ($G_1$) and main bursts ($G_2$) within the selected PE+ME GRB sample. We focus on four key observables that are widely utilized in GRB classification: the $T_{90}$ duration (and related timing properties), the spectral hardness ratio, the minimum variability timescale, and the spectral lag ($\tau_{\rm lag}$). Our primary goal is to determine whether the precursors ($G_1$) and main bursts ($G_2$) of a given burst exhibit values that are internally consistent with both being Type II (collapsar-like), or if one episode falls robustly into the Type I (merger-like) regime. Such a difference would imply a change in the physical central engine or emission mechanism between the two distinct phases. For context, we reference the established bivariate distribution of long versus short bursts drawn from the literature. When appropriate, we display the location of our individual $G_1$ and $G_2$ measurements relative to these distributions, often utilizing scatter plots overlaid with the $1\sigma$ and $2\sigma$ confidence ellipses for each canonical class, as defined in Section \ref{sec:BND}.

\subsection{Duration and Temporal Structure} \label{sec:duration}

\begin{figure*}
\includegraphics[angle=0,scale=0.32]{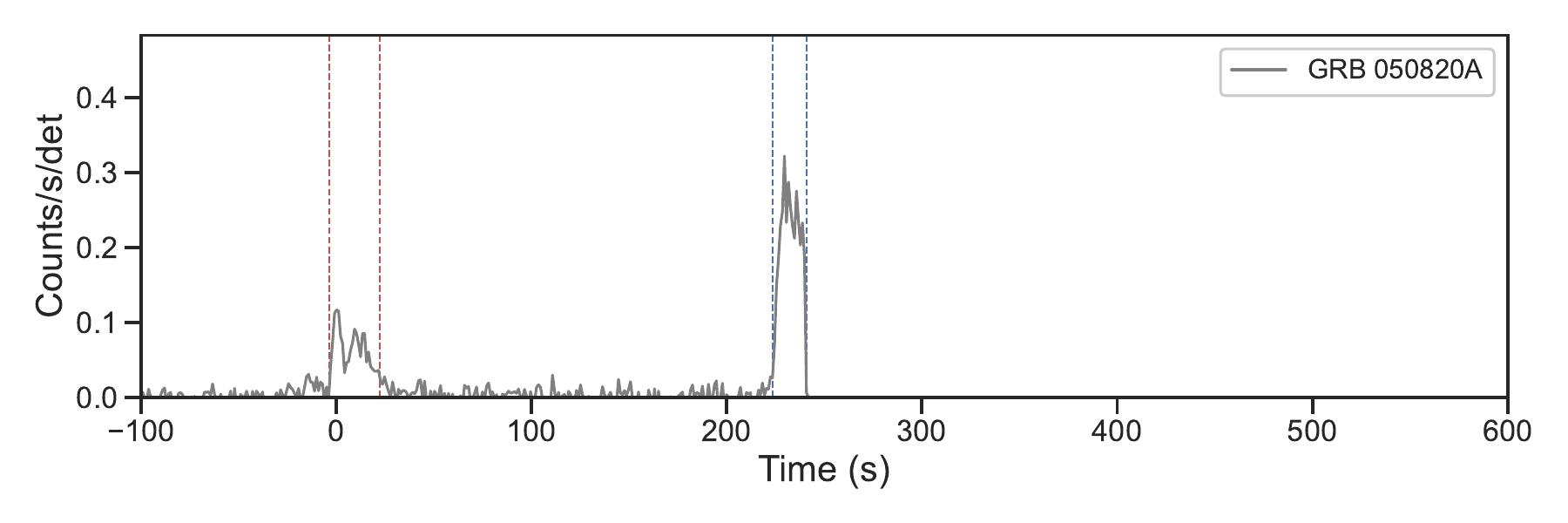}
\includegraphics[angle=0,scale=0.32]{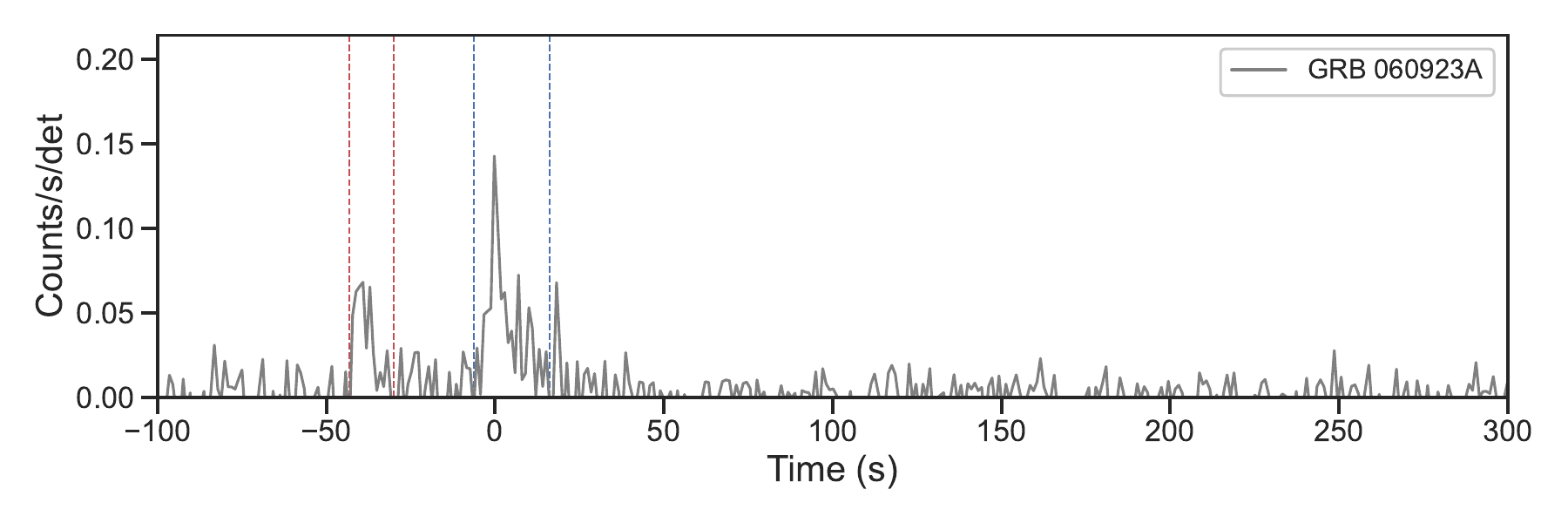}
\includegraphics[angle=0,scale=0.32]{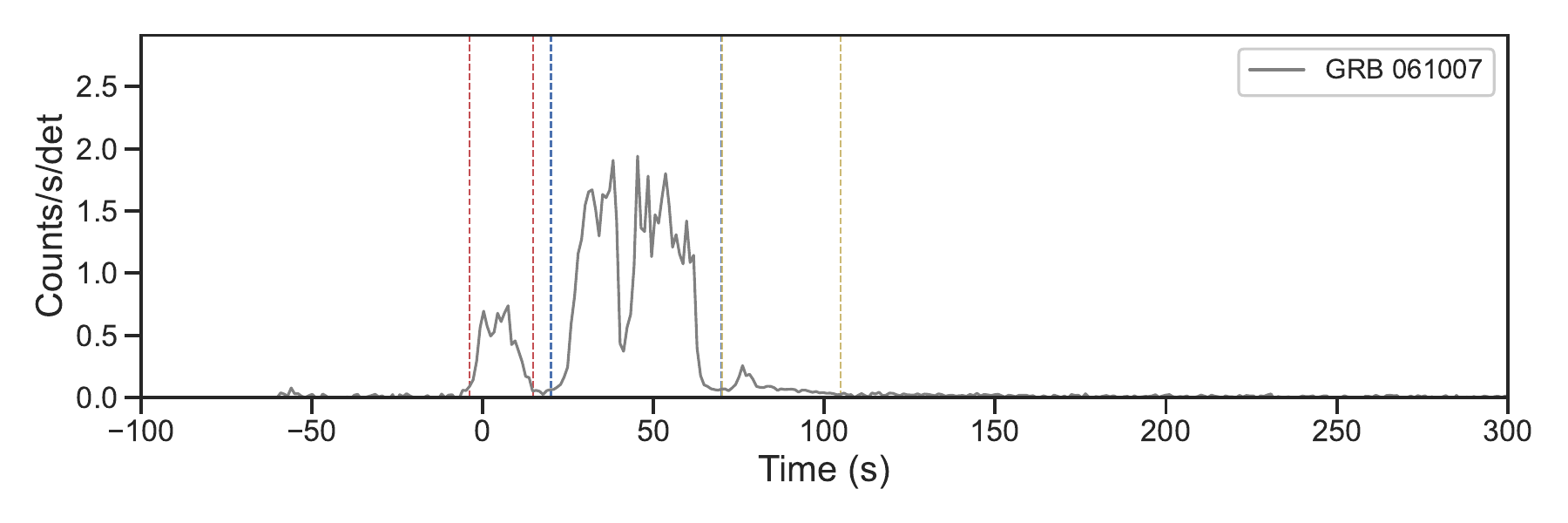}
\includegraphics[angle=0,scale=0.32]{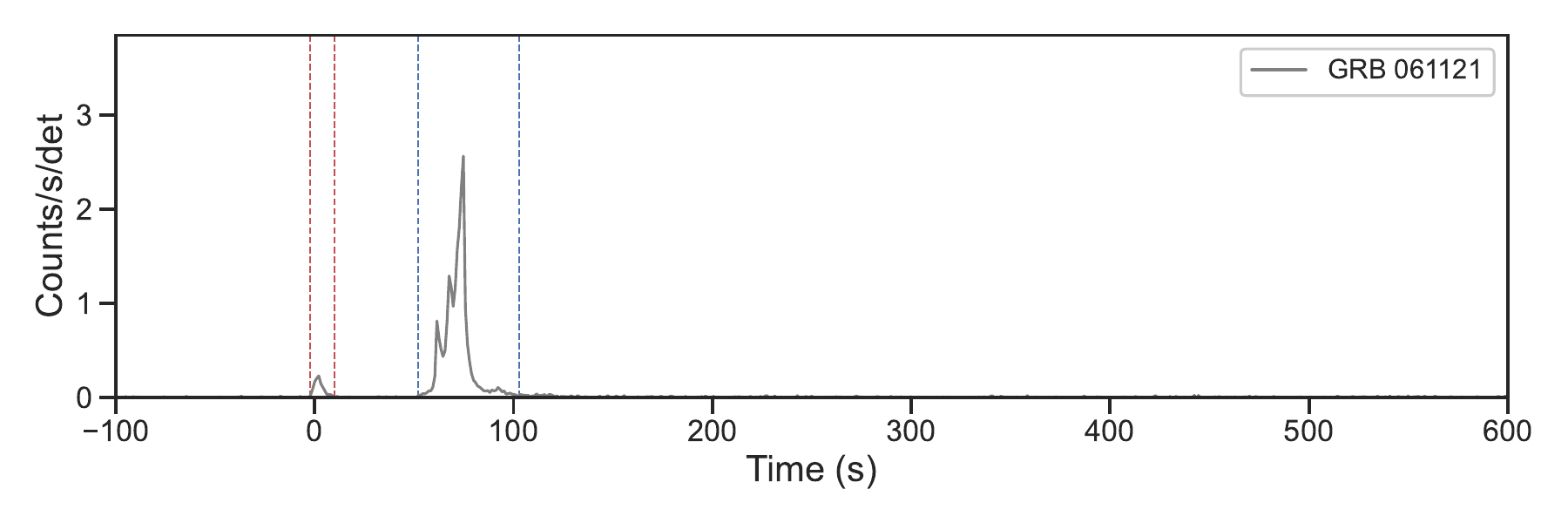}
\includegraphics[angle=0,scale=0.32]{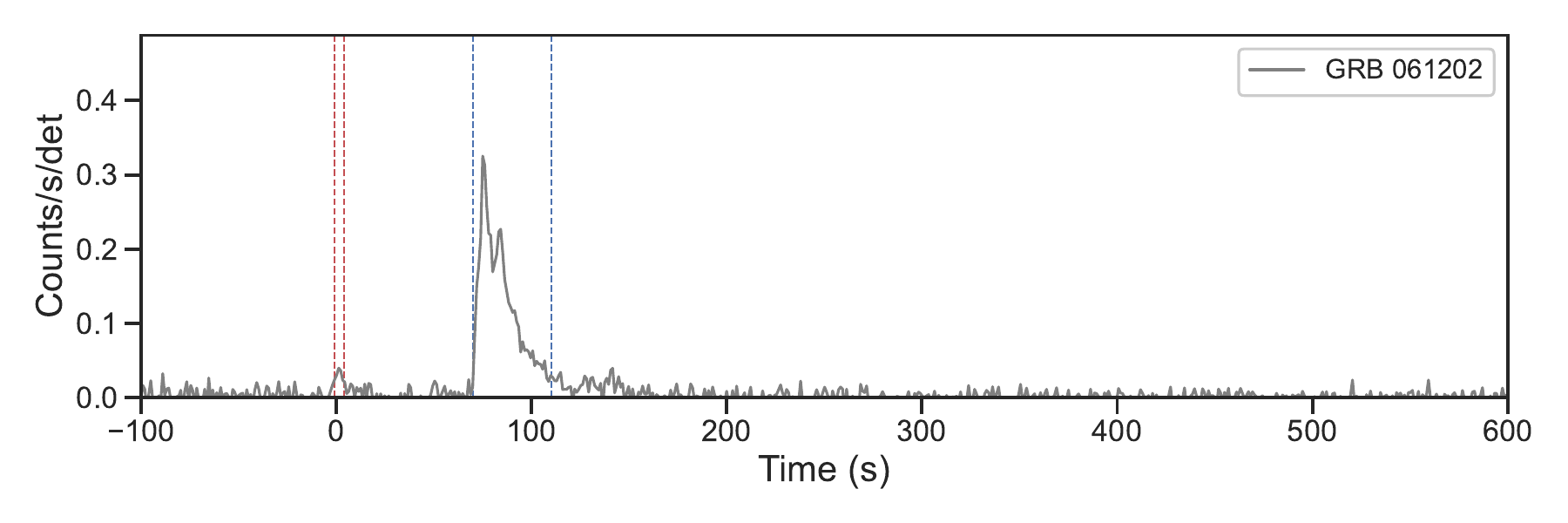}
\includegraphics[angle=0,scale=0.32]{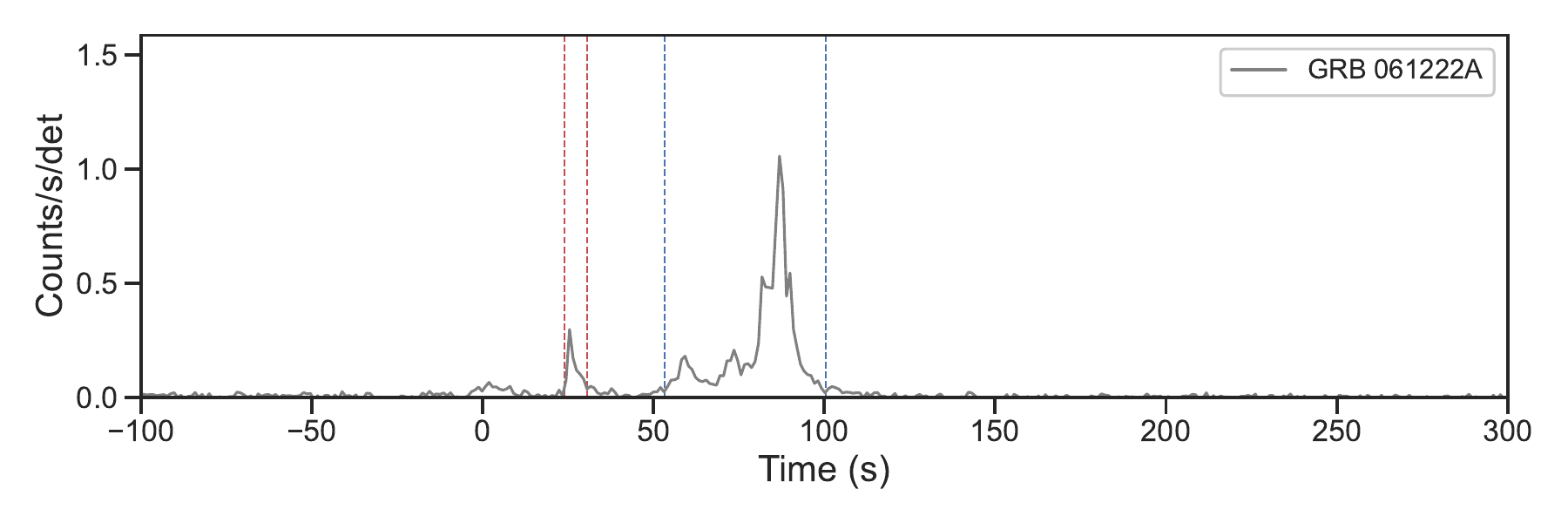}
\includegraphics[angle=0,scale=0.32]{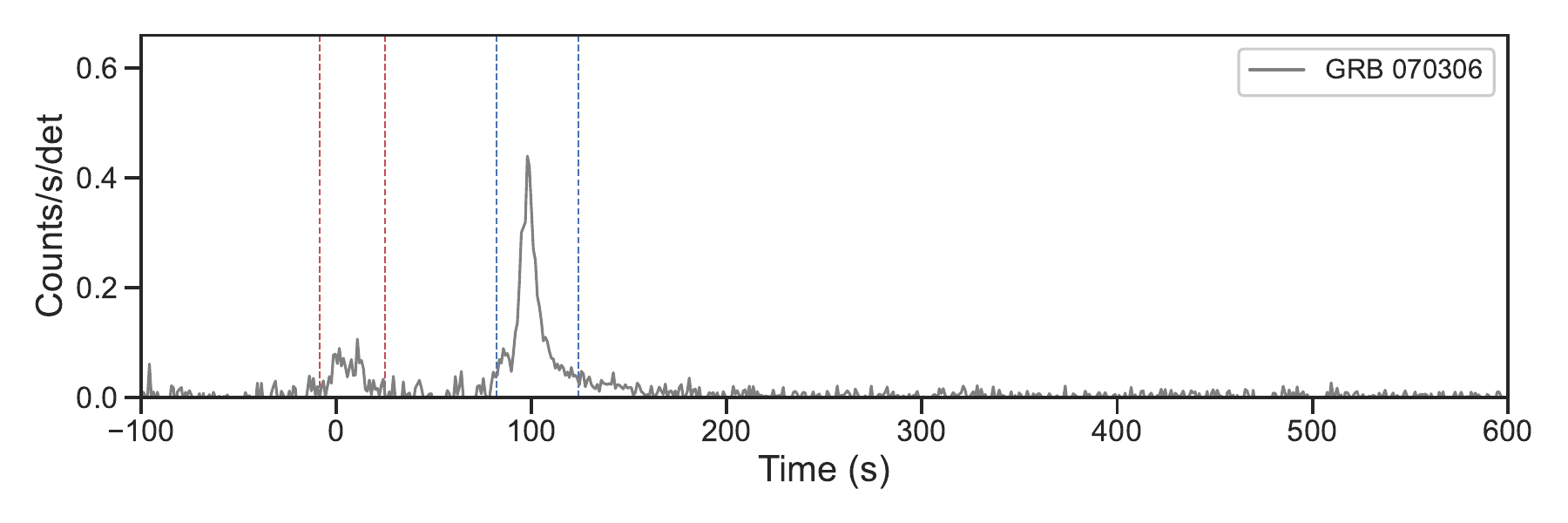}
\includegraphics[angle=0,scale=0.32]{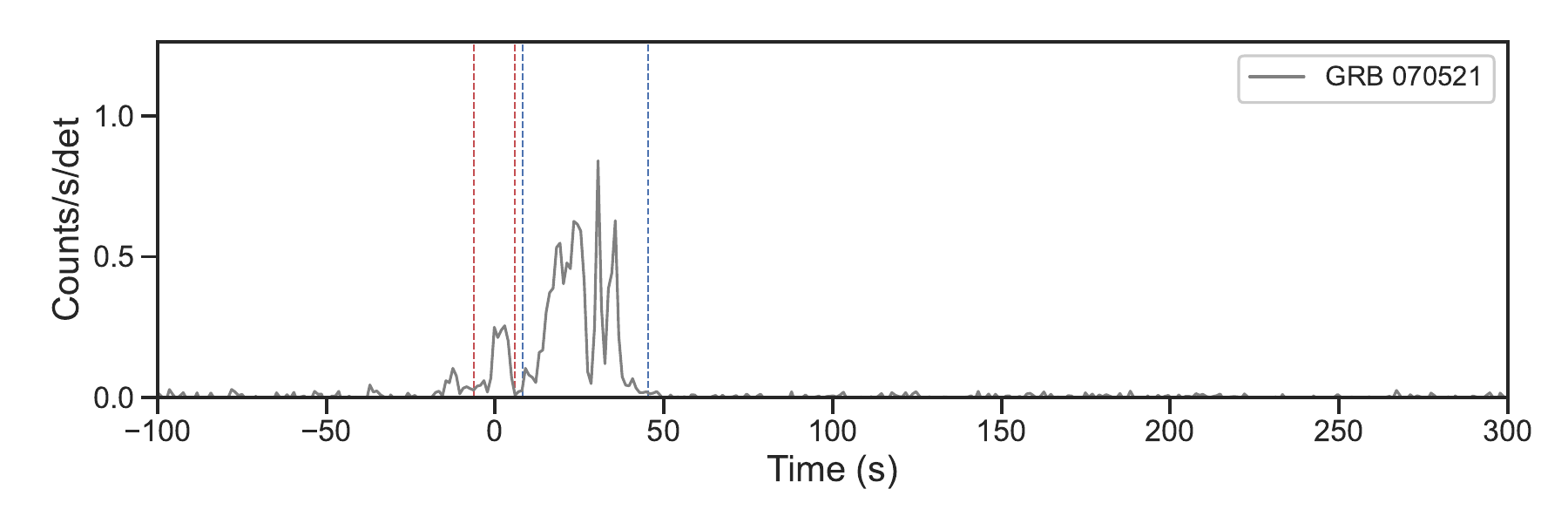}
\includegraphics[angle=0,scale=0.32]{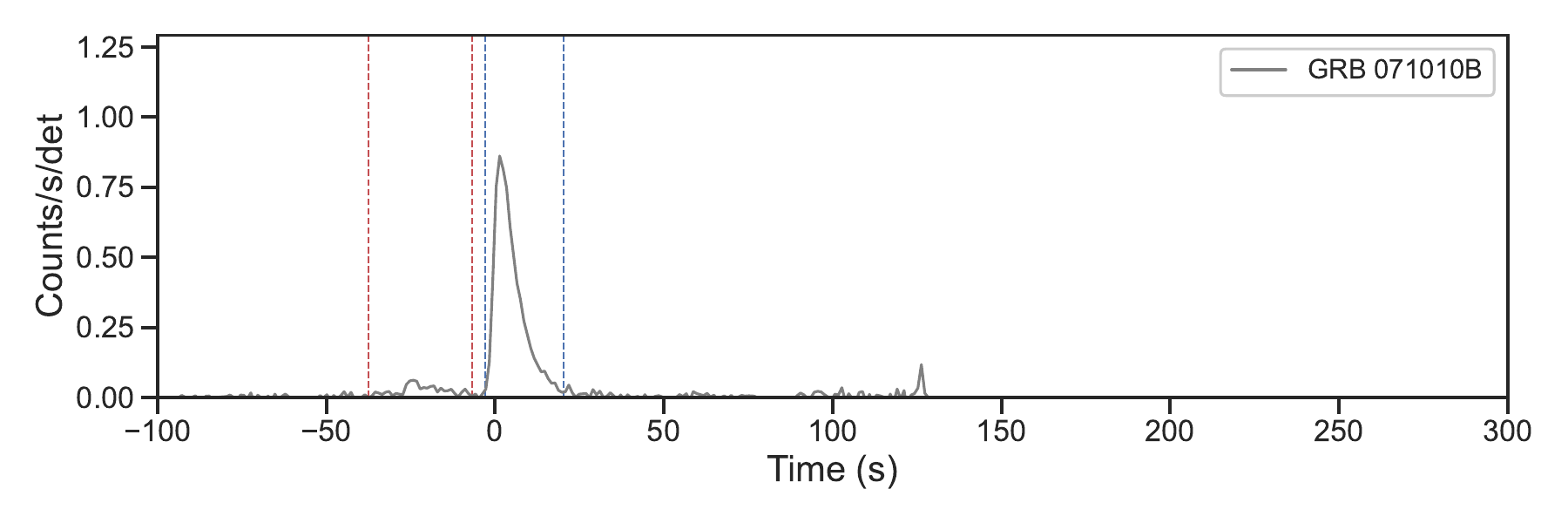}
\includegraphics[angle=0,scale=0.32]{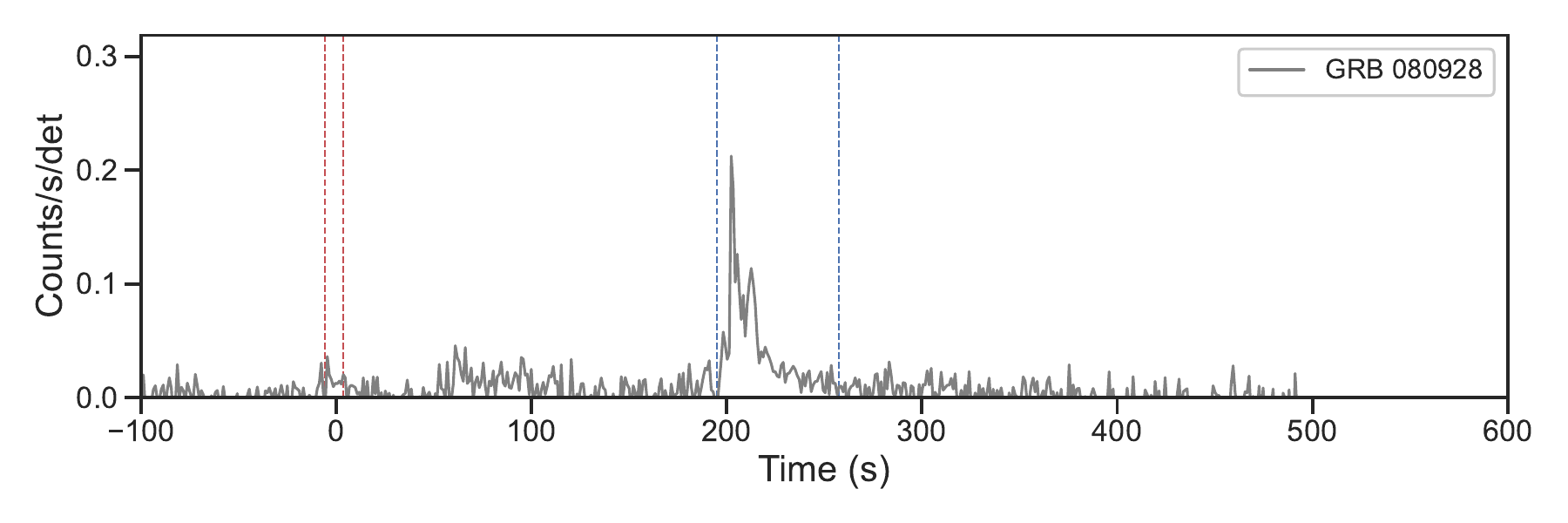}
\caption{Mask-weighted \emph{Swift}/BAT light curves for the GRBs in our sample. Each panel shows a prompt emission consisting of a precursor ($G_1$, indicated by red dashed lines) followed by a main emission ($G_2$, blue dashed lines). The quiescent interval between the two components is clearly visible and corresponds to a return to background-level count rates. These double-pulse structures define our Gold sample and motivate the statistical analyses in this work.}
\label{fig:lc_sample}
\end{figure*}

We begin our analysis by examining the timing properties of the precursor ($G_1$) and main emission ($G_2$) pulses. By construction, all GRBs in our PE+ME sample exhibit a well-defined two-episode structure separated by a quiescent interval during which the count rate returns to background levels (see Figure~\ref{fig:lc_sample}).

Table~\ref{tab:pulse} summarizes the \emph{episode–resolved} timing properties for the two prompt emission sub–bursts ($G_{1}$ and $G_{2}$) of each GRB in our sample. For every episode, we list the BAT time–bin resolution (ms), the determined start and end times that define the analysis time window ($t_{1} \sim t_{2}$), the total pulse width ($t_{\rm pulse}$), the peak time $t_{\rm p}$, and the mask–weighted peak count rate $F_{\rm p}$ (with $1\sigma$ errors). These tabulated quantities rigorously define the temporal windows used for all subsequent analysis in this paper. They provide the necessary inputs for the following analysis, including the crucial quiescent–gap estimate $t_{\rm gap}=t_{1}(G_{2})-t_{2}(G_{1})$ and the episode–level comparisons of hardness, variability, and lag. The rightmost column records the episode label (PE or ME) used in subsequent figures and tables. 

The precursors ($G_{1}$) in our sample exhibit $T_{90}$ durations ranging from $2.8^{+1.0}_{-0.8}$ s (GRB~121128A) to $35.1^{+8.2}_{-6.1}$ s (GRB~070306), with a median of $\sim9.6$ s. The main bursts ($G_2$) span a wider range, from $4.6$ s to $81.4$ s, with a median of $\sim22.8$ s (Table~\ref{tab:t90}). While precursors are typically shorter than their main burst counterparts by a factor of a few, the sample exhibits considerable diversity, for instance, GRB~050820A has a precursor duration that is nearly 1.5 times longer than its main burst. Crucially, all measured precursor $T_{90}$ values exceed the canonical 2-second threshold for long GRBs, consistent with a priori Type II classification for every individual pulse in our sample. Only GRB~121128A approaches the long/short boundary with a $T_{90}^{\rm PE} = 2.8^{+1.0}_{-0.8}$ s. Therefore,  based on purely on the $T_{90}$ criterion, we find no event in our PE+ME sample where the precursor is classified as Type I while the main burst is Type II.

The quiescent intervals ($\Delta t_{\rm gap}$) separating $G_1$ and $G_2$ range from $\sim 0.2$ s (GRB~090618A) to an extended $201.2 \pm 3.4$ s (GRB~050820A). The median gap is $\sim17$ s in the observer frame, corresponding to the rest-frame median of $\sim5.7$ s (see Table~\ref{tab:Compara}). We find no statistically significant correlation between $\Delta t_{\rm gap}$ and either $T_{90}^{\rm PE}$ or $T_{90}^{\rm ME}$, suggesting that the temporal separation time is not merely a consequence of adjacent pulse durations.

We calculate the ratios of the precursor emission to the main burst emission for duration $T_{90}$, peak flux $F_{\rm p}$, hardness ratio, MVT, and spectral lag $\tau$. These characteristic ratios are defined as:
\begin{eqnarray}
 R_{t_{90}} &\equiv & \frac{t_{90}^{\mathrm{PE}}}{t_{90}^{\mathrm{ME}}},\\
 R_{F_{\rm p}} &\equiv& \frac{F_{\rm p}^{\mathrm{PE}}}{F_{\rm p}^{\mathrm{ME}}},\\
 R_{\rm HR} &\equiv& \frac{HR^{\mathrm{PE}}}{HR^{\mathrm{ME}}},\\
 R_{\rm MVT} &\equiv& \frac{MVT^{\mathrm{PE}}}{MVT^{\mathrm{ME}}},\\
 R_{\tau} &\equiv&\frac{\tau^{\mathrm{PE}}}{\tau^{\mathrm{ME}}}.
\end{eqnarray}
Table \ref{tab:Compara} summarizes the resulting values.

\subsection{Hardness Ratio} \label{sec:hardness_ratio}

Spectral hardness is a widely used diagnostic in GRB classification, with short GRBs characteristically exhibiting harder spectra than long GRBs. To quantitatively assess the spectral properties of each sub-burst, we utilized the count-based hardness ratio. This ratio is defined between the 50–100 keV and 25–50 keV energy bands of the \emph{Swift}-BAT:
\begin{equation}
\mathrm{HR} = \frac{C_{50-100}}{C_{25-50}}.
\end{equation}
where $C$ denotes the background-subtracted fluence in each band (integrated over the specific sub-burst interval $[t_1, t_2]$ in each respective band). This band selection is optimal as it falls within the BAT's optimal sensitivity range and effectively divides the energy spectrum near the typical peak energy ($E_{\rm p}$) of long GRBs.

The calculated HRs for the precursor and main burst components of each GRB are presented in Table~\ref{tab:HR}. Precursors ($G_1$) exhibit HRs ranging from $0.16$ to $1.16$, with a median value of $\sim0.70$. In contrast, the main bursts ($G_2$) show a slightly narrower range of $0.52$ to $1.09$, with a median of $\sim0.80$. Within the quoted statistical uncertainties, the two components possess comparable spectral hardness across our sample. Notably, none of the observed precursor exhibits a hardness ratio consistent with the canonical Type I (short) population, which is typically characterized by HR $> 1.5$ in the BAT channels.

\begin{figure*}
\includegraphics[width=0.5\textwidth]{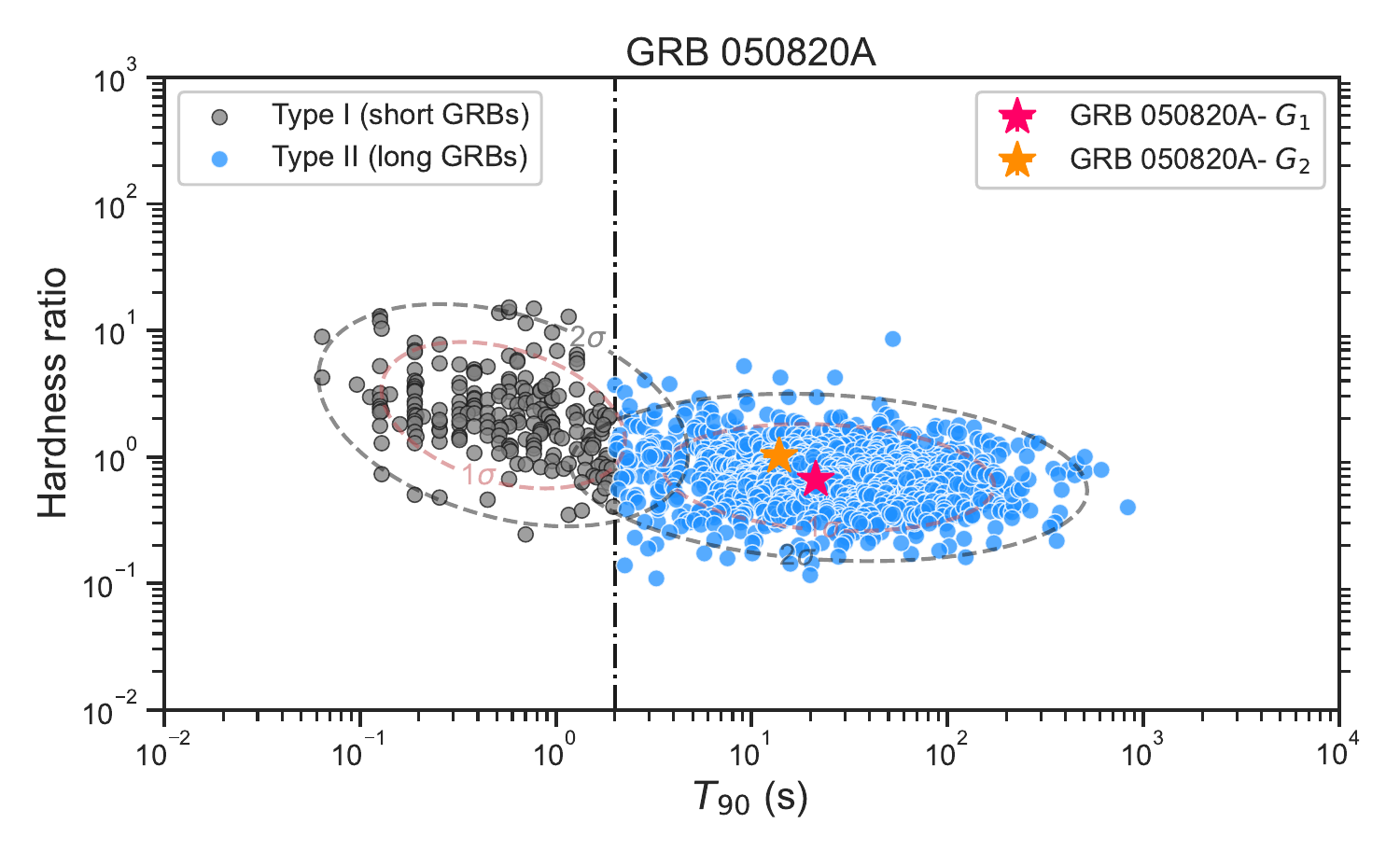}
\includegraphics[width=0.5\textwidth]{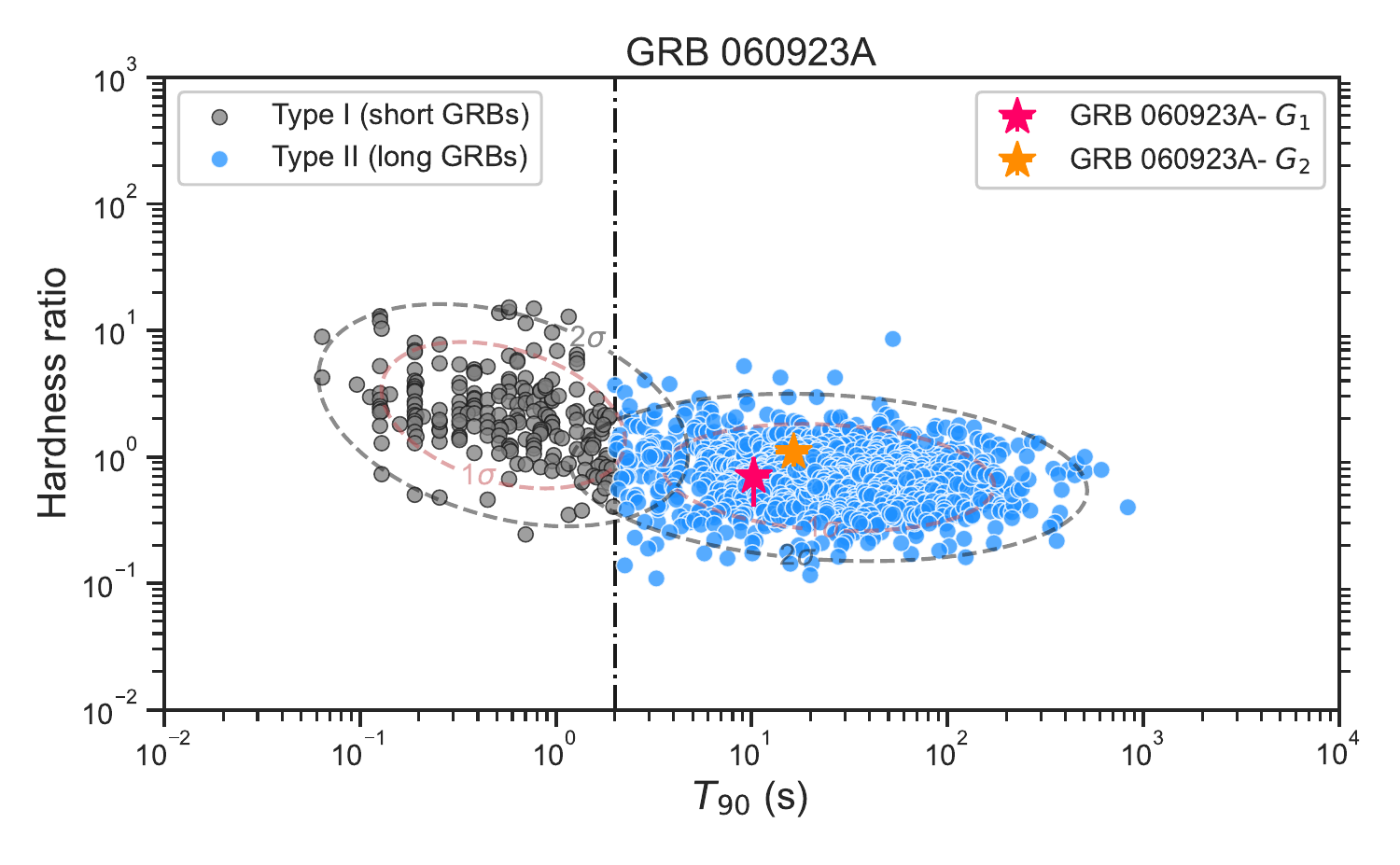}
\includegraphics[width=0.5\textwidth]{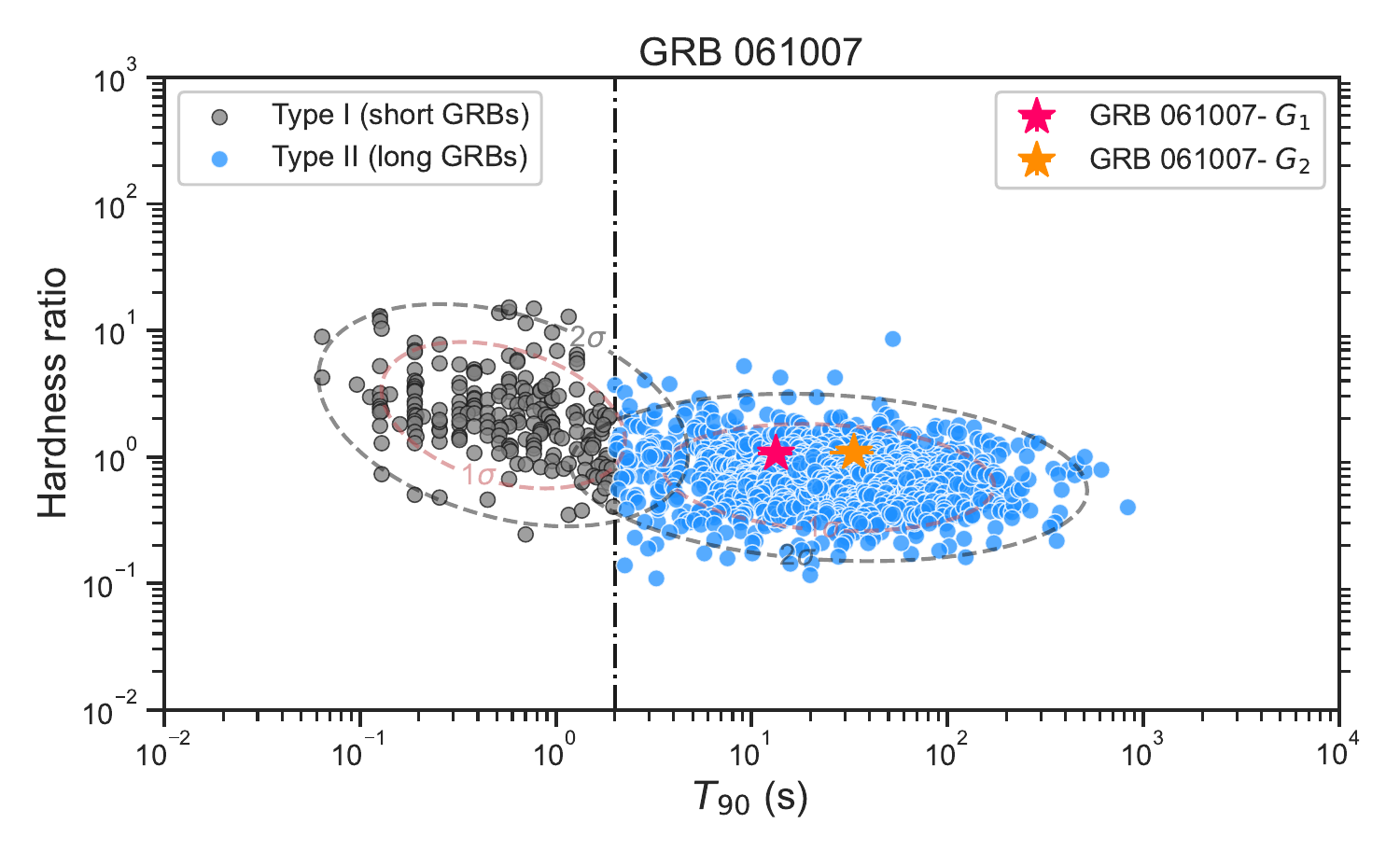}
\includegraphics[width=0.5\textwidth]{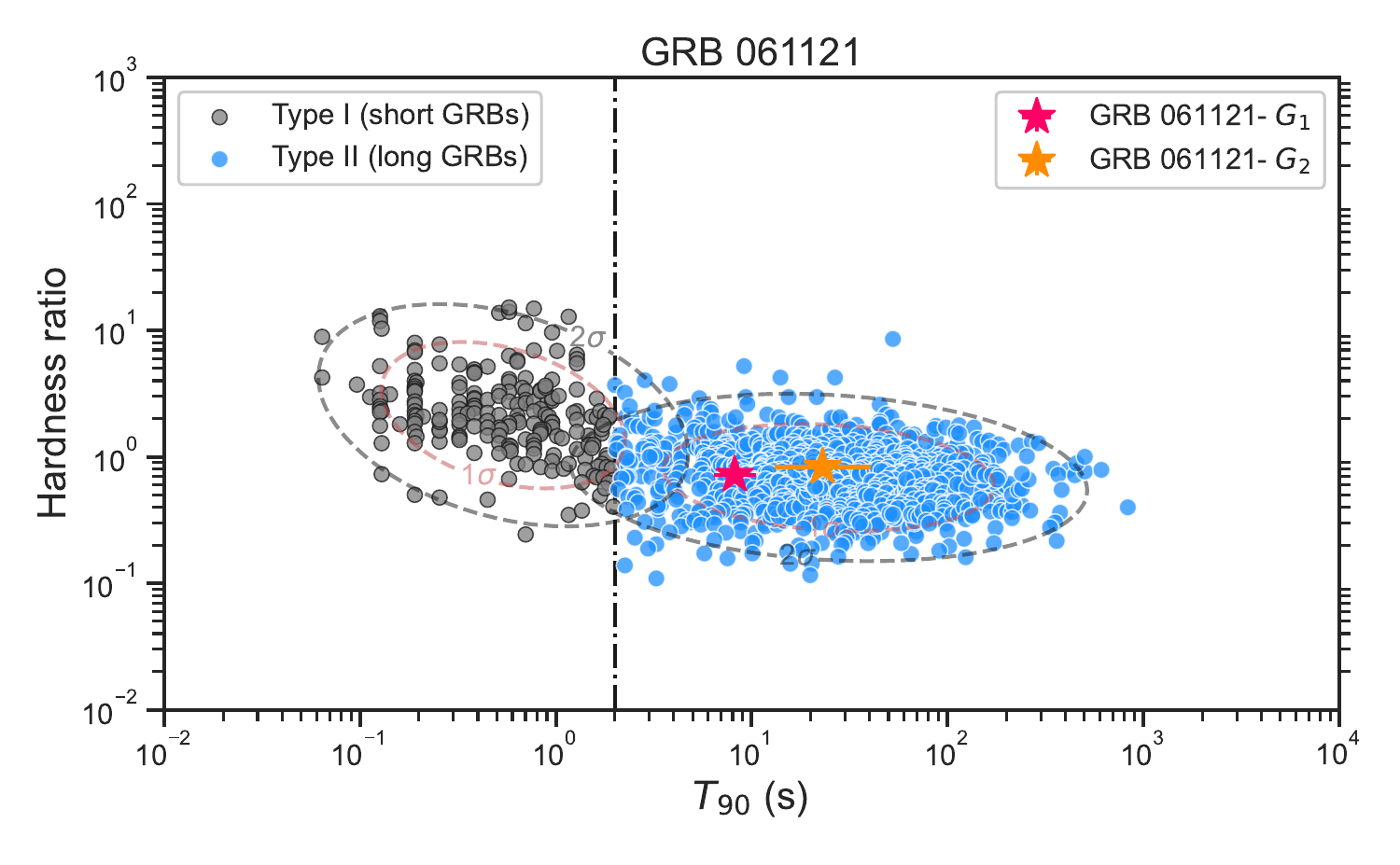}
\includegraphics[width=0.5\textwidth]{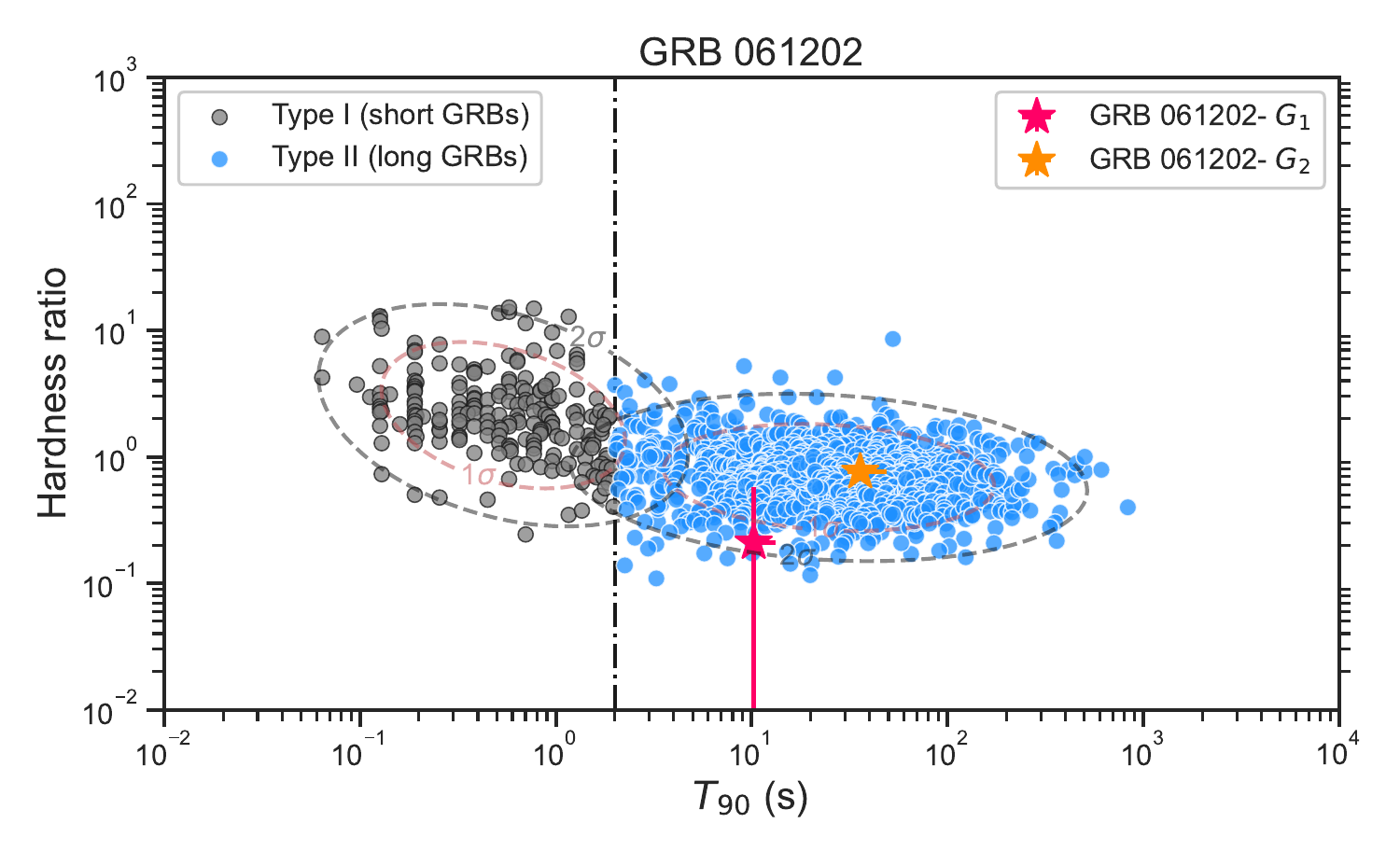}
\includegraphics[width=0.5\textwidth]{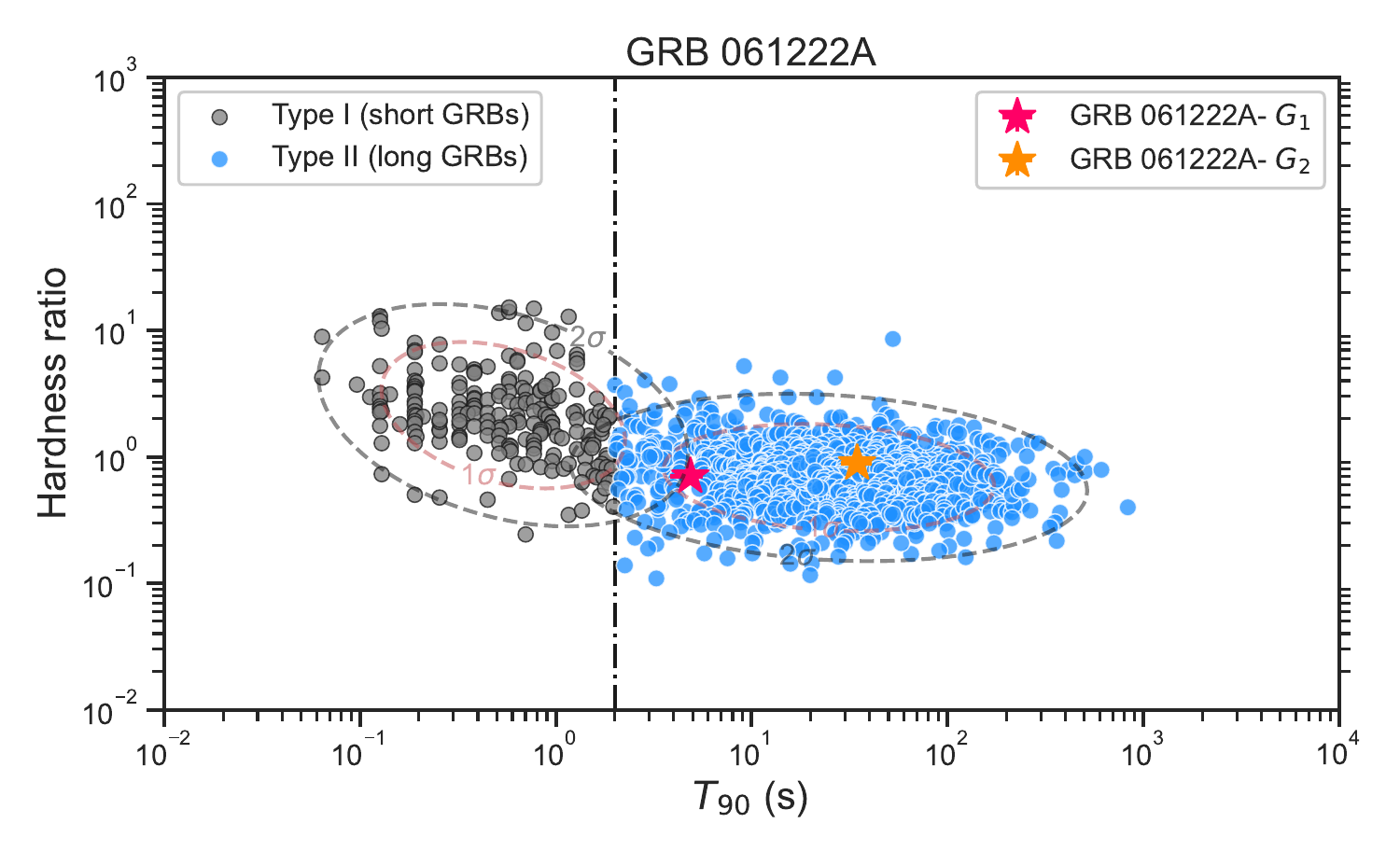}
\caption{Duration–hardness ratio diagrams for eleven representative GRBs in our sample. Each panel overlays the precursor ($G_1$, magenta star) and main emission ($G_2$, orange star) atop the broader \emph{Swift} GRB population from \citet{Horvath2010}, color-coded by Type I (short; black) and Type II (long; blue). The 1$\sigma$ ellipses correspond to the bivariate normal fits for each class (see Section~\ref{sec:BND}). Both components of each GRB lie within the long GRB region, showing comparable spectral hardness consistent with collapsar-type events.}
\label{fig:HR}
\end{figure*}

Figure~\ref{fig:HR} illustrates the traditional duration–hardness diagram for 22 representative GRBs in our sample. In each panel, the precursor ($G_1$; magenta star) and main burst ($G_2$; orange star) are plotted relative to the broader \emph{Swift} GRB population. Crucially, both components cluster well within the 1$\sigma$ ellipse of the Type II (long GRB) distribution. The only minor deviation is observed in GRB~121128A, whose precursor lies near the $T_{90}=2$ s boundary and presents a soft HR of $0.16\pm0.34$, however, even this soft spectrum does not encroach upon the short GRB regime. This similarity in hardness ratio between $G_1$ and $G_2$ significantly strengthens the hypothesis that both episodes arise from the same underlying physical mechanism. For several bright events (e.g., GRB~050820A and GRB~140512A), detailed time-resolved spectral analysis further supports this conclusion. 

In summary, our spectral hardness analysis yields no evidence of significant differences between the precursor and main emission episodes in our sample. Both components consistently fall within the expected hardness range for collapsar-like (Type II) GRBs. This result complements the duration analysis presented in Section~\ref{sec:duration} and strongly supports a unified physical origin for both sub-bursts.

\subsection{Minimum Variability Timescale} \label{sec:MVT}

The minimum variability timescale ($\Delta t_{\min}$) serves as a robust proxy for the smallest emitting region or fastest energy dissipation timescale in a gamma-ray burst. Empirically, $\Delta t_{\min}$ is a strong classifier: short GRBs exhibit rapid variability, typically $\Delta t_{\min}\sim$10–30 ms, while long GRBs are characterized by longer timescales of 100–300 ms \citep{Golkhou2015}. We employed the Haar wavelet method \citep{Golkhou2014} on background-subtracted \emph{Swift}/BAT light curves, binned down to 1 ms where the signal-to-noise allowed, to rigorously extract $\Delta t_{\min}$ for each sub-burst in our sample.

Table~\ref{tab:MVT} lists the minimum variability timescales (MVTs) derived for the precursor ($G_{1}$) and main emission ($G_{2}$) of each GRB in the sample, measured using the Haar wavelet method applied to the \emph{Swift}/BAT light curves. In all cases, both episodes fall within the long–GRB (Type~II) classification on the $T_{90}$–MVT plane. The results, compiled in Table~\ref{tab:MVT}, reveal a clear and systematic trend: precursors generally exhibit significantly longer MVTs than their associated main burst emissions. Across the entire sample, $\Delta t_{\min}$ for precursors $G_1$ span from $\sim36$ ms up to $\sim4312$ ms, whereas main emission pulses ($G_2$) show much shorter timescales, ranging from $\sim33$ ms to $\sim2476$ ms. In the majority of cases, the ratio $\Delta t_{\min}^{\rm PE} / \Delta t_{\min}^{\rm ME}$ ranges from 2 to 10, quantitatively indicating a significantly smoother temporal structure inherent to the precursor pulses. The median ${\rm MVT}_{G_{1}}$ is roughly an order of magnitude greater than ${\rm MVT}_{G_{2}}$, indicating that precursor emission is temporally smoother and less variable. For instance, GRB~061121 shows ${\rm MVT}_{G_{1}}\!\simeq\!411\pm43$~ms compared with ${\rm MVT}_{G_{2}}\!\simeq\!50\pm17$~ms, while GRB~070306 exhibits an even larger contrast ($2802\pm1288$~ms vs.\ $155\pm24$~ms).

\begin{figure*}
\includegraphics[width=0.5\textwidth]{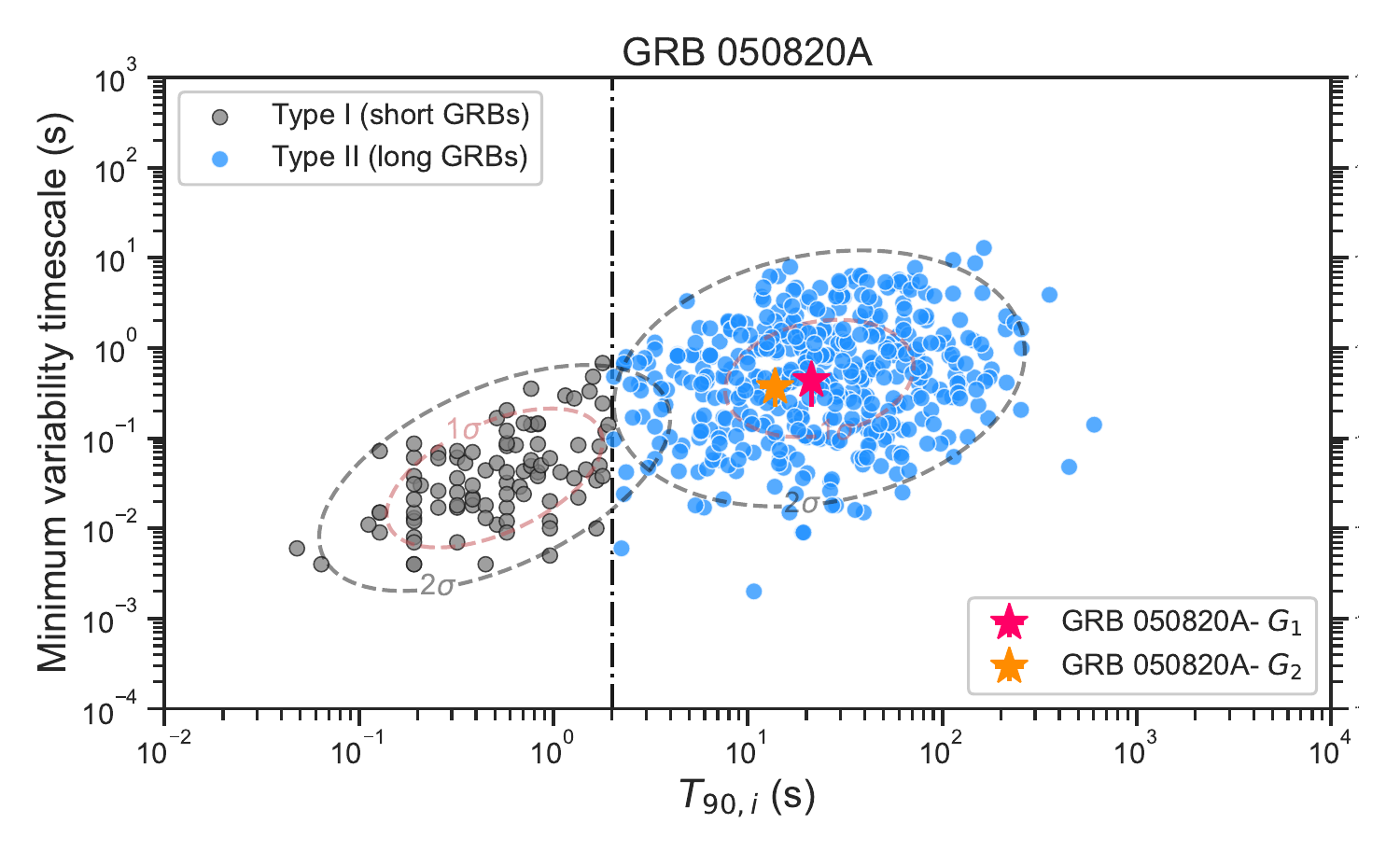}
\includegraphics[width=0.5\textwidth]{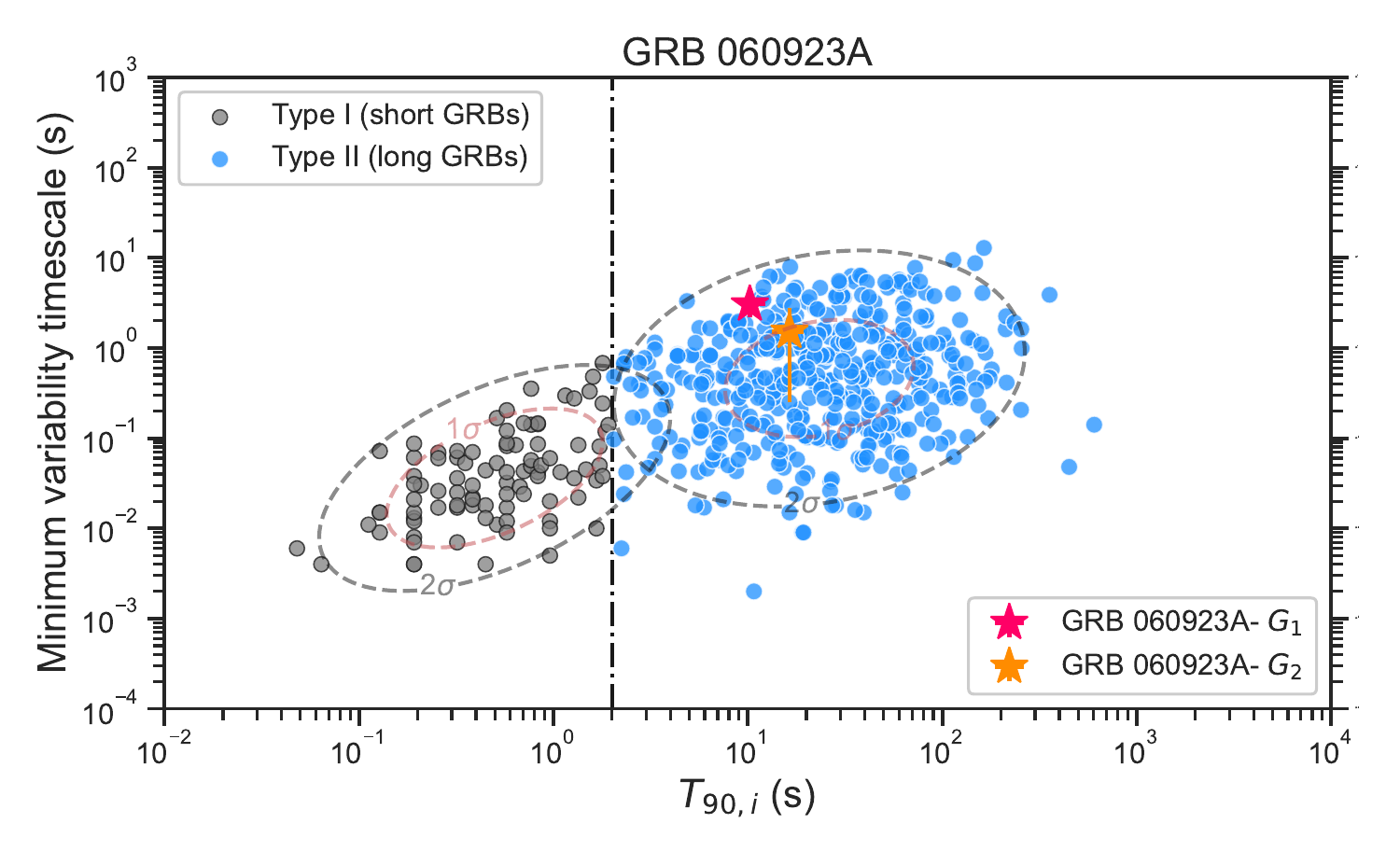}
\includegraphics[width=0.5\textwidth]{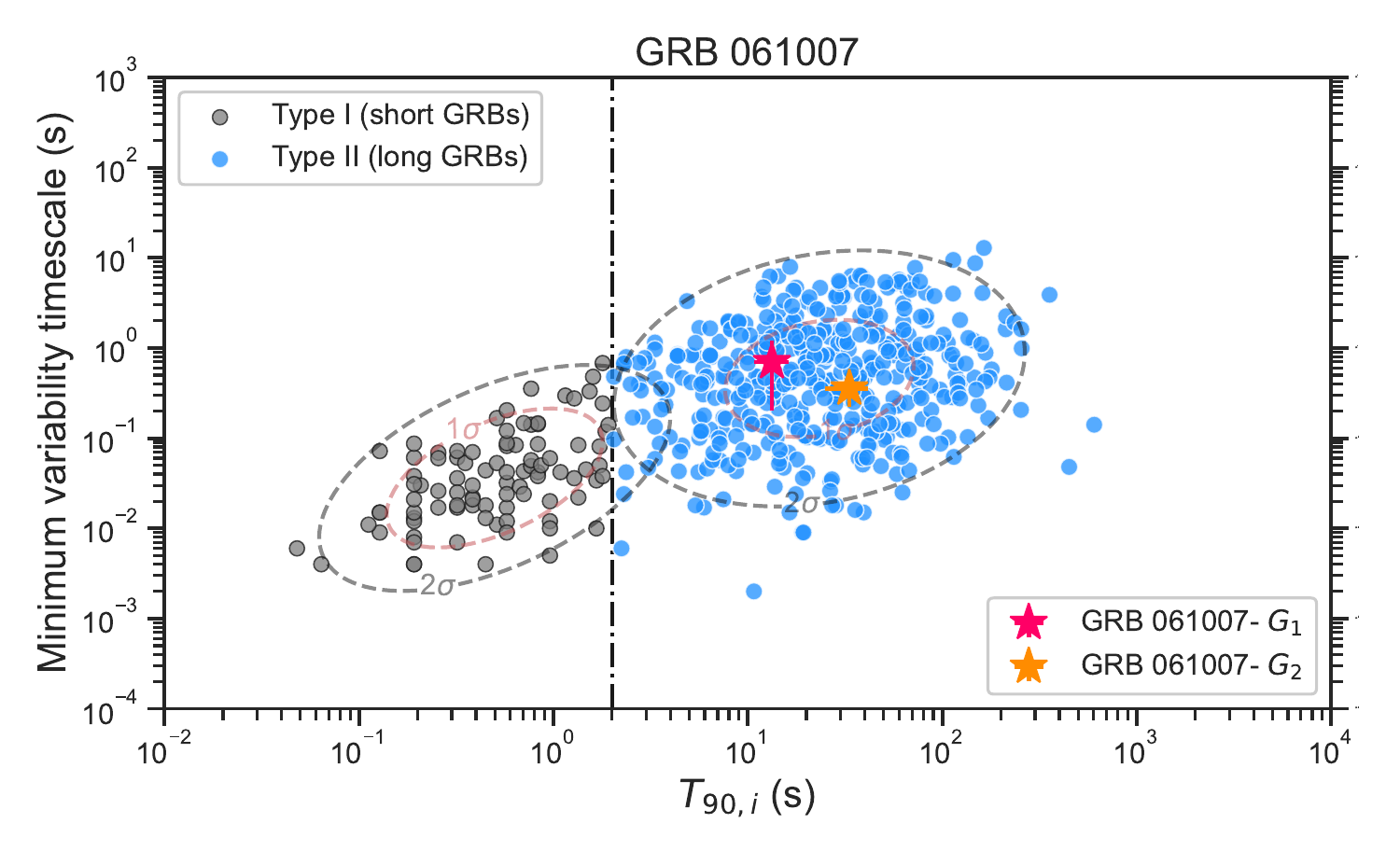}
\includegraphics[width=0.5\textwidth]{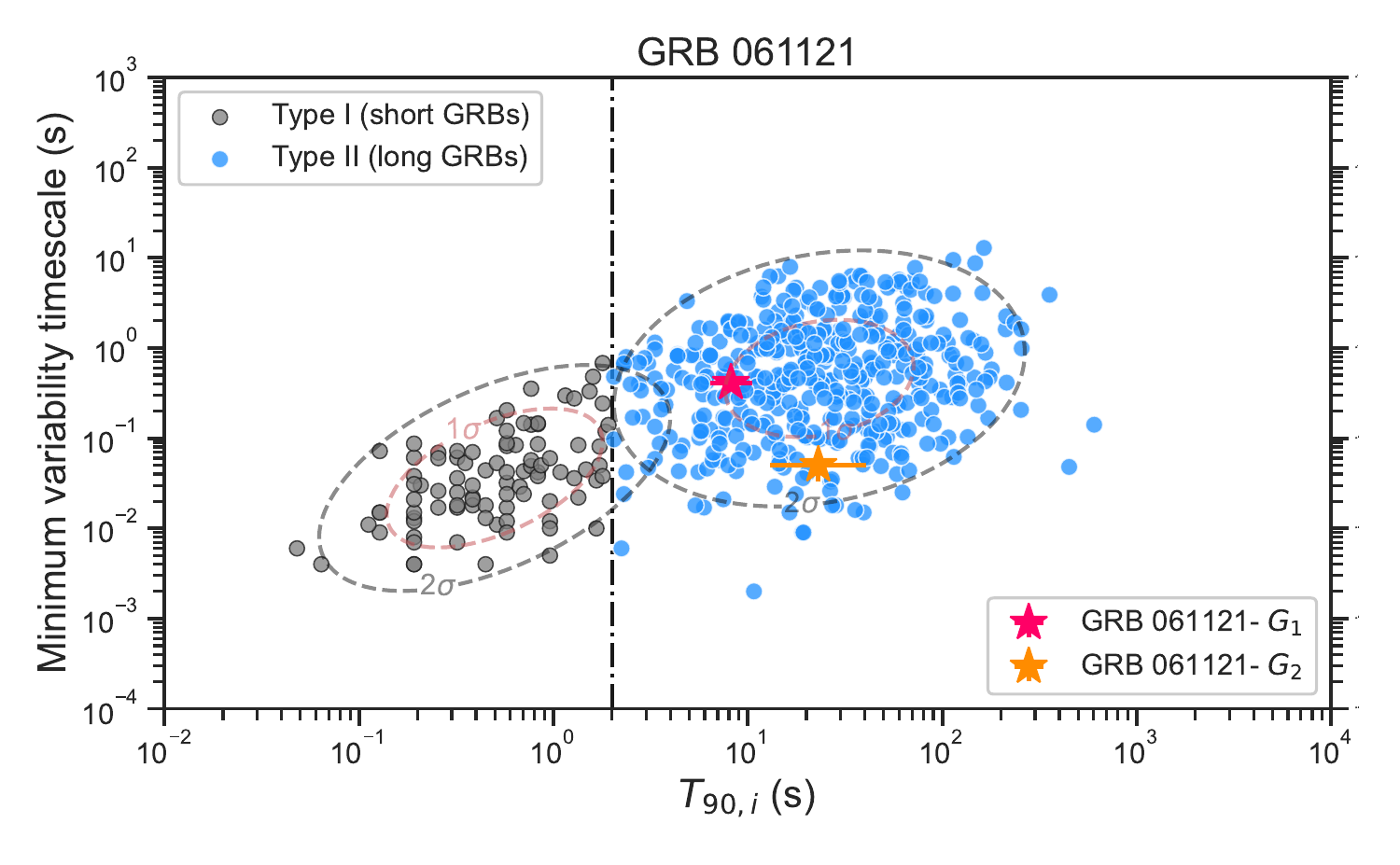}
\includegraphics[width=0.5\textwidth]{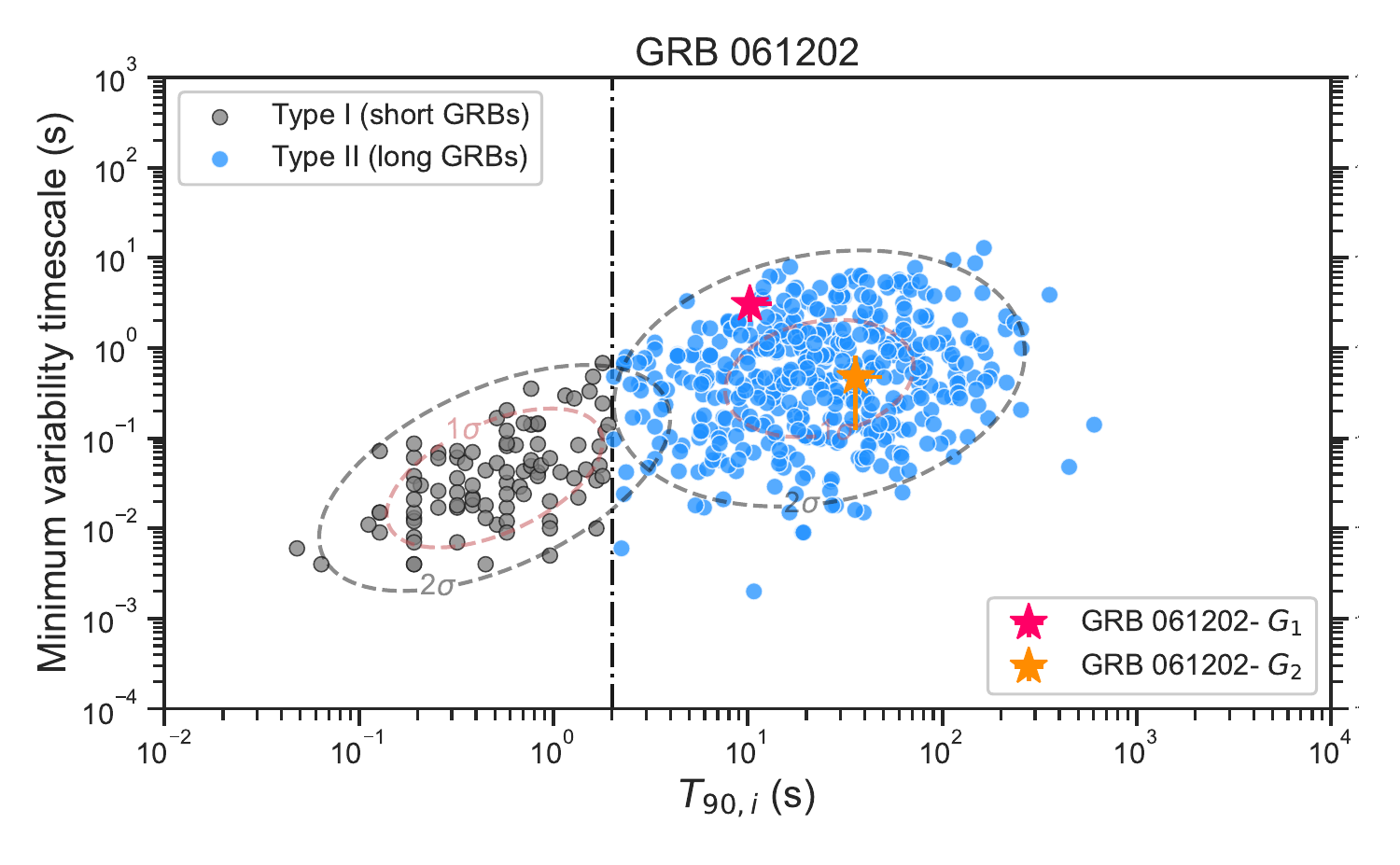}
\includegraphics[width=0.5\textwidth]{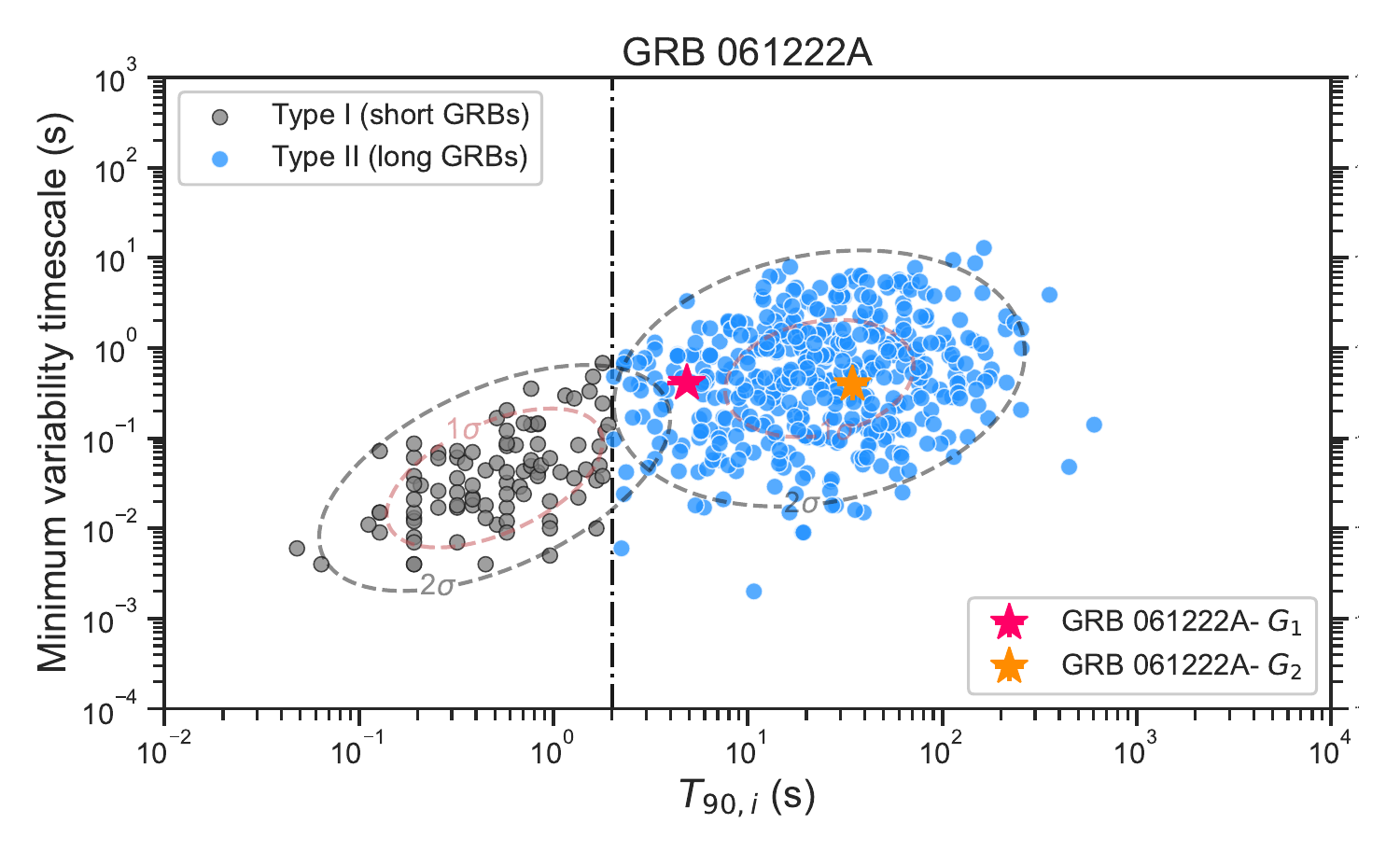}
\caption{Duration ($T_{90}$) versus minimum variability timescale (MVT) for precursor ($G_1$, red star) and main emission ($G_2$, orange star) pulses in the sample. Gray and blue points show the distributions of Type I (short) and Type II (long) GRBs, respectively, from \citet{Golkhou2015}. The ellipses represent 1$\sigma$ clustering regions for each population. While main emissions lie squarely within the Type II region, precursors typically exhibit longer MVTs, indicating smoother temporal structure, yet remain consistent with the collapsar (Type II) population.}
\label{fig:MVT_T90}
\end{figure*}

Figure~\ref{fig:MVT_T90} places these $\Delta t_{\min}$ values within the broader context of GRB classification via the $T_{90}$-$\Delta t_{\min}$ plane. 
Each panel displays the two components ($G_1$ and $G_2$) overplotted on the established distribution of Type I and Type II bursts from \citet{Golkhou2015}. All main emission components ($G_2$) are well-confined within the long GRB cluster, exhibiting typical $\Delta t_{\min}$ values commensurate with their durations. In contrast, precursors ($G_1$) lie at systematically higher $\Delta t_{\min}$ values, with several events (e.g., GRB~061202 and GRB~150323A) placing them on the upper edge of the Type II distribution.

Despite the observed systematic difference, it is crucial to note that none of the precursor MVTs fall into the short-GRB regime, which is typically defined by $\Delta t_{\min} \lesssim 20$ ms. Even the shortest precursor MVT in our sample (GRB~180728A, $\sim33$ ms) remains within the lower edge of long burst population. This result reinforces the conclusion that precursors, while temporally extended and smoother, are inconsistent with a compact-merger central engine. 

In conclusion, the $\Delta t_{\min}$ analysis further validates the consistent picture emerging from duration and hardness ratio: precursor pulses are systematically longer, smoother, and softer than main emissions, yet both sub-bursts remain fully consistent with a Type II (collapsar) classification. 
The physical implications of this systematic MVT difference are discussed in Section~\ref{sec:discussion}.

\subsection{Spectral Lag} \label{sec:lag}

The spectral lag ($\tau_{\rm lag}$) of a GRB quantifies the time delay between the arrival of high-energy and low-energy photons. This observable is a sensitive classification measurement that correlates with peak luminosity (the Norris relation). Long GRBs (Type II) typically exhibit significant positive lags, often in the hundreds of milliseconds, whereas short GRBs (Type I) are generally characterized by negligible or zero lag \citep{Yi2006, Bernardini2015}.

We measured the spectral lag for each sub-burst using the standard cross-correlation function (CCF) method \citep{Cheng1995, Band1997, Norris2000, Ukwatta2010}. Specifically, we compared the background-subtracted light curves between the 25–50 keV (hard band) and 15–25 keV (soft band) BAT channels. Given the relatively short duration of our individual pulses, we utilized the modified CCF formulation introduced by \citet{Band1997}, which is better suited for isolated transient events.

For two discretely sampled light curves $x_i$ (hard band) and $y_i$ (soft band), the modified unnormalized cross-correlation at lag $k$ (in bins) is given by:
\begin{equation}
{\rm CCF}(k \Delta t; x, y)=\frac{\sum_{i={\rm max}(1,1-k)}^{{\rm min}(N,N-k)} x_{i}y_{(i+k)}}{\sqrt{\sum_{i}x_{i}^{2}\sum_{i}y_{i}^{2}}},
\end{equation}
where $\Delta t$ is the time bin duration, $k \Delta t$ represents the time delay $\tau$, and the summation is performed over the overlapping time indices.
$x_{i}$ and $y_{i}$ are the count rates in energy bands $E_{1}$ and $E_{2}$. 
We then determined the spectral lag $\tau_{\rm peak}$ by fitting a smooth function (typically a Gaussian) to the peak of the computed CCF. Positive values of $\tau_{\rm lag}$ indicate that the hard-band photons lead the soft-band photons. Uncertainties on $\tau_{\rm lag}$ were rigorously estimated via a Monte Carlo bootstrap resampling approach, where the standard deviation of the resulting lag distribution was adopted as the $1\sigma$ error.

Table~\ref{tab:lag} presents the measured spectral lags between for the precursor ($G_1$) and main emission ($G_2$) episodes. Positive values indicate that the higher–energy photons (25–50~keV) peak earlier than the softer band, whereas negative lags correspond to soft photons leading the hard emission. Across the sample, both episodes fall within the long–GRB (Type~II) classification by duration, but they show distinct timing behaviors. In most bursts the lag of $G_{2}$ is either comparable to or larger than that of $G_{1}$, indicating stronger hard–to–soft evolution during the extended emission. For example, GRB~050820A shows $(480\pm28)$~ms for $G_{1}$ versus $(640\pm57)$~ms for $G_{2}$, while GRB~061202 changes from $-480\pm28$~ms to $+274\pm17$~ms, demonstrating a reversal in lag sign between the two episodes. Several events exhibit negative lags during one or both episodes (e.g., GRB~061121, GRB~080928, GRB~140512A, and GRB~150323A), implying complex spectral–temporal evolution and possibly multi–zone emission within the prompt phase. The overall pattern—moderate, positive lags for the main pulse and larger or sign–rever.

Our results reveal a striking contrast in temporal evolution between the two components. Main emissions ($G_2$) consistently exhibit positive lags, ranging from $\sim$18 ms up to $\sim$640 ms (observer frame). These values are fully consistent with the characteristic spectral lag of Type II GRBs. 
In contrast, precursors ($G_1$) exhibit a broad range of lags from -1984 ms to +710 ms with a median of -178 ms, showing substantial diversity that includes both negative and positive values, whereas main bursts consistently show positive lags with a median of $\sim$43 ms. For example, GRB~121128A exhibits a positive precursor lag of $710 \pm 21$ ms (this value likely needs a unit correction or review, given the context of ``negligible" and the range of 18 ms to 640 ms for the main bursts, suggesting a possible typo in the source text), while its main burst lag is a short $26 \pm 14$ ms. Conversely, GRB~061121 shows a main lag of $39 \pm 14$ ms but a longer precursor lag of $-106 \pm 15$ ms. This complex pattern, where $G_1$ often shows significantly smaller or near-zero lags compared to $G_2$, suggests distinct emission dynamics. Furthermore, while the general pattern holds, several events exhibit complex or negative lags (e.g., GRB~061121, GRB~080928), possibly indicating multi-zone or reverse-shock emission processes.

\section{Discussion}\label{sec:discussion}

Our pulse-resolved analysis reveals both similarities and systematic differences between precursor and main emission episodes in long GRBs. We summarize the key observational findings before discussing their physical implications: (1) Both $G_1$ and $G_2$ exhibit $T_{90}$ durations and spectral hardness ratios consistent with Type II (collapsar-origin) GRBs, with no precursor falling into the Type I regime. (2) Precursors show systematically longer minimum variability timescales than main bursts (median MVT ratio $\sim$ 5-10, with high statistical significance $D=0.91$, $p=6.2\times10^{-5}$). (3) Precursors show diverse lags with near-zero median values, while main bursts display positive lags characteristic of standard Type II behavior. These observational contrasts, despite the shared Type II classification, motivate the physical interpretations presented below.

A pulse-resolved analysis of long GRBs featuring precursor emission components reveals a combination of shared and contrasting properties between the precursor pulses and the main burst prompt emission. Both components exhibit durations and spectral hardness that align with the established phenomenology of Type II (collapsar-like) GRBs. However, significant differences are observed in their temporal characteristics, particularly in the MVT and the spectral lag.

\begin{figure*}[htbp]
\includegraphics[width=1.0\textwidth]{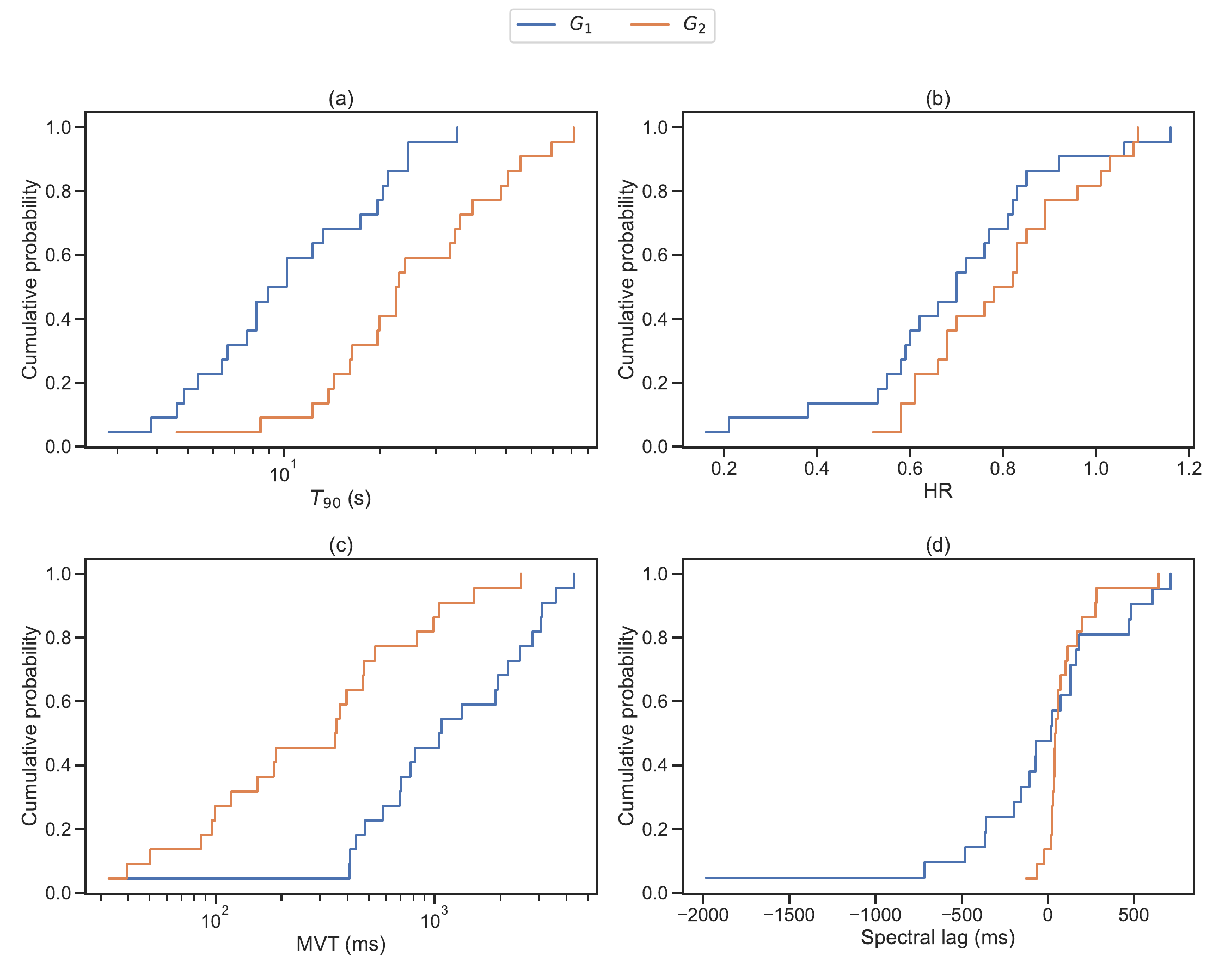}
\caption{Cumulative distribution functions (CDFs) for episode–level observables in the precursor emission ($G_1$) and main emission ($G_2$) phases in our long-duration GRB PE+ME sample. 
\textbf{(a),} individual $T_{90}$ durations (15–150 keV) of precursor and main burst for each GRB using values from Table~\ref{tab:t90};
\textbf{(b),} hardness ratio ($S_{50\text{--}100}/S_{25\text{--}50}$) using values from Table~\ref{tab:HR};
\textbf{(c),} minimum variability timescale (MVT) using values from Table~\ref{tab:MVT};
\textbf{(d),} spectral lag between 25–50 and 15–25~keV using values from Table~\ref{tab:lag}.
For each parameter, the $G_1$ and $G_2$ distributions are plotted separately to highlight systematic contrasts: $G_2$ episodes are consistently smoother (longer MVT), spectrally softer (lower HR), and exhibit more diverse lags than $G_1$.} \label{fig:cdf_kde_results}
\end{figure*}

To assess whether the precursor and main emission pulses of long GRBs systematically differ in key classification measurements, we examined their empirical cumulative distribution functions (CDFs) for three observables (HR, MVT, and spectral lag). As shown in Figure \ref{fig:cdf_kde_results}, the most significant difference is found in the MVT comparison, where precursors exhibit systematically longer variability timescales than the main prompt emission, yielding a Kolmogorov–Smirnov (KS) statistic of $D=0.91$ and a highly significant $p$-value = $6.2 \times 10^{-5}$. This statistical distinction supports the interpretation that precursors originate from smoother emission regions or more gradual dissipation processes compared to the spiky, rapidly varying main bursts. In contrast, the spectral lag shows a moderate discrepancy with $D=0.55$, and $p=0.075$, suggesting a tendency for precursors to exhibit smaller (often near-zero) lags, though this difference is not statistically conclusive at the 5\% confidence level. Furthermore, the spectral hardness distributions are statistically consistent, with the HR comparison yielding $D=0.36$ and $p=0.48$, indicating that both components share a similar intrinsic spectral character consistent with the Type II (collapsar-origin) GRB class.

\subsection{Minimum variability timescale as a radius diagnostic for $G_1$ vs.\ $G_2$}

The systematic difference in MVT between precursors and main bursts implies distinct dissipation regimes within the same GRB event. The smooth, long-MVT structure of precursors suggests emission from a less variable dissipation regime, possibly linked to a larger emission radius or lower Lorentz factor. Conversely, the spiky, short-MVT main burst emission reflects more efficient and rapidly varying energy release once the jet fully develops.

Our analysis reveals a highly significant difference in MVT between the precursor and main emission components ($D=0.91$, $p=6.2\times10^{-5}$), with the empirical cumulative distribution functions show that precursors have systematically longer MVTs than the main bursts (Figure \ref{fig:cdf_kde_results}). We determine the MVT ($\Delta t_{\min}$) objectively using a Haar wavelet-based structure function (Haar-SF) analysis, identifying it as the break point where the initial linear rise transitions to a noise-dominated regime. Assuming variability reflects causality-limited emission regions, we estimate the emission radius $R_{\rm em}$ using the constraint $R_{\rm em} \lesssim c\,\Gamma^2\,\frac{\Delta t_{\min}}{1+z}$, where $z$ is the burst redshift. Applying characteristic sample values, $\Delta t_{\min}^{G_1} \sim 0.5$--$1$~s for precursors and $\Delta t_{\min}^{G_2} \sim 0.1$--$0.5$~s for the main pulses, and adopting a typical bulk Lorentz factor $\Gamma \sim 10$ for precursors \citep{Nakar2012} and $\Gamma \sim 300$ for main bursts \citep{Peer2007, Zou2010, Liang2010}, we infer an order-of-magnitude separation in the characteristic emission radii: 
\begin{equation}
R^{G_1}_{\rm em} > 1.5 \times10^{12}\,(1+z)^{-1}~\text{cm},
\end{equation}
for the precursor emission and 
\begin{equation}
R^{G_2}_{\rm em} > 2.7\times10^{14}\,(1+z)^{-1}~\text{cm},
\end{equation}
for the main burst emission. 

This result suggests that precursors originate from a emission region roughly an order of magnitude larger in radius, which naturally explains the observed light curve morphology: the smooth (long MVT) precursor emission stems from a larger, more extended emitting region, whereas the spiky (short MVT) main emission is produced in a more compact region. 
This pronounced contrast in the characteristic emission radius fits well into a two-phase collapsar (long GRB) scenario. In the early phase, the cocoon shock breakout \citep{Gottlieb2025} (possibly a photospheric, quasi-thermal component) emanates from a broad region at large radius, leading to the precursor $G_1$ pulses with longer durations and fewer short-timescale fluctuations. In the later phase, the jet’s internal dissipation processes (such as internal shocks or magnetic reconnection events) dominate at a smaller radius, giving rise to the $G_2$ main pulses with rapid, intermittent flux variations. This interpretation is economical in terms of parameters, essentially only $R$ and $\Gamma$ are needed to explain the timescales via $R_{\rm em} \lesssim c\,\Gamma^2\,\frac{\Delta t_{\min}}{1+z}$, and it aligns with broader GRB population studies that find typical source-frame MVT values of $\sim$0.1--1~s for long GRBs, with the shortest variability timescales observed at higher (harder) photon energies.

\subsection{Spectral lag: weak spectral evolution in $G_1$ vs.\ pronounced evolution in $G_2$}

Our pulse-resolved analysis also reveals a systematic difference in the spectral lag ($\tau_{\rm lag}$) between the precursor and main emission episodes. While the spectral lag, defined as the delay between the arrival of high-energy and lower-energy photons, exhibits diverse values with a near-zero median for the $G_1$ precursors, it is distinctly positive and appreciable for many $G_2$ main pulses, which typically show lags on the order of $\sim 0.03-0.6$~s (with lower-energy photons arriving later, as shown in Table \ref{tab:lag}). The empirical cumulative distribution functions reflect this trend, with a K–S test yielding $D=0.55$ and $p=0.075$, indicating a moderate, though not statistically overwhelming, distinction between the two groups compared to the MVT results.

A positive spectral lag generally implies that the burst's spectral content evolves and softens over time due to processes such as cooling, or the delayed arrival of photons from the periphery of a curved emitting surface (the curvature effect). The magnitude of the lag can be constrained by physical parameters through relations such as:
\begin{equation}
\tau_{\rm lag} \propto \frac{1+z}{\Gamma}\,B'^{-3/2}\,\big(\nu_1^{-1/2} -\nu_2^{-1/2}\big),
\end{equation}
where $B'$ is the comoving magnetic-field strength, $\Gamma$ is the bulk Lorentz factor, and $\nu_1$ and $\nu_2$ are the characteristic photon frequencies. In this model, a larger lag is expected from emission regions with a smaller $\Gamma$ (less time dilation) or weaker comoving magnetic fields $B'$ (slower synchrotron cooling, allowing spectral evolution to occur over a longer timescale). Conversely, a minimal lag results if the spectrum remains nearly stationary in time or if $\Gamma$ is extremely high.

The observed pattern, diverse lags (both negative and positive) with a near-zero median for $G_1$ and consistently positive lags for $G_2$, is consistent with the two-phase scenario. The $G_1$ precursors, arising from a larger, more extended environment such as cocoon shock breakout or quasi-thermal photospheric emission, exhibit diverse lag behaviors possibly reflecting different emission geometries or multi-zone dissipation, resulting in a broad lag distribution with a near-zero median. In contrast, the $G_2$ main pulses, produced by internal shocks or magnetic reconnection at smaller radii, undergo rapid spectral softening and cooling over the pulse duration, resulting in the observed positive lags. Therefore, the spectral lag appears to trace the radiative evolution of the prompt emission, while the MVT is primarily linked to the geometric scale of the emission region. Together, these two timing diagnostics robustly reinforce the idea of two distinct physical regimes operating within a single collapsar-driven event.

\section{Conclusion}\label{sec:conclusion}

We have conducted a systematic, pulse-by-pulse comparative study of 22 long-duration GRBs detected by \emph{Swift}, all of which exhibit distinct precursor ($G_1$) emission episodes preceding the main burst ($G_2$). By rigorously analyzing key observables, including duration, spectral hardness, minimum variability timescale, and spectral lag, we draw several major conclusions regarding the physical origins and classification of these complex events. First, both the precursor ($G_1$) and the main burst ($G_2$) consistently display observational signatures characteristic of Type II (collapsar-origin) GRBs, such as moderate to long durations and soft hardness ratios, confirming that the precursor and main burst originate from the same central engine rather than representing distinct progenitor events. Second, despite this shared origin, the precursor episodes differ systematically and significantly from the main emission in two key temporal measurement: 
precursors ($G_1$) exhibit significantly longer MVTs (by factors of 3–10) and show diverse spectral lags with a near-zero median, whereas main burst ($G_2$) pulses show shorter MVTs and positive spectral lags consistent with standard Type II behavior. This pronounced temporal dichotomy strongly suggests different dissipation scales or emission radii within the same jet system. The long, smooth nature of many precursors aligns physically with emission from cocoon shock breakout occurring at large radii ($R \sim 10^{11}$--$10^{12}$ cm), an interpretation reinforced by the observed lack of variability and lag. Conversely, short and hard precursors may result from early-stage, magnetically dominated outflows, further supporting a unified two-stage jet emission process. 

In summary, our results suggest that the $G_1$ and $G_2$ components may originate from distinct physical regimes, likely involving different radiative efficiencies, emission region scales, or dissipation mechanisms, within the same collapsing-star central engine. The precursors are typically generated at larger radii with minimal spectral evolution, while the main pulses arise from smaller radii with significant spectral evolution. Our study highlights the critical importance of a pulse-wise classification scheme in GRBs, as simple duration-based categorization ($T_{90}$) can be misleading for multi-episode bursts. We recommend using a combination of duration, hardness, MVT, and spectral lag for each pulse to accurately assess physical origin. Future studies will extend this pulse-level analysis to other complex temporal structures, such as short GRBs with extended emission, and incorporate $E_{\rm p}$ measurements and broader-band spectral fits (e.g., from Fermi/GBM or HXMT) will allow a more subtlety test of energy correlations.

An important question beyond the scope of this work is whether the observed differences between precursor and main burst continue evolving in later emission episodes, such as the second main burst (Component IIb) or extended emission (Component III). Addressing this question requires separate sample selection, as many precursor GRBs lack clear later components. We plan to investigate the temporal evolution of emission properties across multiple episodes systematically in forthcoming papers in this series, examining whether precursors represent a unique initial phase or one end of a continuum that evolves throughout the GRB event.
Finally, a few additional comments are worth mentioning regarding the physical interpretation and methodological limitations inherent in separating the precursor ($G_1$) and main burst ($G_2$) phases. First, the long minimum variability timescale ($\Delta t_{\min}$) and near-zero spectral lag ($\tau \sim 0$) observed for the precursor component are highly consistent with the jet shock breakout model. In this scenario, the precursor signal arises as the relativistic jet punches through the stellar envelope, resulting in quasi-thermal emission from large radii ($R \sim 10^{11}$--$10^{12}$ cm). Alternatively, the observed dichotomy could stem from a two-stage jet launching scenario, where a magnetically dominated outflow, emitting a hard, lag-free precursor via magnetic reconnection, precedes a baryonic jet that generates the internal shocks responsible for the main, lag-bearing emission \citep{Zhang2018NA}. Such a transition would naturally account for the distinct temporal and spectral lag properties observed. Second, we must acknowledge that a portion of the observed MVT difference is attributable to signal-to-noise ratio (SNR) selection effects. Our paired within-burst analysis reveals a strong anti-correlation between $\Delta\log_{10}\mathrm{MVT}$ and $\Delta\log_{10}F_{\rm p}$ ($\beta\simeq-1.07$, $R^2\simeq0.75$), implying that the lower flux and poorer SNR of precursors naturally inflate their measurable MVT. However, the existence of a robust subset of events featuring intrinsically bright and short-timescale precursors confirms that the observed MVT separation is not purely an artifact of measurement sensitivity. Third, the relative energetics further support the precursor's peripheral role: pulses carry only a small fraction of the total prompt fluence, and no distinct afterglow component is observed exclusively in association with the precursor. This is consistent with a relatively low-energy, possibly radiation-dominated outflow that fails to drive a strong external shock. By contrast, the main emission accounts for the bulk of the energy release and drives the standard long GRB afterglow. Finally, we caution that our physical radius estimates rely on an assumed, common bulk Lorentz factor $\Gamma$, and since $R \propto \Gamma^2$, the absolute values are sensitive to this choice. Moreover, the statistical distinction in spectral lag is only moderate ($p$=0.075) given the current sample size, necessitating future confirmation from a larger set of events. A major limitation is the reliance on the \textit{Swift}/BAT band for spectral analysis. Joint \textit{Swift}+{\it Fermi} observations covering a broader energy range will be crucial to accurately constrain the spectral peak energy ($E_{\rm p}$) and $\Gamma$ for each phase. 

\acknowledgments

We thank Wang Xiao, Cui Zhi-Li, Xiao Cheng-Long, and Li Wen for carefully verifying the sample data in this study. We also thank Bing Zhang for many useful discussions. LL is supported by the Natural Science Foundation of China (grant No. 11874033), the KC Wong Magna Foundation at Ningbo University, and made use of the High Energy Astrophysics Science Archive Research Center (HEASARC) Online Service at the NASA/Goddard Space Flight Center (GSFC). YFH is supported by the National Natural Science Foundation of China (Grant No. 12233002), by the National Key R\&D Program of China (2021YFA0718500). YFH also acknowledges the support from the Xinjiang Tianchi Program. The computations were supported by the high performance computing center at Ningbo University.

\vspace{5mm}
\facilities{{\it Swift}}
\software{
{\tt 3ML} \citep{Vianello2015}, 
{\tt matplotlib} \citep{Hunter2007}, 
{\tt NumPy} \citep{Harris2020,Walt2011}, 
{\tt SciPy} \citep{Virtanen2020}, 
{\tt $lmfit$} \citep{Newville2016}, 
{\tt astropy} \citep{AstropyCollaboration2013},
{\tt pandas} \citep{Reback2022},
{\tt emcee} \citep{Foreman-Mackey2013},
{\tt seaborn} \citep{Waskom2017}}  

\vspace{5mm}
\bibliography{lGRBs.bib}

\setcounter{table}{0}
\begin{table*}
\centering
\small
\setlength{\tabcolsep}{4pt}
\caption{Temporal properties of the precursor ($G_1$) and main emission ($G_2$) sub-bursts in our GRB sample. Each entry lists the time bin resolution used, the start and end times ($t_1 \sim t_2$), total pulse width, $T_{90}$ duration with asymmetric errors, pulse peak time ($t_{\rm p}$), and peak count rate ($F_{\rm p}$) with $1\sigma$ uncertainties. These values form the basis for the duration and gap analyses presented in Section~\ref{sec:duration}.\label{tab:pulse}}
\begin{tabular}{lccccccc}
\hline\hline
GRB & Sub-burst & Time bin & $t_1 \sim t_2$ & $T_{\text{pulse}}$ & $t_{\rm p}$ & $F_{\rm p}$& Classified as\\
&&(ms)&(s)&(s)&(s)&&(PE or ME)\\
\hline
050820A & $G_1$ & 256 & $-3.35\sim22.50$ & 25.86 & 0.49 & $0.16 \pm 0.02$ & PE\\
        & $G_2$ & 256 & $223.72\sim241.13$ & 17.41 & 229.61 & $0.38 \pm 0.04$ & ME\\
060923A & $G_1$ & 1024 & $-43.22\sim-29.91$ & 13.31 & -39.13 & $0.07 \pm 0.02$ & PE\\
        & $G_2$ & 1024 & $-6.36\sim16.17$ & 22.53 & -0.22 & $0.14 \pm 0.02$ & ME\\
061007 & $G_1$ & 256 & $-3.78\sim14.90$ & 18.69 & 0.31 & $0.83 \pm 0.07$ & PE\\
        & $G_2$ & 256 & $20.02\sim69.94$ & 49.92 &45.88 & $2.12 \pm 0.05$ & ME\\
061121 & $G_1$ & 256 & $-2.28\sim10.01$ & 12.29 & 2.84 & $0.25 \pm 0.02$ & PE\\
       & $G_2$ & 256 & $51.99\sim102.68$ & 50.69 & 74.78 & $3.15 \pm 0.05$ & ME\\ 
061121 & $G_1$ & 256 & $-2.28\sim10.01$ & 12.29 & 2.84 & $0.25 \pm 0.02$ &PE\\
       & $G_2$ & 256 & $51.99\sim102.68$ & 50.69 & 74.78 & $3.15 \pm 0.05$ & ME\\
061202 & $G_1$ & 1024 & $-2.76\sim11.58$ & 14.34 & 1.34 & $0.04 \pm 0.01$ & PE\\
       & $G_2$ & 1024 & $66.87\sim121.14$ & 54.27 & 75.06 & $0.33 \pm 0.01$ & ME\\
061222A & $G_1$ & 256 & $23.92\sim30.58$ & 6.66 & 25.71 & $0.35 \pm 0.03$ & PE\\
        & $G_2$ & 256 & $53.36\sim100.46$ & 47.10 & 88.43 & $1.35 \pm 0.04$ & ME\\
070306 & $G_1$ & 1024 & $-8.56\sim25.23$ & 33.79 & 10.9 & $0.11 \pm 0.02$ & PE\\
       & $G_2$ & 1024 & $78.48\sim157.33$ & 78.85 & 97.94 & $0.44 \pm 0.01$ & ME\\
070521 & $G_1$ & 256 & $-6.36\sim5.93$ & 12.29 & 0.30 & $0.31 \pm 0.03$ & PE\\
       & $G_2$ & 256 & $8.23\sim45.35$ & 37.12 & 31.02 & $1.10 \pm 0.04$ & ME\\
071010B & $G_1$ & 1024 & $-37.54\sim-6.82$ & 30.72 & -24.23 & $0.06 \pm 0.01$ & PE\\
        & $G_2$ & 256 & $-2.98\sim20.31$ & 23.30 & 2.14 & $0.89 \pm 0.03$ & ME\\
080928 & $G_1$ & 1024 & $-5.53\sim3.69$ & 9.22 & -4.5 & $0.04 \pm 0.01$ & PE\\
       & $G_2$ & 1024 & $195.18\sim257.64$ & 62.46 & 202.34 & $0.21 \pm 0.01$ & ME\\
090618 & $G_1$ & 256 & $-7.63\sim41.78$ & 49.41 & 9.52 & $0.79 \pm 0.04$ & PE\\
       & $G_2$ & 256 & $42.03\sim159.79$ & 117.76 & 65.84 & $4.65 \pm 0.07$ & ME\\
091208B & $G_1$ & 256 & $-0.72\sim4.40$ & 5.12 & 0.30 & $0.59 \pm 0.09$ & PE\\
        & $G_2$ & 256 & $5.17\sim13.36$ & 8.19 & 8.24 & $2.16 \pm 0.14$ & ME\\
121128A & $G_1$ & 256 & $-1.57\sim3.04$ & 4.61 & 0.22 & $0.11 \pm 0.03$ & PE\\
        & $G_2$ & 256 & $9.18\sim41.70$ & 32.51 & 21.73 & $1.80 \pm 0.05$ & ME\\
140206A & $G_1$ & 256 & $-1.44\sim24.93$ & 26.37 & 6.24 & $1.42 \pm 0.04$ & PE\\
        & $G_2$ & 256 & $50.02\sim90.72$ & 40.70 & 61.02 & $3.44 \pm 0.06$ & ME\\
\hline
\end{tabular}
\end{table*}
\setcounter{table}{0}
\begin{table*}
\caption{--- continued}
\begin{tabular}{lcccccccc}
\hline\hline
GRB & Sub-burst & Time bin & $t_1 \sim t_2$ & $T_{\text{pulse}}$ & $t_{\rm p}$ & $F_{\rm p}$& Classified as\\
&&(ms)&(s)&(s)&(s)&&(PE or ME)\\
\hline
140512A & $G_1$ & 256 & $-16.50\sim10.13$ & 26.62 & 1.68 & $0.52 \pm 0.05$ & PE\\
        & $G_2$ & 256 & $91.02\sim166.29$ & 75.26 & 122.51 & $0.97 \pm 0.04$ & ME\\
140629A & $G_1$ & 1024 & $-9.13\sim9.30$ & 18.43 & 0.09 & $0.19 \pm 0.02$ & PE\\
        & $G_2$ & 1024 & $10.33\sim30.81$ & 20.48 & 12.38 & $0.51 \pm 0.03$ & ME\\
150323A & $G_1$ & 1024 & $-5.53\sim24.17$ & 29.7 & 2.66 & $0.12 \pm 0.02$ & PE\\
        & $G_2$ & 256 & $121.45\sim155.75$ & 34.3 & 131.94 & $0.58 \pm 0.03$ & ME\\
160131A & $G_1$ & 256 & $-2.18\sim7.81$ & 9.98 & 5.50 & $0.16 \pm 0.03$ & PE\\
        & $G_2$ & 256 & $8.06\sim114.05$ & 105.98 & 16.77 & $0.88 \pm 0.05$ & ME\\
160703A & $G_1$ & 256 & $-26.98\sim-18.53$ & 8.45 & -25.18 & $0.12 \pm 0.02$ & PE\\
        & $G_2$ & 256 & $-14.43\sim45.22$ & 59.65 & 27.30 & $0.77 \pm 0.03$ & ME\\
180325A & $G_1$ & 256 & $-0.93\sim8.29$ & 9.22 & 1.12 & $0.31 \pm 0.03$ & PE\\
        & $G_2$ & 256 & $72.29\sim103.01$ & 30.72 & 80.74 & $1.45 \pm 0.05$ & ME\\
180728A & $G_1$ & 256 & $-1.74\sim6.97$ & 8.7 & 0.82 & $0.49 \pm 0.04$ & PE\\
        & $G_2$ & 16 & $11.10\sim23.43$ & 12.34 & 12.87 & $17.46 \pm 0.71$ & ME\\
190324A & $G_1$ & 256 & $-0.66\sim5.22$ & 5.89 & 0.62 & $0.17 \pm 0.04$ & PE\\
        & $G_2$ & 256 & $16.49\sim43.62$ & 27.14 & 24.17 & $1.65 \pm 0.07$ & ME\\
\hline
\end{tabular}
\end{table*}

\clearpage
\setcounter{table}{1}
\setlength{\tabcolsep}{0.15em}
\begin{deluxetable*}{ccccccccc}
\tablewidth{700pt}
\tabletypesize{\scriptsize}
\tablecaption{Comparative analysis of precursor and main emission properties for each GRB in the sample. \label{tab:Compara}}
\tablehead{
\colhead{GRB}
&\colhead{$z$}
&\colhead{$t_{\mathrm{gap}}$}
&\colhead{$t_{\mathrm{gap}}/(1+z)$}
&\colhead{$R_{t_{90}}$}
&\colhead{$R_{F_{\rm p}}$}
&\colhead{$R_{\rm HR}$}
&\colhead{$R_{\rm MVT}$}
&\colhead{$R_{\tau}$}\\
\hline
&&(s)&(s)&&&&&\\
}
\colnumbers
\startdata
\hline
050820A & 2.612 & $201.2 \pm 3.4$ & $55.7 \pm 1.0$ & $1.54 \pm 0.26$ & $0.42 \pm 0.07$ & $0.65 \pm 0.12$ & $1.19 \pm 0.75$ & $0.75 \pm 0.08$ \\
060923A & 2.800 & $23.6 \pm 3.7$ & $6.2 \pm 1.0$ & $0.63 \pm 0.15$ & $0.50 \pm 0.16$ & $0.64 \pm 0.31$ & $2.01 \pm 1.77$ & $0.67 \pm 0.03$  \\
061007 & 1.260 & $\sim 5.1$ & $2.3 \pm 3.8$ & $0.40 \pm 0.13$ & $0.39 \pm 0.03$ & $0.98 \pm 0.04$ & $1.96 \pm 4.28$ &  $4.00 \pm 0.80$ \\
061121 & 1.314 & $42.0 \pm 10.3$ & $18.1 \pm 4.4$ & $0.36 \pm 0.23$ & $0.08 \pm 0.01$ & $0.87 \pm 0.11$ & $8.15 \pm 2.88$ & $-2.69 \pm 1.02$ \\
061202 &2.253 & $55.3 \pm 6.0$ & $17.0 \pm 1.8$ & $0.29 \pm 0.09$ & $0.12 \pm 0.03$ & $0.28 \pm 0.49$ & $6.46 \pm 5.35$ & $-1.75 \pm 0.15$ \\
061222A& 2.088 & $22.8 \pm 5.9$ & $7.4 \pm 1.9$ & $0.14 \pm 0.04$ & $0.26 \pm 0.02$ & $0.79 \pm 0.07$ & $1.03 \pm 1.92$ & $-1.64 \pm 0.30$  \\
070306 & 1.496 & $53.2 \pm 8.8$ & $21.3 \pm 3.5$ & $0.44 \pm 0.13$ & $0.25 \pm 0.05$ & $1.35 \pm 0.45$ & $18.05 \pm 8.77$ & $-1.30 \pm 0.51$ \\
070521 & 2.087 & $\sim 2.3$ & $0.7 \pm 2.0$ & $0.40 \pm 0.15$ & $0.28 \pm 0.03$ & $0.85 \pm 0.07$ & $1.96 \pm 2.46$ & $3.15 \pm 0.43$ \\
071010B & 0.947 & $\sim 3.8$ & $2.0 \pm 2.1$ & $1.71 \pm 0.58$ & $0.07 \pm 0.01$ & $1.06 \pm 0.17$ & $3.61 \pm 1.15$ & $-3.14 \pm 2.39$ \\
080928 & 1.690 & $191.5 \pm 7.4$ & $71.2 \pm 2.7$ & $0.16 \pm 0.03$ & $0.19 \pm 0.05$ & $0.91 \pm 1.40$ & $4.30 \pm 1.31$ & $-11.92 \pm 1.21$ \\
090618 & 0.540 & $\sim 0.2$ & $0.2 \pm 9.6$ & $0.51 \pm 0.20$ & $0.17 \pm 0.01$ & $1.21 \pm 0.03$ & $2.33 \pm 4.19$ & $8.22 \pm 2.65$ \\
091208B & 1.063 & $\sim 0.8$ & $0.4 \pm 1.1$ & $0.83 \pm 0.35$ & $0.27 \pm 0.05$ & $0.85 \pm 0.32$ & $1.66 \pm 3.53$ & $0.55 \pm 0.38$ \\
121128A & 2.200 & $\sim 6.1$ & $1.9 \pm 1.4$ & $0.13 \pm 0.05$ & $0.06 \pm 0.02$ & $0.28 \pm 0.59$ & $13.34 \pm 3.45$ & $27.75 \pm 14.94$ \\
140206A & 2.730 & $25.1 \pm 8.4$ & $6.7 \pm 2.2$ & $0.88 \pm 0.52$ & $0.41 \pm 0.01$ & $0.99 \pm 0.05$ & $0.37 \pm 0.15$ & $0.44 \pm 0.29$ \\
140512A & 0.725 & $80.9 \pm 8.9$ & $46.9 \pm 5.2$ & $0.39 \pm 0.12$ & $0.54 \pm 0.06$ & $1.36 \pm 0.19$ & $12.16 \pm 8.07$ & $-11.79 \pm 1.92$ \\
140629A & 2.275 & $\sim 1.0$ & $0.3 \pm 1.0$ & $1.00 \pm 0.42$ & $0.37 \pm 0.04$ & $0.94 \pm 0.35$ & $1.29 \pm 0.51$ &  $1.46 \pm 0.23$ \\
150323A & 0.593 & $97.3 \pm 5.8$ & $61.1 \pm 3.7$ & $0.85 \pm 0.26$ & $0.21 \pm 0.04$ & $1.36 \pm 0.38$ & $23.37 \pm 15.30$ & $2.80 \pm 1.26$ \\
160131A & 0.972 & $\sim 0.3$ & $0.1 \pm 2.8$ & $0.10 \pm 0.02$ & $0.18 \pm 0.04$ & $0.85 \pm 0.24$ & $0.87 \pm 1.59$ & \nodata \\
160703A & 1.500 & $\sim 4.1$ & $1.6 \pm 4.5$ & $0.17 \pm 0.05$ & $0.16 \pm 0.03$ & $0.93 \pm 0.34$ & $4.11 \pm 8.24$ & $12.25 \pm 4.20$ \\
180325A & 2.248 & $64.0 \pm 3.6$ & $19.7 \pm 1.1$ & $0.32 \pm 0.11$ & $0.21 \pm 0.02$ & $0.57 \pm 0.10$ & $6.57 \pm 1.72$ & $1.75 \pm 1.64$ \\
180728A & 0.117 & $3.3 \pm 2.2$ & $3.0 \pm 2.0$ & $0.64 \pm 0.47$ & $0.03 \pm 0.00$ & $0.62 \pm 0.18$ & $21.21 \pm 3.17$ & $3.18 \pm 0.66$ \\
190324A & 1.171 & $11.3 \pm 2.4$ & $5.2 \pm 1.1$ & $0.29 \pm 0.11$ & $0.10 \pm 0.02$ & $0.77 \pm 0.46$ & $22.16 \pm 16.27$ & $-5.44 \pm 0.53$ \\
\enddata 
\vspace{3mm}
\tablecomments{Listed quantities include the redshift, observed-frame quiescent gap duration and its rest-frame corrected value, and the ratios of precursor to main emission in terms of $T_{90}$, $F_{\rm p}$, HR, MVT, and spectral lag. These values support the classification and consistency testing discussed in Sections~\ref{sec:duration}–\ref{sec:lag}.} 
\end{deluxetable*}

\clearpage
\begin{table*}[htbp]
\centering
\footnotesize
\caption{Individual $T_{90}$ durations (15–150 keV) of precursor and main emission sub-bursts for each GRB. All components in the sample satisfy the long-burst ($T_{90} > 2$~s) classification. This supports the conclusion that both precursor and main emission episodes generally belong to the Type II GRB population based on duration alone. \label{tab:t90}}
\centering
\begin{tabular}{c|c|c|c|c}
\hline
 \multicolumn{1}{c|}{GRB} & \multicolumn{2}{c|}{Precursor} & \multicolumn{2}{c}{Main emission} \\
& \multicolumn{2}{c|}{($G_1$)} & \multicolumn{2}{c}{($G_2$)} \\
\hline
&Value&Classification&Value&Classification\\
\hline
050820A & 21.2$^{+2.6}_{-2.0}$~[s] & Long & 13.8$^{+1.3}_{-2.3}$~[s] & Long \\
060923A  & 10.2$^{+2.0}_{-1.0}$~[s] & Long & 16.4$^{+3.1}_{-3.1}$~[s] & Long \\
061007  & 13.3$^{+2.8}_{-2.6}$~[s] & Long & 33.3$^{+8.4}_{-8.2}$~[s] & Long \\
061121  & 8.2$^{+2.3}_{-1.8}$~[s] & Long & 23.0$^{+17.7}_{-10.0}$~[s] & Long \\
061202  & 10.2$^{+3.1}_{-1.0}$~[s] & Long & 35.8$^{+13.3}_{-5.1}$~[s] & Long \\
061222A  & 4.9$^{+0.8}_{-1.0}$~[s] & Long & 34.6$^{+6.7}_{-5.9}$~[s] & Long \\
070306  & 24.6$^{+5.1}_{-4.1}$~[s] & Long & 55.3$^{+16.4}_{-7.2}$~[s] & Long \\
070521  & 9.0$^{+1.3}_{-2.0}$~[s] & Long & 22.5$^{+8.7}_{-5.9}$~[s] & Long \\
071010B  & 24.6$^{+3.1}_{-3.1}$~[s] & Long & 14.3$^{+6.1}_{-2.8}$~[s] & Long \\
080928  & 7.7$^{+0.5}_{-1.0}$~[s] & Long & 48.1$^{+10.2}_{-4.1}$~[s] & Long \\
090618  & 35.1$^{+8.2}_{-6.1}$~[s] & Long & 69.4$^{+36.1}_{-12.3}$~[s] & Long \\
091208B  & 3.8$^{+0.8}_{-0.5}$~[s] & Long & 4.6$^{+1.5}_{-2.0}$~[s] & Long \\
121128A  & 2.8$^{+1.0}_{-0.8}$~[s] & Long & 22.5$^{+5.6}_{-4.3}$~[s] & Long \\
140206A & 17.4$^{+7.2}_{-1.8}$~[s] & Long & 19.7$^{+16.6}_{-4.3}$~[s] & Long \\
140512A  & 19.7$^{+2.8}_{-4.1}$~[s] & Long & 50.7$^{+16.1}_{-8.4}$~[s] & Long \\
140629A & 12.3$^{+3.1}_{-3.1}$~[s] & Long & 12.3$^{+7.2}_{-1.0}$~[s] & Long \\
150323A  & 20.5$^{+5.1}_{-4.1}$~[s] & Long & 24.1$^{+7.4}_{-2.8}$~[s] & Long \\
160131A  & 8.2$^{+1.0}_{-0.8}$~[s] & Long & 81.4$^{+19.2}_{-5.4}$~[s] & Long \\
160703A  & 6.7$^{+1.0}_{-0.8}$~[s] & Long & 39.2$^{+9.2}_{-11.3}$~[s] & Long \\
180325A  & 6.4$^{+1.8}_{-1.0}$~[s] & Long & 20.0$^{+7.7}_{-3.1}$~[s] & Long \\
180728A  & 5.4$^{+1.8}_{-1.5}$~[s] & Long & 8.4$^{+10.0}_{-1.3}$~[s] & Long \\
190324A  & 4.6$^{+0.5}_{-0.8}$~[s] & Long & 16.1$^{+8.7}_{-2.3}$~[s] & Long \\
\hline
\end{tabular}
\end{table*}

\begin{table*}[htbp]
\centering
\footnotesize
\caption{Hardness ratios [HR = $\frac{S_{50–100~\mathrm{keV}}}{S_{25–50~\mathrm{keV}}}$] for each precursor ($G_1$) and main emission ($G_2$) pulse in the sample, computed from background-subtracted fluences in the specified BAT bands. Classification is based on the $T_{90}$–HR location in the duration–hardness diagram. All episodes are consistent with Type II (long) GRBs in both duration and spectral hardness.\label{tab:HR}}
\centering
\begin{tabular}{c|c|c|c|c}
\hline
 \multicolumn{1}{c|}{GRB} & \multicolumn{2}{c|}{Precursor} & \multicolumn{2}{c}{Main emission}\\
& \multicolumn{2}{c|}{($G_1$)} & \multicolumn{2}{c}{($G_2$)}\\
\hline
&Value&Classification&Value&Classification\\
\hline
050820A & 0.66$\pm$0.10 & Long & 1.01$\pm$0.10 & Long\\
060923A & 0.70$\pm$0.30 & Long & 1.09$\pm$0.25 & Long\\
061007 & 1.06$\pm$0.04 & Long & 1.08$\pm$0.01 & Long\\
061121 & 0.72$\pm$0.09 & Long & 0.83$\pm$0.01 & Long\\
061202 & 0.21$\pm$0.37 & Long & 0.76$\pm$0.05 & Long\\
061222A & 0.70$\pm$0.06 & Long & 0.89$\pm$0.01 & Long\\
070306 & 0.92$\pm$0.30 & Long & 0.68$\pm$0.05 & Long\\
070521 & 0.76$\pm$0.06 & Long & 0.89$\pm$0.02 & Long\\
071010B & 0.55$\pm$0.09 & Long & 0.52$\pm$0.01 & Long\\
080928 & 0.53$\pm$0.81 & Long & 0.58$\pm$0.09 & Long\\
090618 & 0.85$\pm$0.02 & Long & 0.70$\pm$0.00 & Long\\
091208B & 0.58$\pm$0.19 & Long & 0.68$\pm$0.12 & Long\\
121128A & 0.16$\pm$0.34 & Long & 0.58$\pm$0.02 & Long\\
140206A & 0.81$\pm$0.04 & Long & 0.82$\pm$0.02 & Long\\
140512A & 1.16$\pm$0.16 & Long & 0.85$\pm$0.03 & Long\\
140629A & 0.62$\pm$0.21 & Long & 0.66$\pm$0.11 & Long\\
150323A & 0.83$\pm$0.23 & Long & 0.61$\pm$0.03 & Long\\
160131A & 0.82$\pm$0.23 & Long & 0.96$\pm$0.03 & Long\\
160703A & 0.77$\pm$0.28 & Long & 0.83$\pm$0.03 & Long\\
180325A & 0.59$\pm$0.10 & Long & 1.03$\pm$0.05 & Long\\
180728A & 0.38$\pm$0.11 & Long & 0.61$\pm$0.01 & Long\\
190324A & 0.60$\pm$0.36 & Long & 0.78$\pm$0.04 & Long\\
\hline
\end{tabular}
\end{table*}

\begin{table*}[htbp]
\centering
\footnotesize
\caption{Minimum variability timescales ($\Delta t_{\min}$) of precursors ($G_1$) and main emissions ($G_2$) for GRBs in the sample. Values were determined using the Haar wavelet method applied to \emph{Swift}/BAT light curves. All episodes are classified as Type II based on MVT–$T_{90}$ distributions, but precursors tend to exhibit significantly longer MVTs than their corresponding main bursts.\label{tab:MVT}}
\centering
\begin{tabular}{c|c|c|c|c}
\hline
 \multicolumn{1}{c|}{GRB} & \multicolumn{2}{c|}{Precursor} & \multicolumn{2}{c}{Main emission} \\
& \multicolumn{2}{c|}{($G_1$)} & \multicolumn{2}{c}{($G_2$)}\\
\hline
&Value&Classification&Value&Classification\\
\hline
050820A & (439$\pm$214)~[ms] & Long & (369$\pm$147)~[ms] & Long \\
060923A & (3056$\pm$854)~[ms] & Long & (1519$\pm$1270)~[ms] & Long\\
061007 & (700$\pm$499)~[ms] & Long & (357$\pm$135)~[ms] & Long\\
061121 & (411$\pm$43)~[ms] & Long & (50$\pm$17)~[ms] & Long \\
061202 & (3075$\pm$1140)~[ms] & Long & (476$\pm$353)~[ms] & Long\\
061222A & (409$\pm$62)~[ms] & Long & (396$\pm$134)~[ms] & Long\\
070306 & (2802$\pm$1288)~[ms] & Long & (155$\pm$24)~[ms] & Long\\
070521 & (1047$\pm$97)~[ms] & Long & (534$\pm$270)~[ms] & Long\\
071010B & (3578$\pm$1140)~[ms] & Long & (990$\pm$20)~[ms] & Long\\
080928 & (814$\pm$233)~[ms] & Long & (189$\pm$19)~[ms] & Long\\
090618 & (2452$\pm$1400)~[ms] & Long & (1051$\pm$105)~[ms] & Long\\
091208B & (581$\pm$209)~[ms] & Long & (350$\pm$135)~[ms] & Long\\
121128A & (1329$\pm$265)~[ms] & Long & (100$\pm$16)~[ms] & Long\\
140206A & (36$\pm$11)~[ms] & Long & (96$\pm$23)~[ms] & Long\\
140629A & (1074$\pm$154)~[ms] & Long & (833$\pm$305)~[ms] & Long\\
140512A & (480$\pm$192)~[ms] & Long & (39$\pm$21)~[ms] & Long\\
150323A & (4312$\pm$1118)~[ms] & Long & (185$\pm$111)~[ms] & Long\\
160131A & (2161$\pm$416)~[ms] & Long & (2476$\pm$1491)~[ms] & Long\\
160703A & (1942$\pm$577)~[ms] & Long & (472$\pm$209)~[ms] & Long\\
180325A & (775$\pm$88)~[ms] & Long & (118$\pm$28)~[ms] & Long\\
180728A & (692$\pm$43)~[ms] & Long & (33$\pm$4)~[ms] & Long\\
190324A & (1901$\pm$294)~[ms] & Long & (86$\pm$62)~[ms] & Long\\
\hline
\end{tabular}
\end{table*}

\begin{table*}[htbp]
\centering
\footnotesize
\caption{Spectral lags between the 25–50~keV and 15–25~keV bands for individual bursts in the sample. Positive values indicate that the 25–50~keV band peaks earlier than the 15–25~keV band (i.e., high energy leads). Uncertainties are $1\sigma$. Columns list the lag for the main emission (G$_1$) and the extended emission (G$_2$), together with the duration-based classification.\label{tab:lag}}
\centering
\begin{tabular}{c|c|c|c|c}
\hline
 \multicolumn{1}{c|}{GRB} & \multicolumn{2}{c|}{Precursor} & \multicolumn{2}{c}{Main emission}\\
& \multicolumn{2}{c|}{($G_1$)} & \multicolumn{2}{c}{($G_2$)}\\
\hline
&Value&Classification&Value&Classification\\
\hline
050820A & (480.0$\pm$27.5)~[ms] & Long & (640.0$\pm$57.0)~[ms] & Long \\
060923A & (131.2$\pm$3.7)~[ms] & Long & (195.2$\pm$5.1)~[ms] & Long \\
061007 & (73.6$\pm$12.5)~[ms] & Long & (18.4$\pm$1.9)~[ms] & Long \\
061121 & (-105.6$\pm$14.7)~[ms] & Long & (39.2$\pm$13.8)~[ms] & Long \\
061202 & (-480.0$\pm$27.5)~[ms] & Long & (273.6$\pm$17.3)~[ms] & Long \\
061222A & (-72.0$\pm$10.7)~[ms] & Long & (44.0$\pm$4.6)~[ms] & Long \\
070306 & (-364.8$\pm$60.2)~[ms] & Long & (281.6$\pm$101.8)~[ms] & Long \\
070521 & (131.2$\pm$10.2)~[ms] & Long & (41.6$\pm$4.6)~[ms] & Long \\
071010B & (-70.4$\pm$25.6)~[ms] & Long & (22.4$\pm$15.0)~[ms] & Long \\
080928 & (-1984.0$\pm$32.0)~[ms] & Long & (166.4$\pm$16.6)~[ms] & Long \\
090618 & (604.8$\pm$12.0)~[ms] & Long & (73.6$\pm$23.7)~[ms] & Long \\
091208B & (19.2$\pm$12.8)~[ms] & Long & (35.2$\pm$8.0)~[ms] & Long \\
121128A & (710.4$\pm$21.1)~[ms] & Long & (25.6$\pm$13.8)~[ms] & Long \\
140206A & (25.6$\pm$15.4)~[ms] & Long & (57.6$\pm$16.0)~[ms] & Long \\
140512A & (-716.8$\pm$34.6)~[ms] & Long & (60.8$\pm$9.4)~[ms] & Long \\
140629A & (163.2$\pm$7.7)~[ms] & Long & (112.0$\pm$16.6)~[ms] & Long \\
150323A & (-358.4$\pm$143.4)~[ms] & Long & (-128.0$\pm$26.2)~[ms] & Long \\
160131A & \nodata & Long & (-22.4$\pm$7.0)~[ms] & Long \\
160703A & (470.4$\pm$13.8)~[ms] & Long & (38.4$\pm$13.1)~[ms] & Long \\
180325A & (179.2$\pm$94.7)~[ms] & Long & (102.4$\pm$79.4)~[ms] & Long \\
180728A & (-198.4$\pm$25.6)~[ms] & Long & (-62.4$\pm$10.1)~[ms] & Long \\
190324A & (-156.8$\pm$12.5)~[ms] & Long & (28.8$\pm$1.6)~[ms] & Long \\
\hline
\end{tabular}
\end{table*}

\begin{figure*}
\includegraphics[angle=0,scale=0.32]{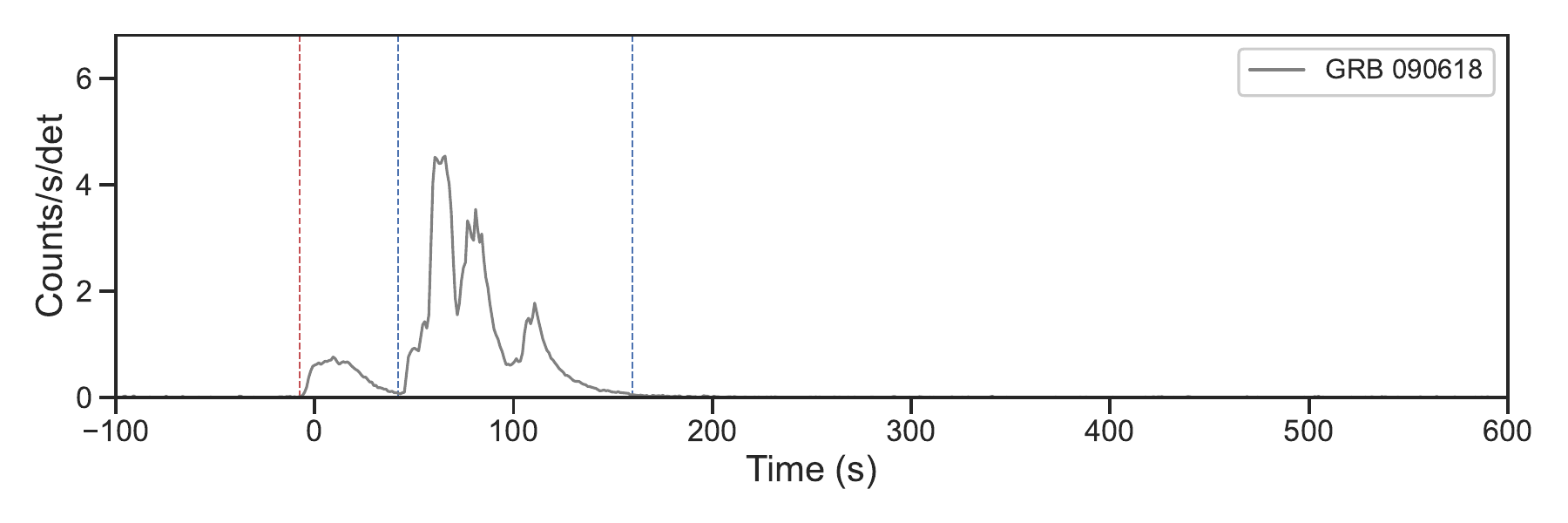}
\includegraphics[angle=0,scale=0.32]{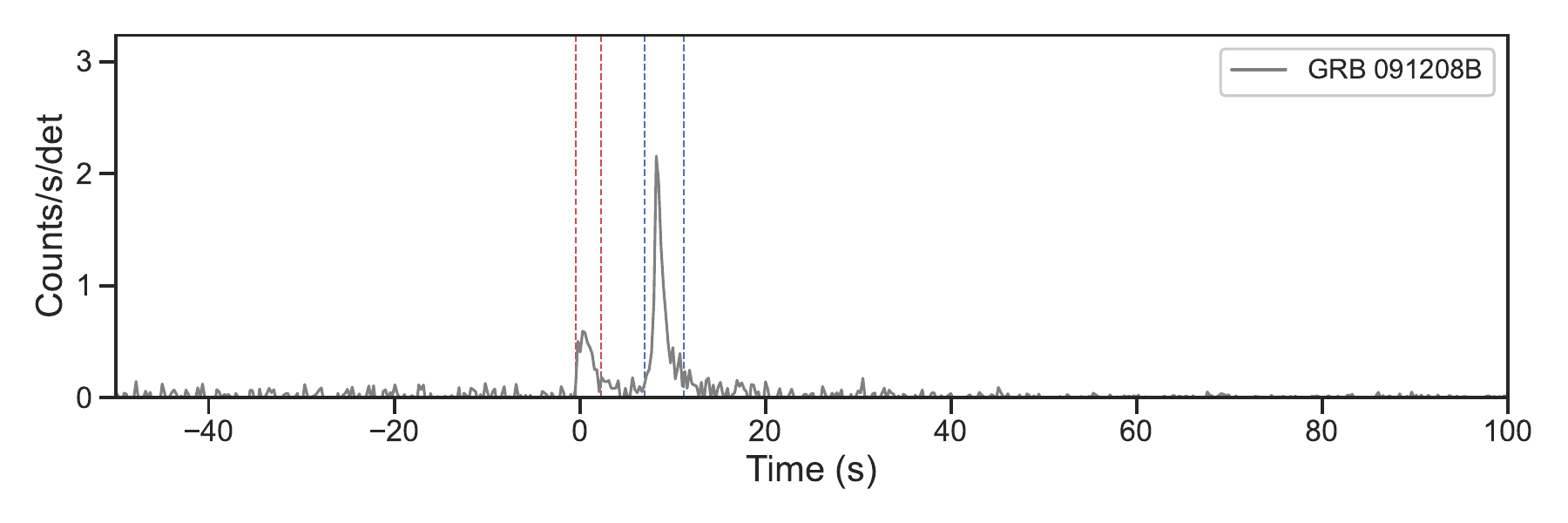}
\includegraphics[angle=0,scale=0.32]{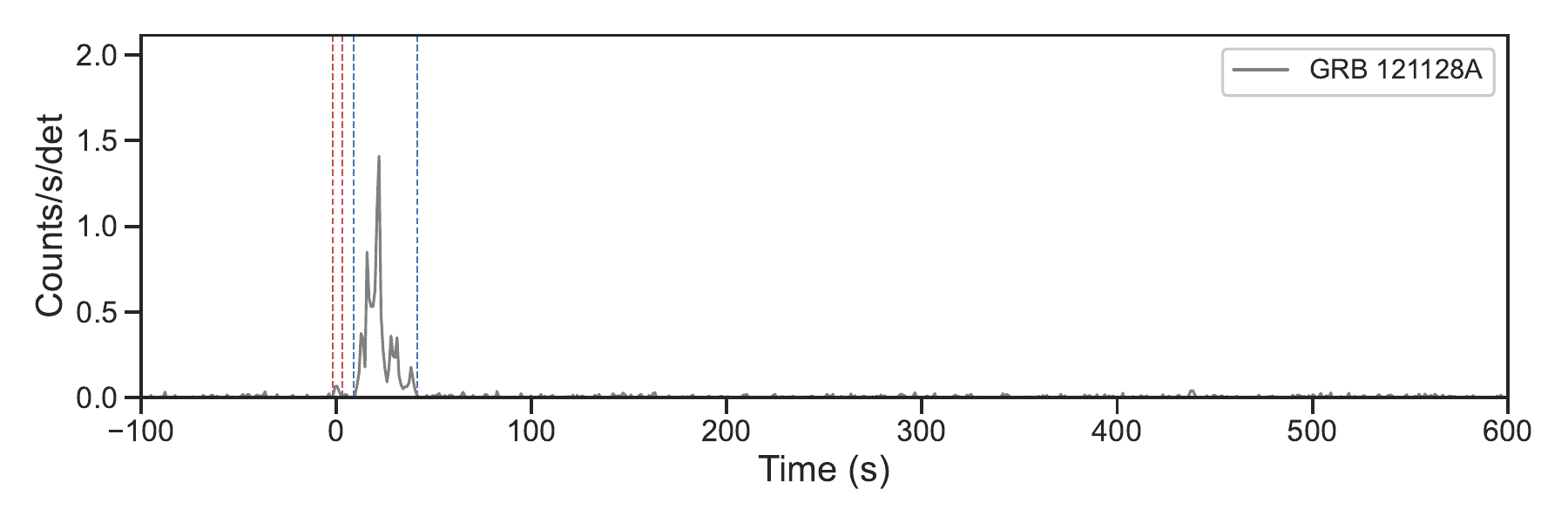}
\includegraphics[angle=0,scale=0.32]{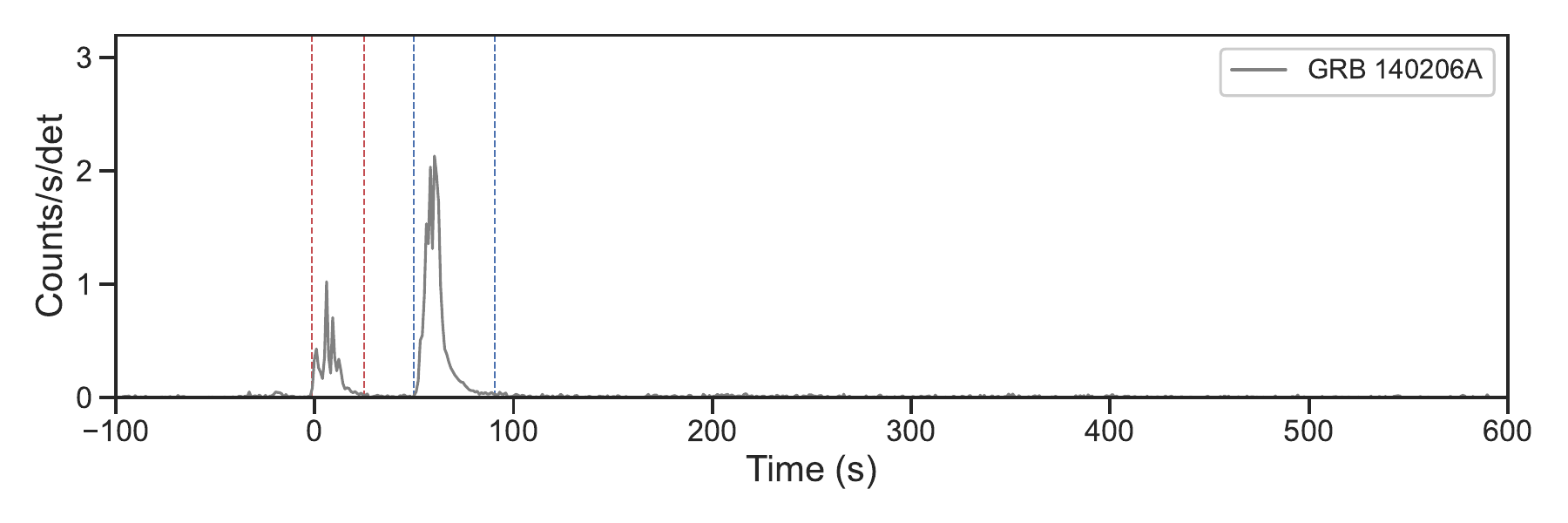}
\includegraphics[angle=0,scale=0.32]{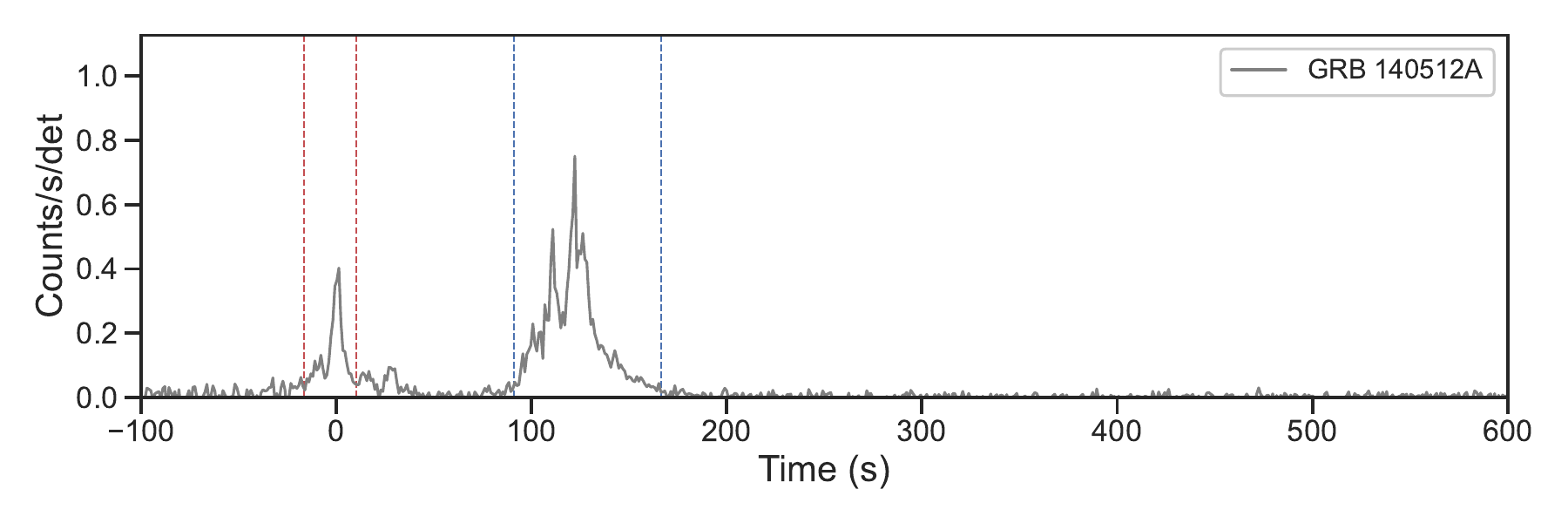}
\includegraphics[angle=0,scale=0.32]{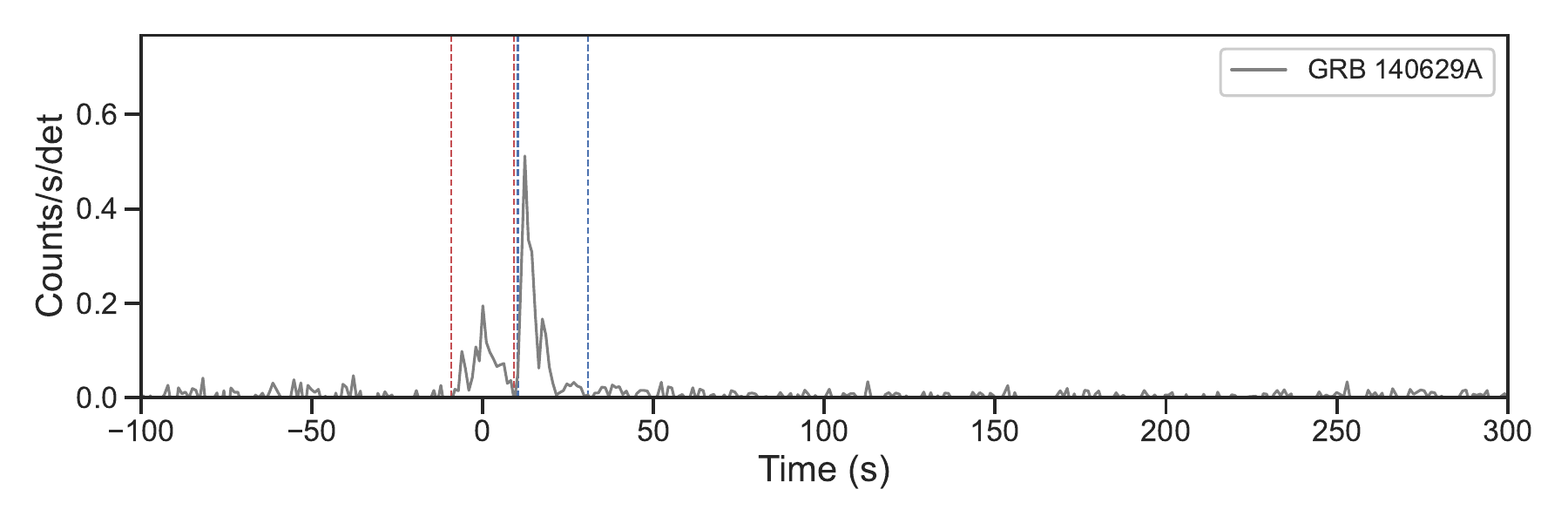}
\includegraphics[angle=0,scale=0.32]{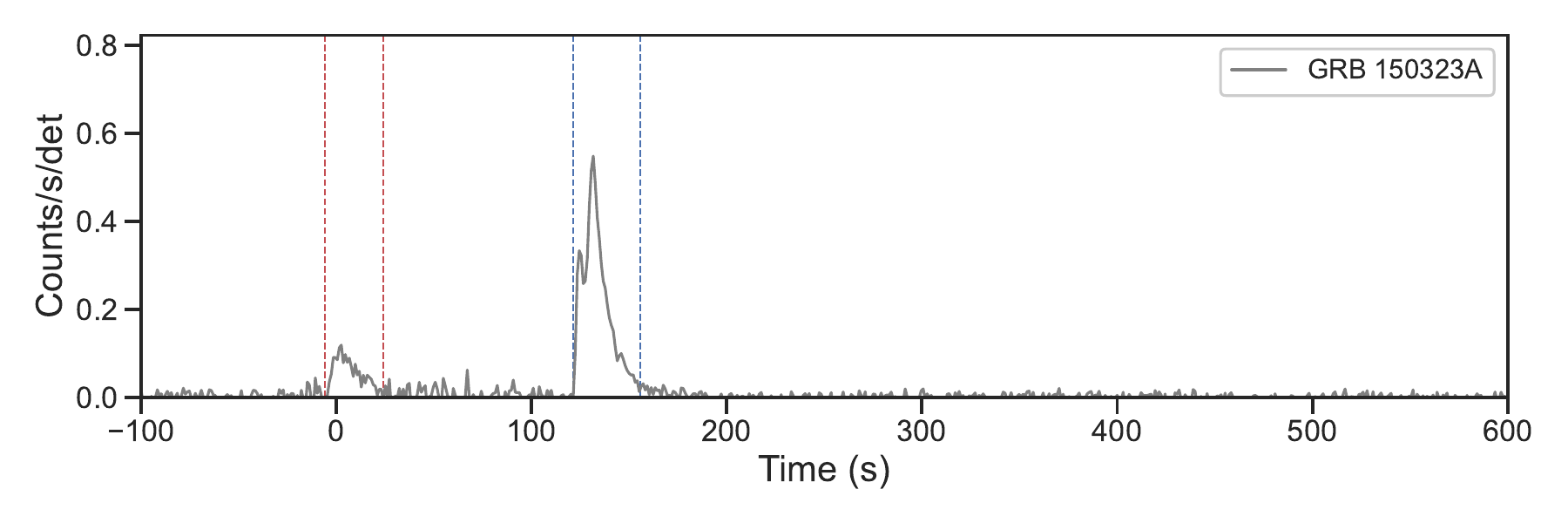}
\includegraphics[angle=0,scale=0.32]{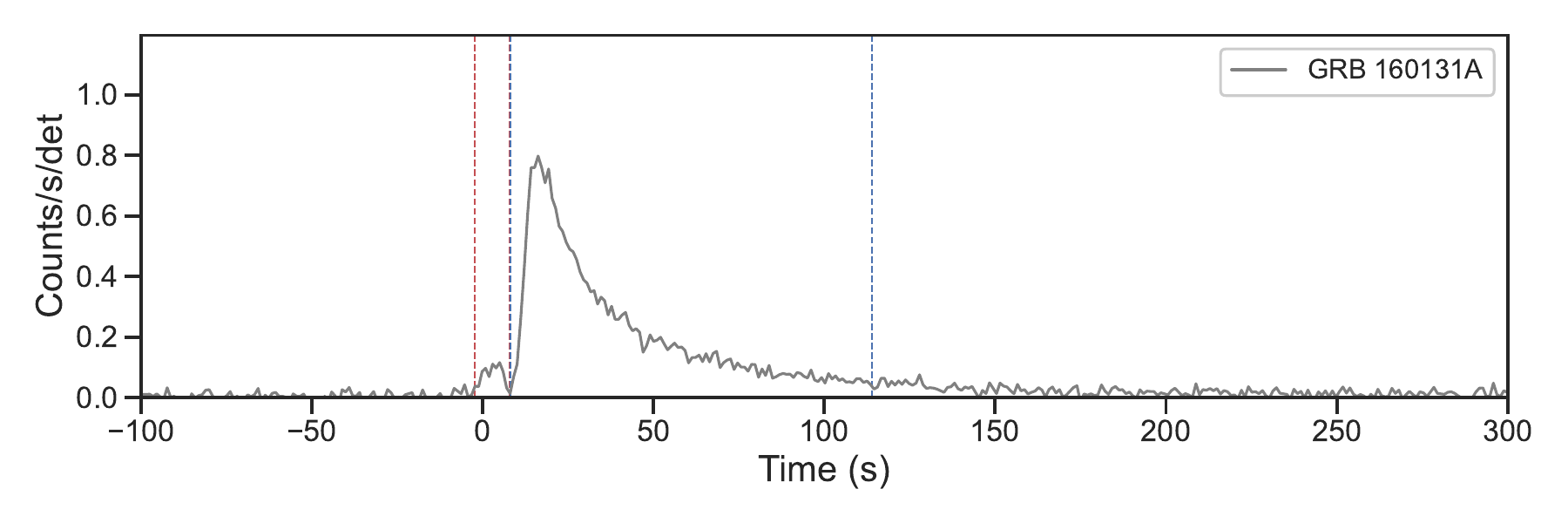}
\includegraphics[angle=0,scale=0.32]{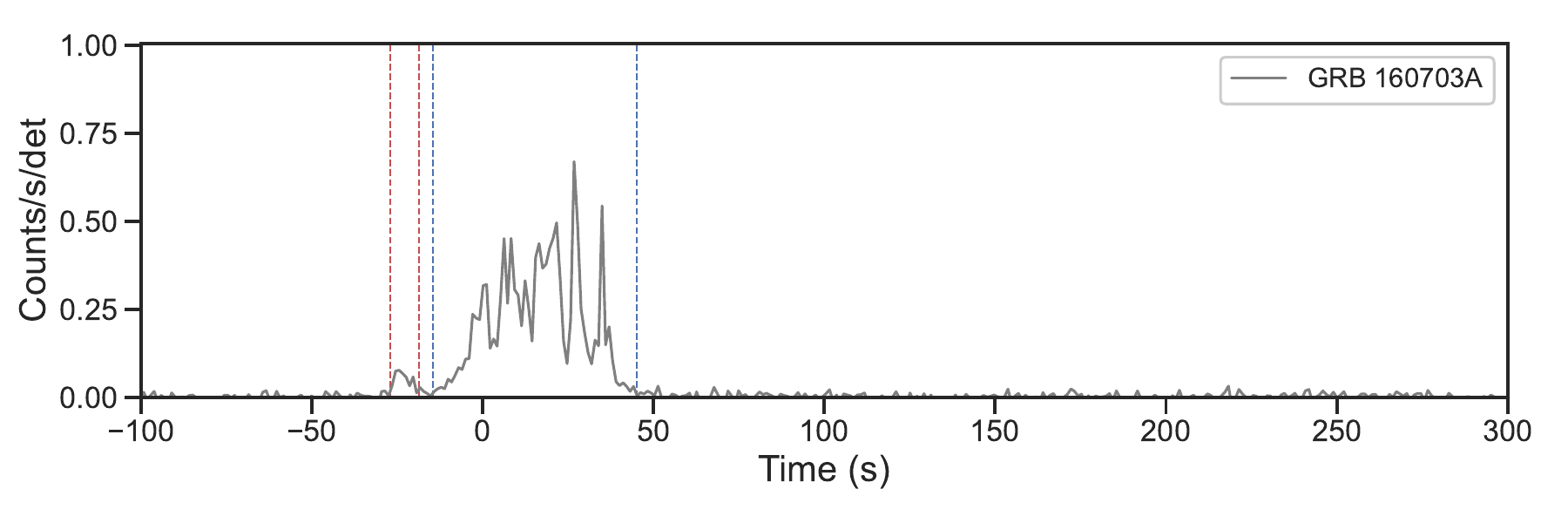}
\includegraphics[angle=0,scale=0.32]{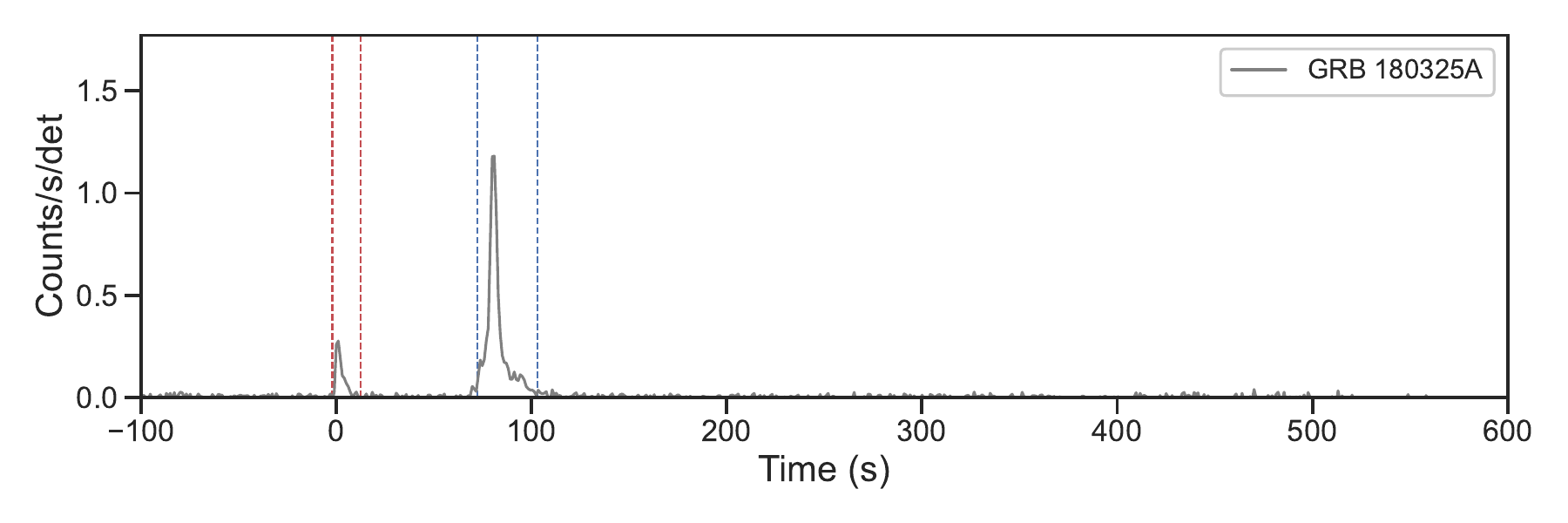}
\includegraphics[angle=0,scale=0.32]{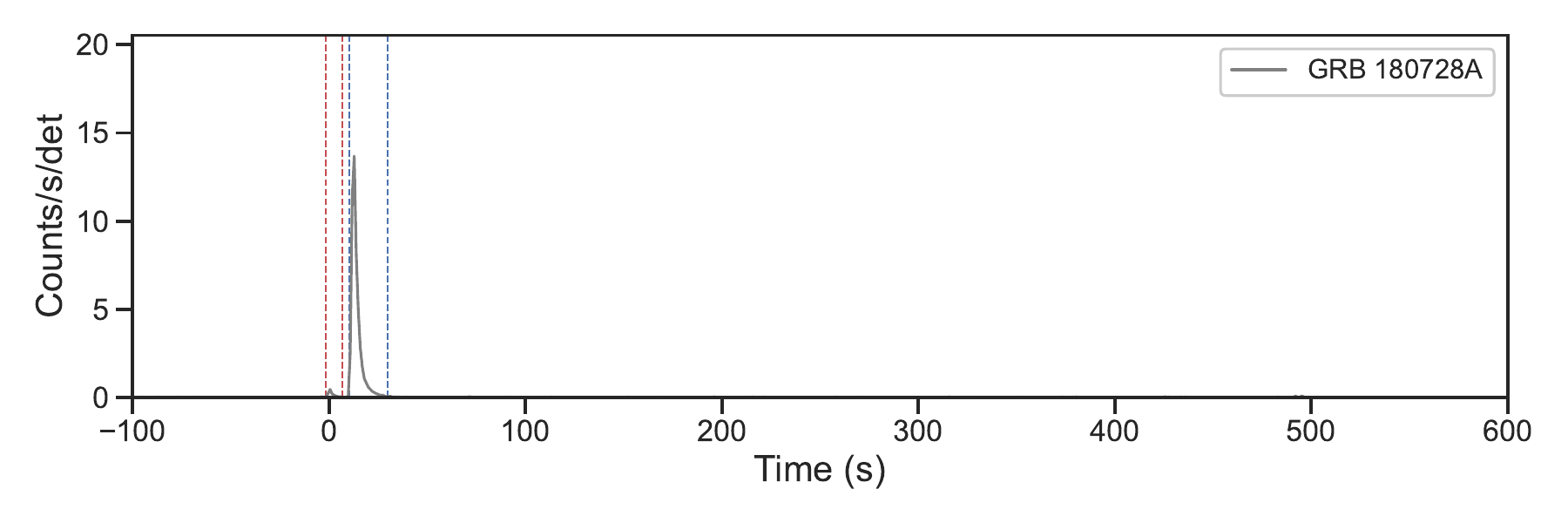}
\includegraphics[angle=0,scale=0.32]{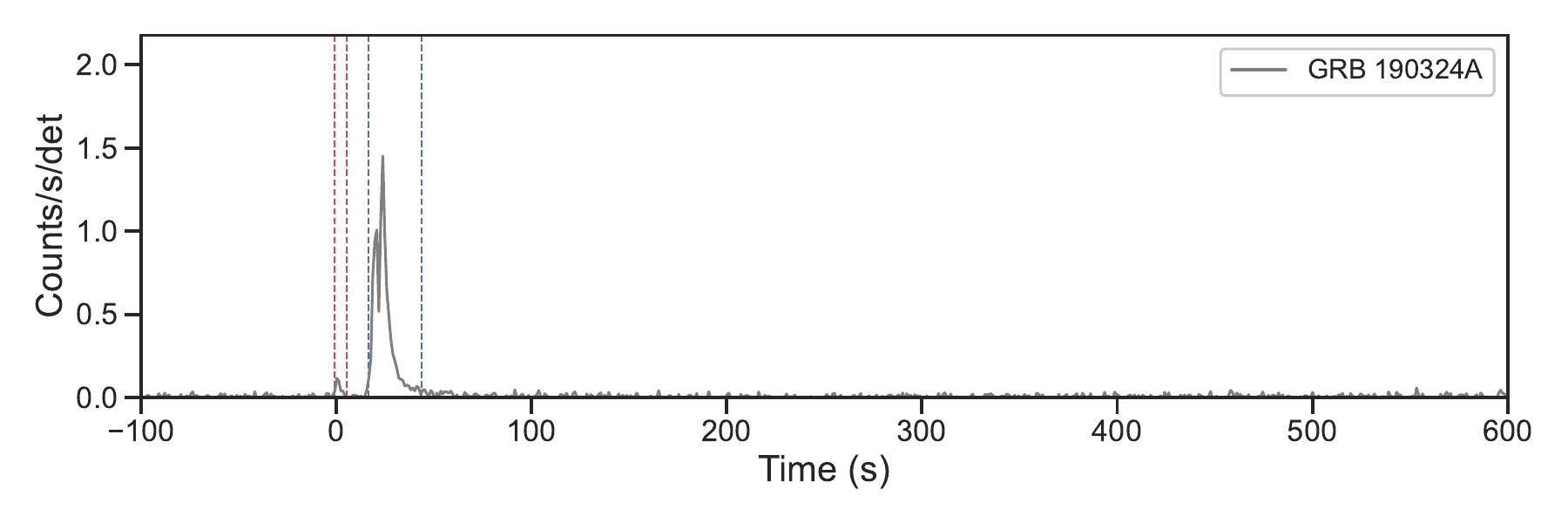}
\center{Figure \ref{fig:lc_sample}--- Continued}
\end{figure*}

\begin{figure*}
\includegraphics[width=0.5\textwidth]{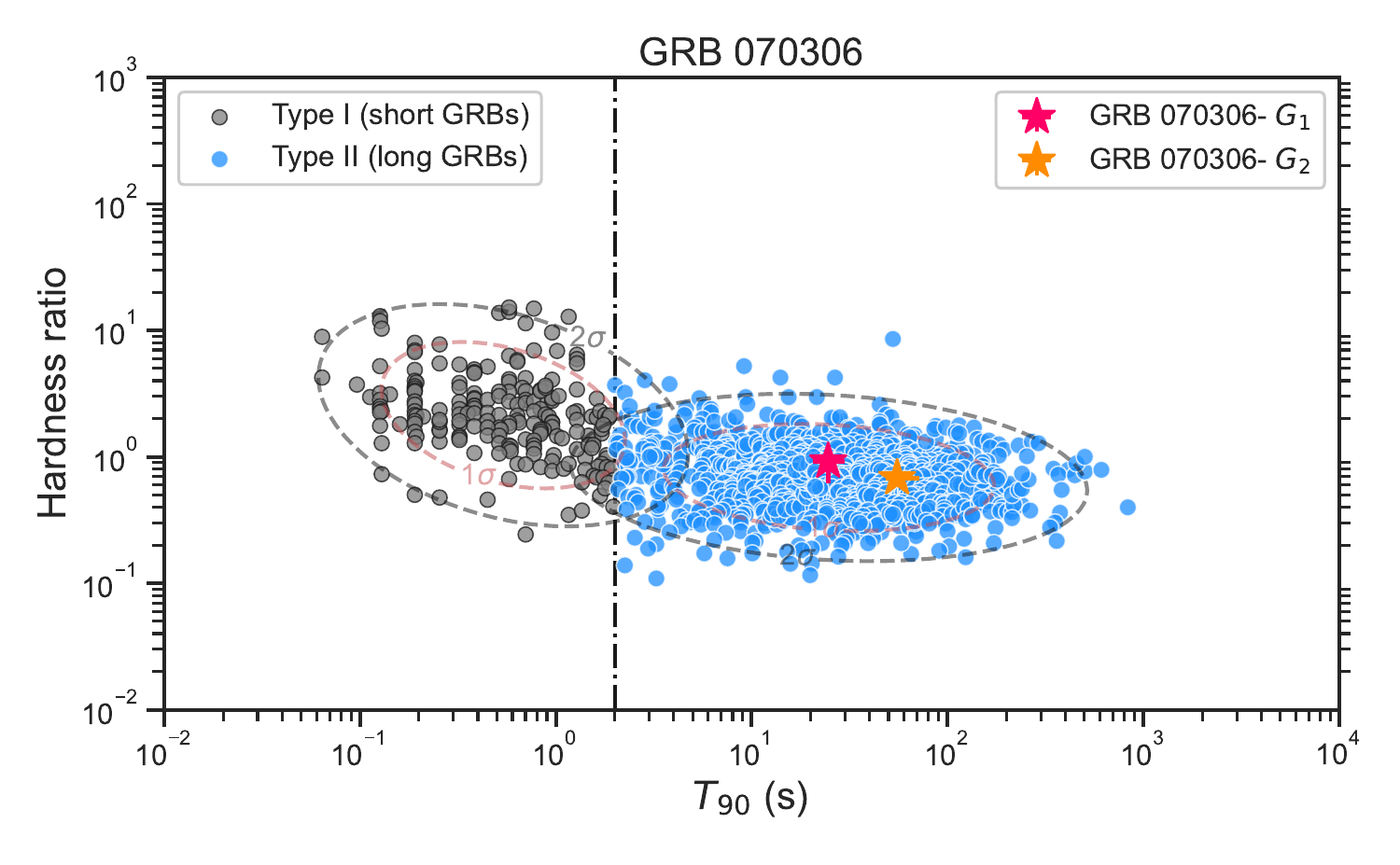}
\includegraphics[width=0.5\textwidth]{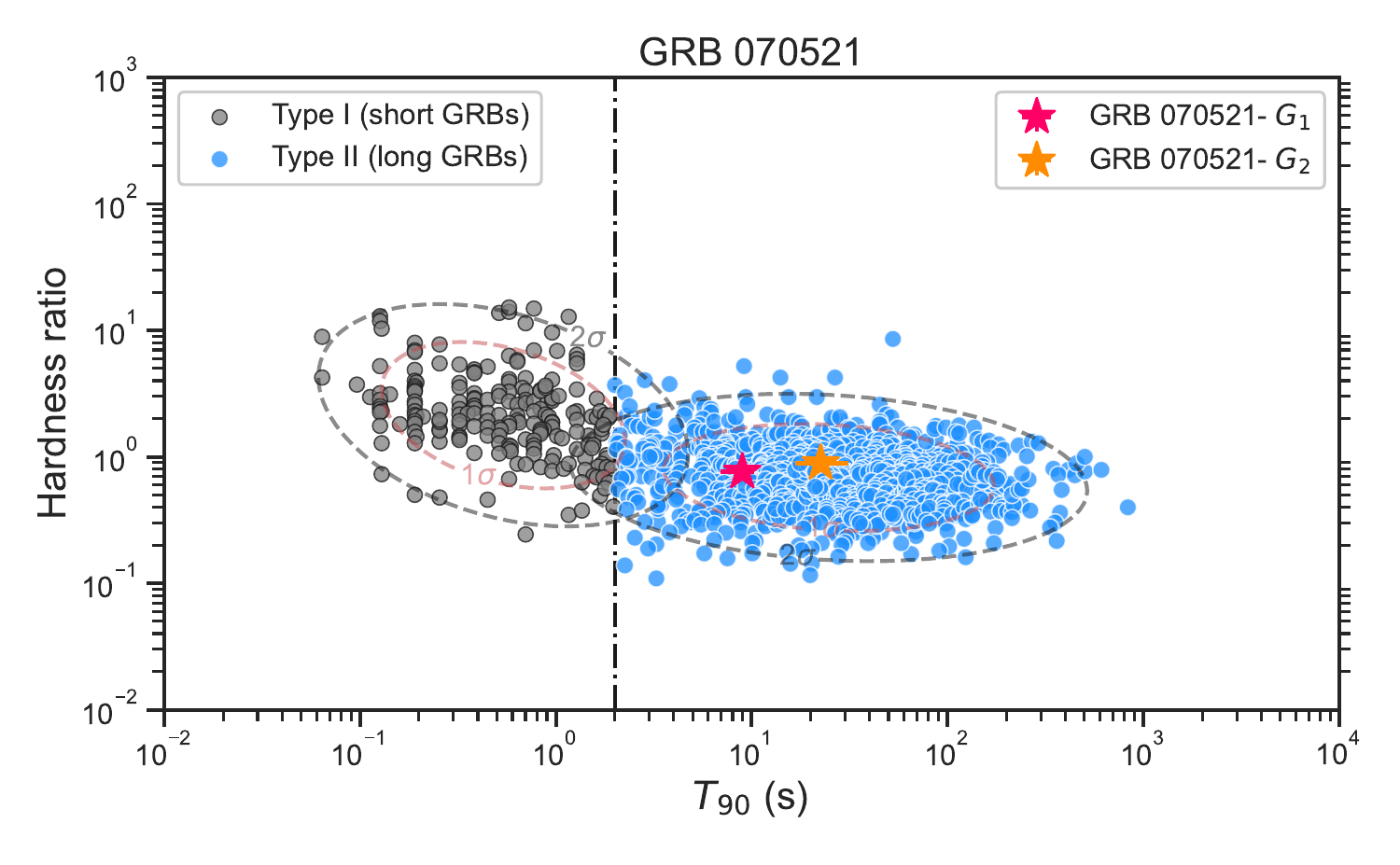}
\includegraphics[width=0.5\textwidth]{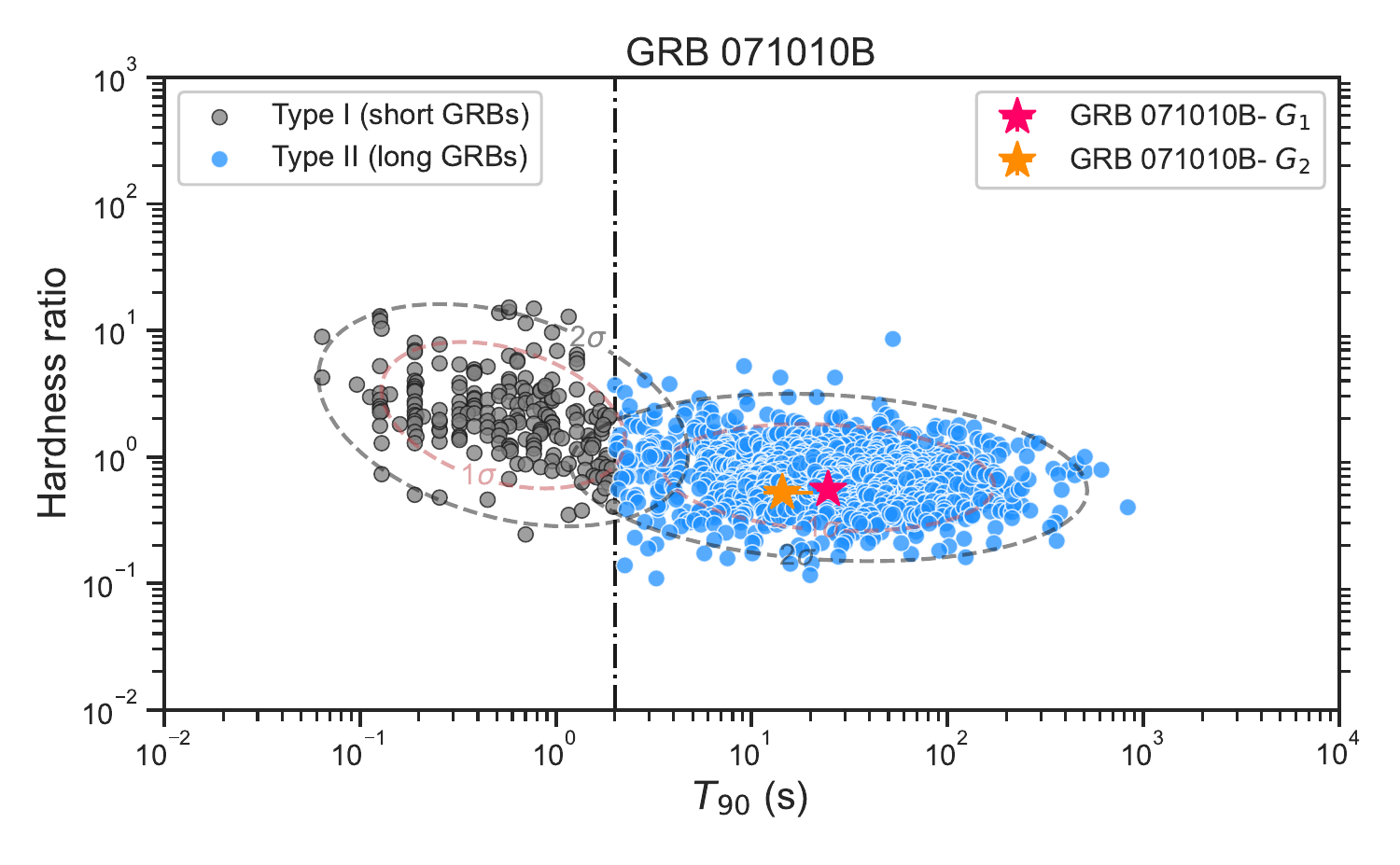}
\includegraphics[width=0.5\textwidth]{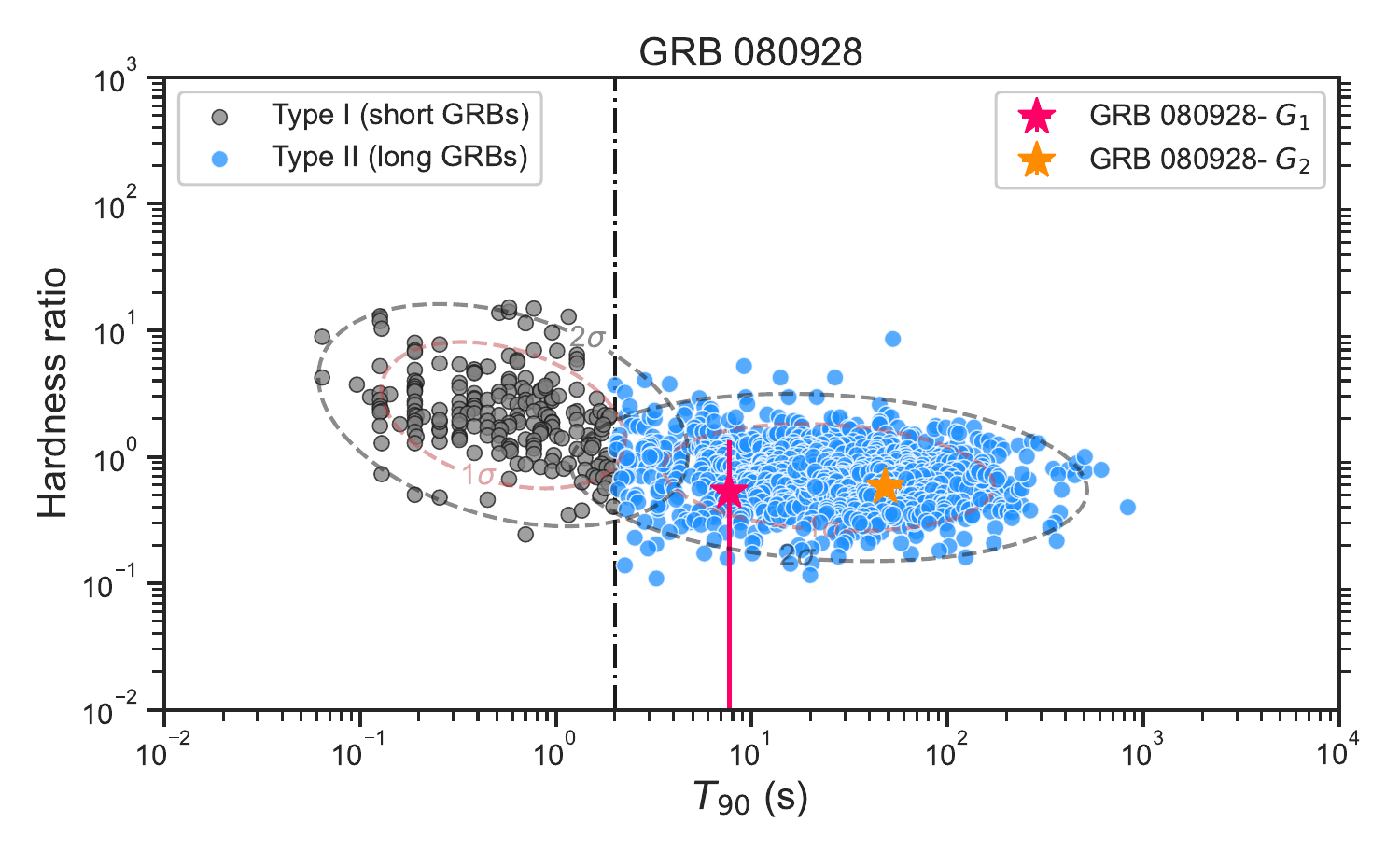}
\includegraphics[width=0.5\textwidth]{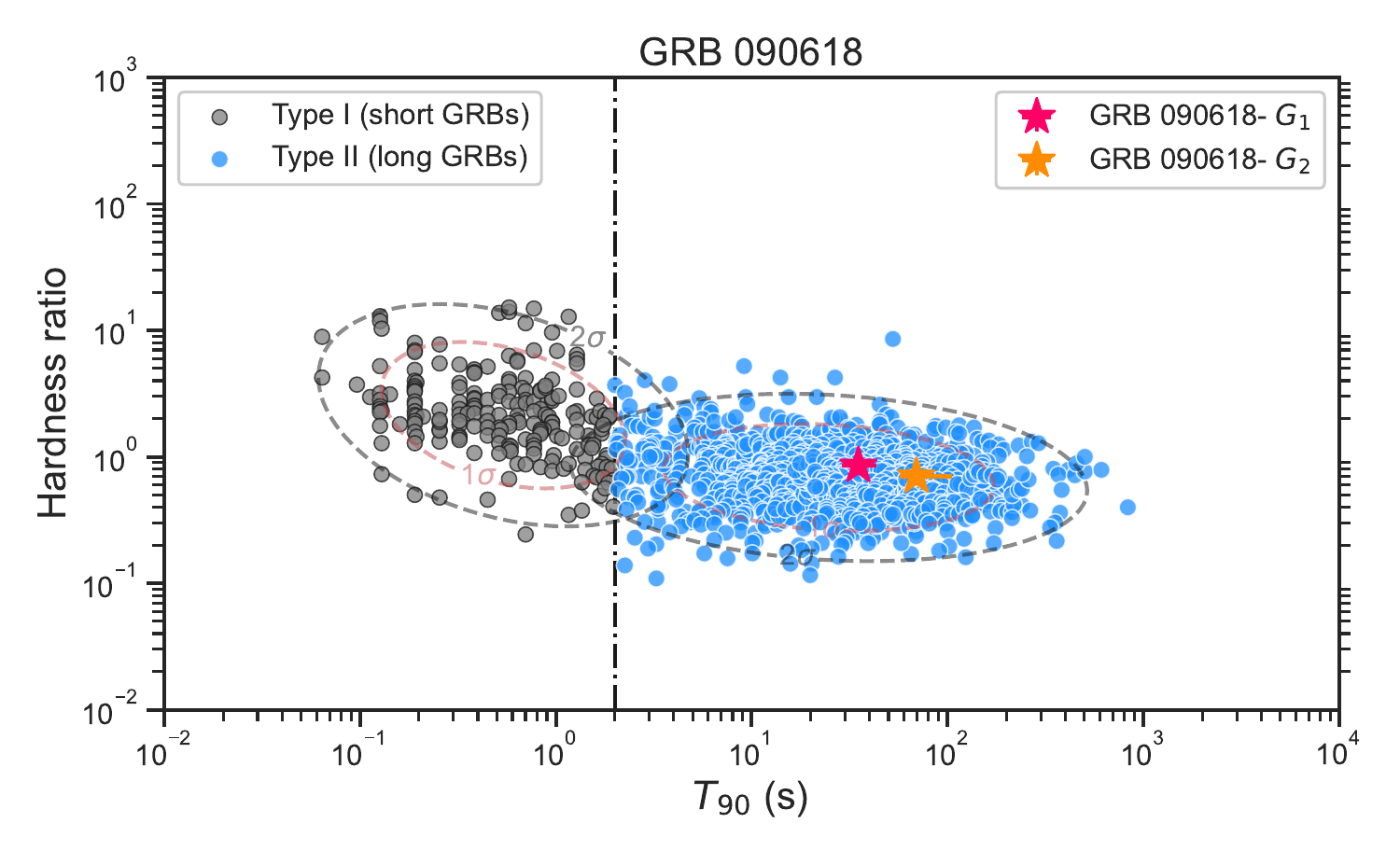}
\includegraphics[width=0.5\textwidth]{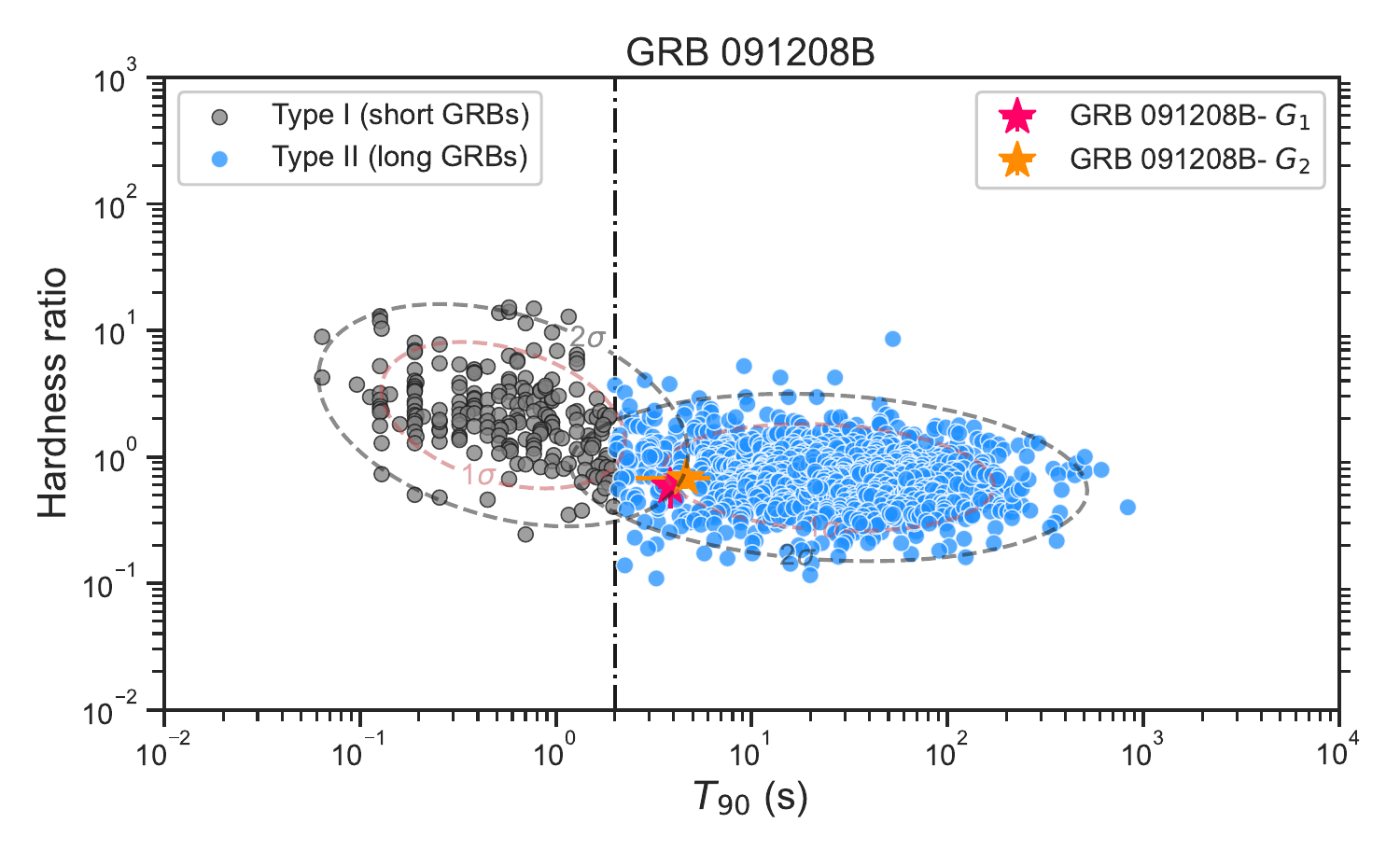}
\center{Figure \ref{fig:HR}--- Continued}
\end{figure*}
\begin{figure*}
\includegraphics[width=0.5\textwidth]{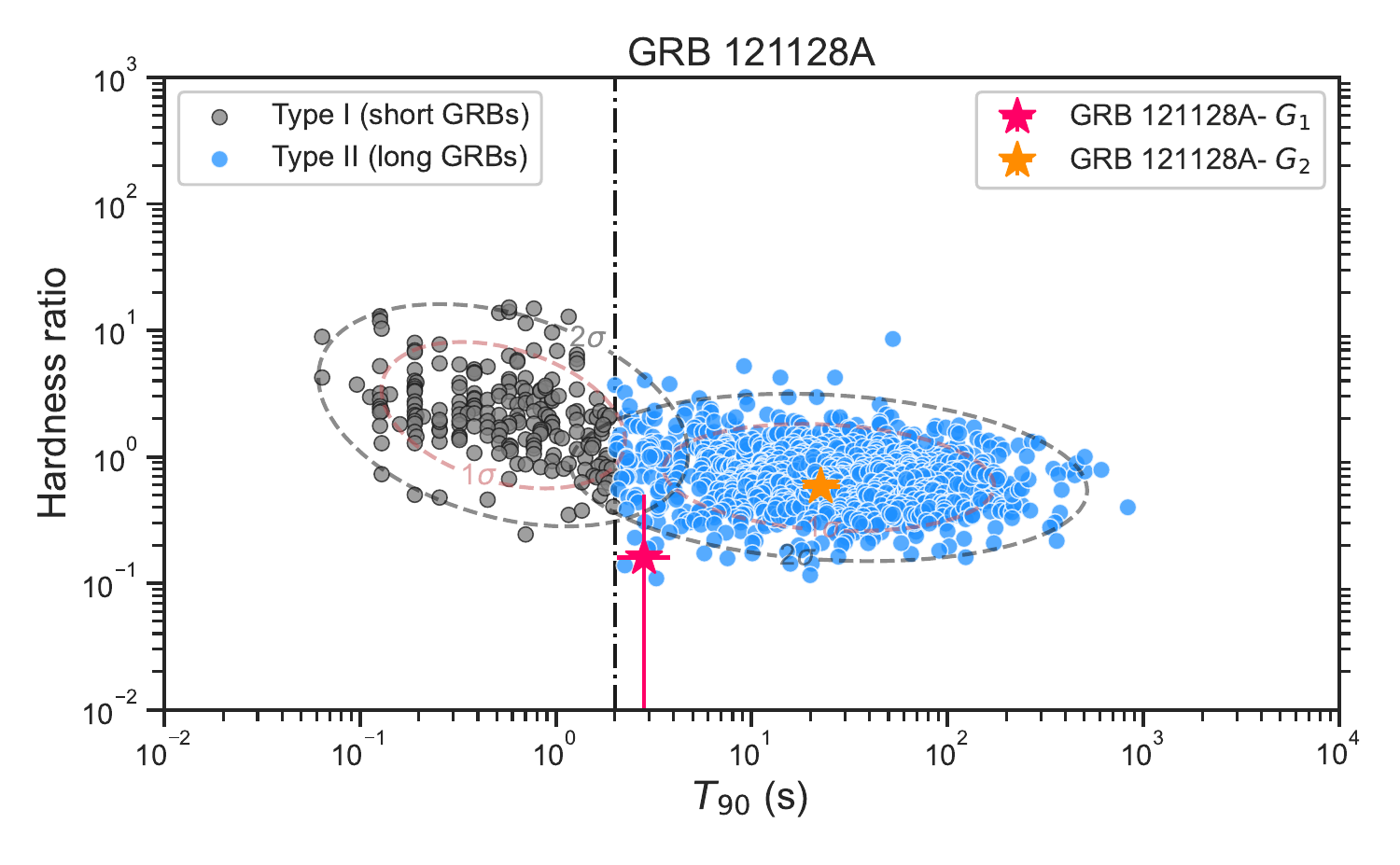}
\includegraphics[width=0.5\textwidth]{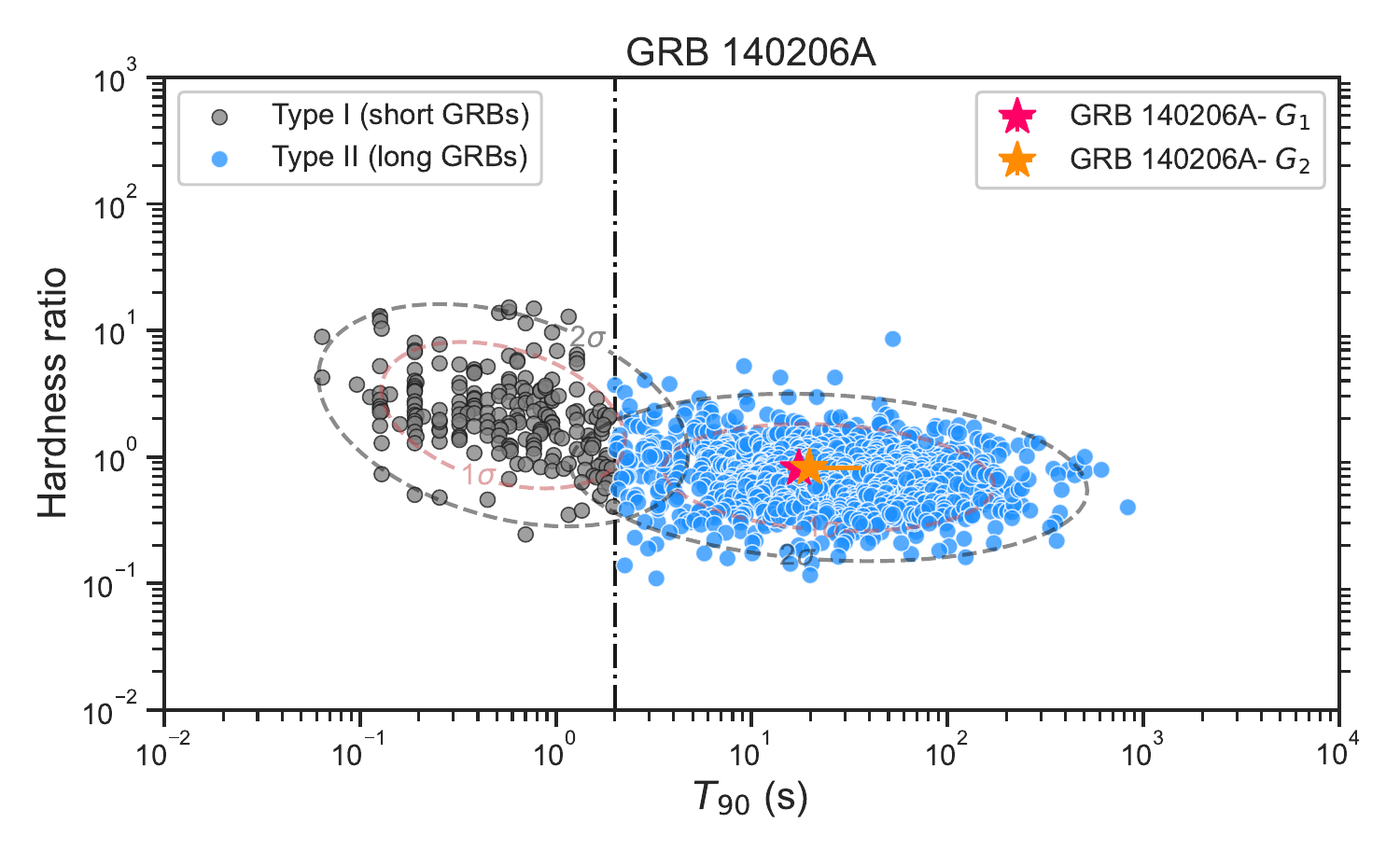}
\includegraphics[width=0.5\textwidth]{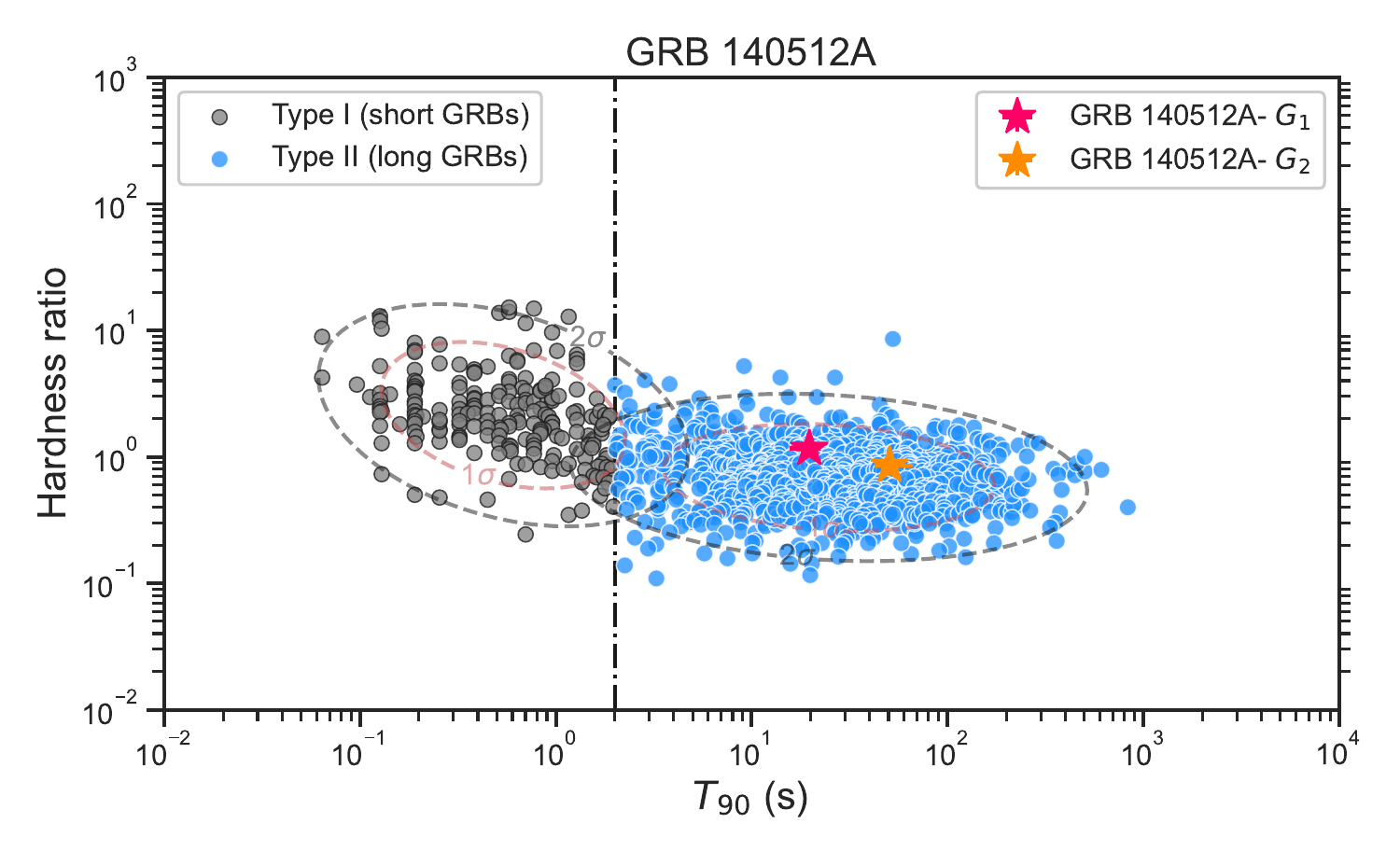}
\includegraphics[width=0.5\textwidth]{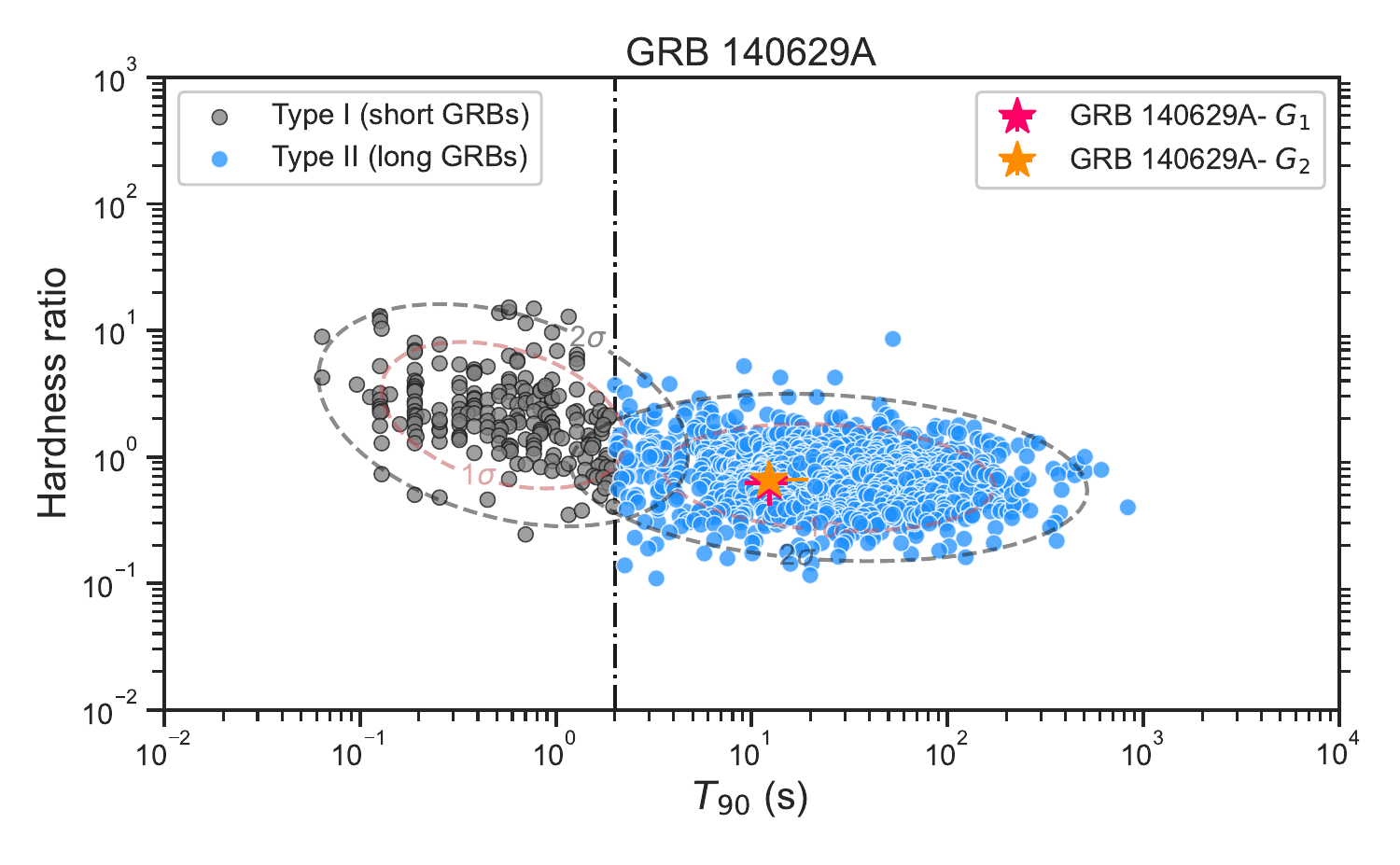}
\includegraphics[width=0.5\textwidth]{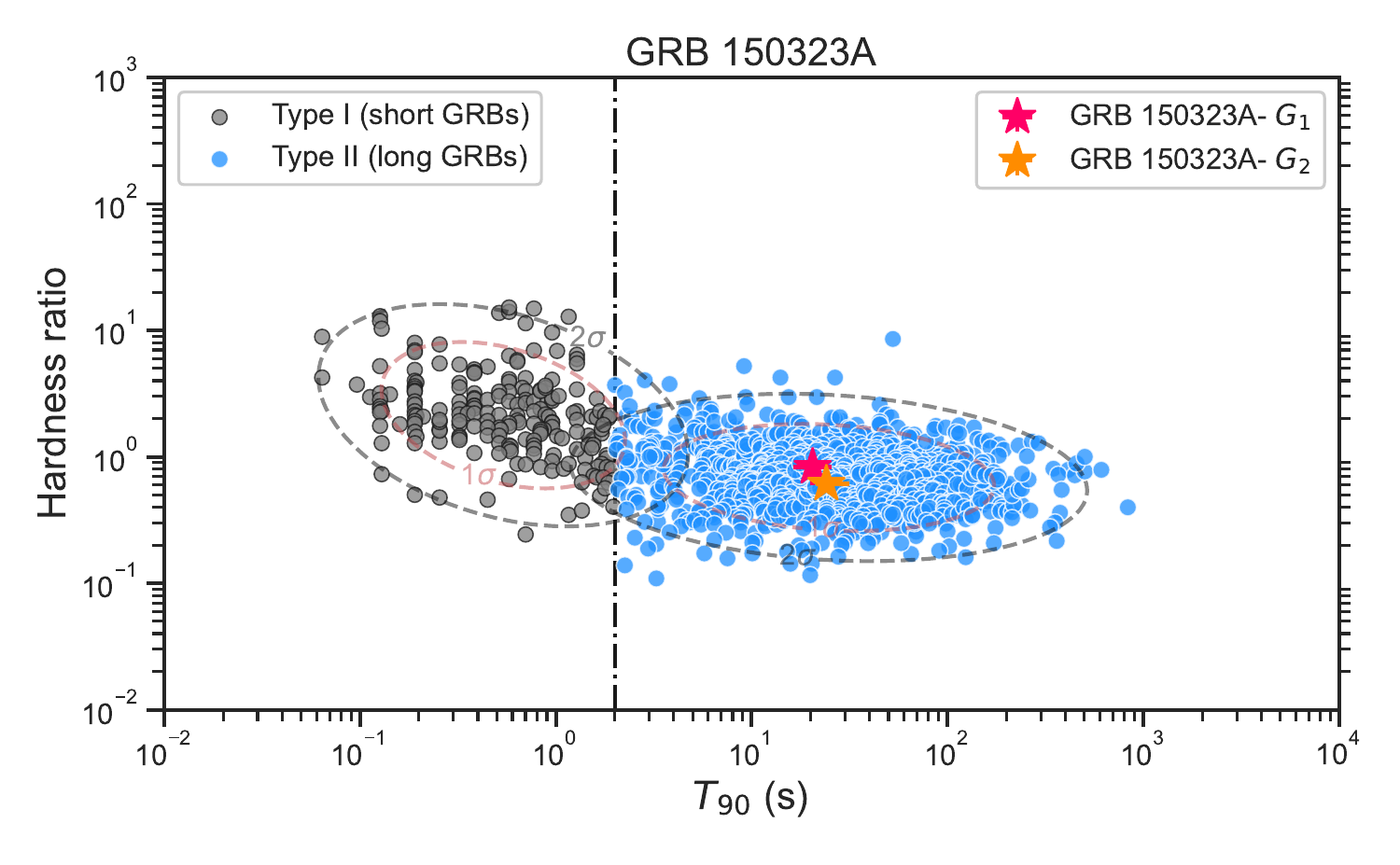}
\includegraphics[width=0.5\textwidth]{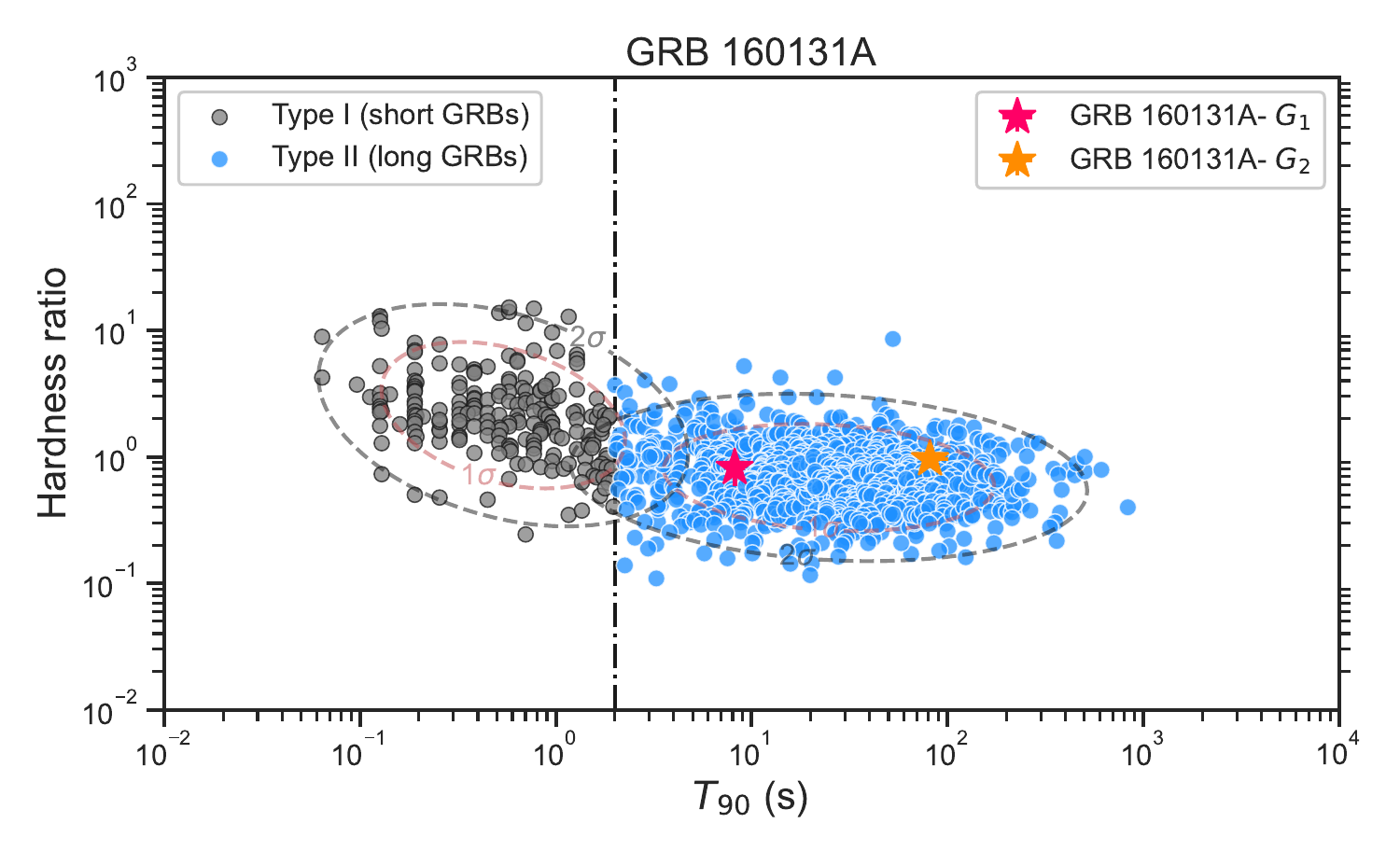}
\center{Figure \ref{fig:HR}--- Continued}
\end{figure*}
\begin{figure*}
\includegraphics[width=0.5\textwidth]{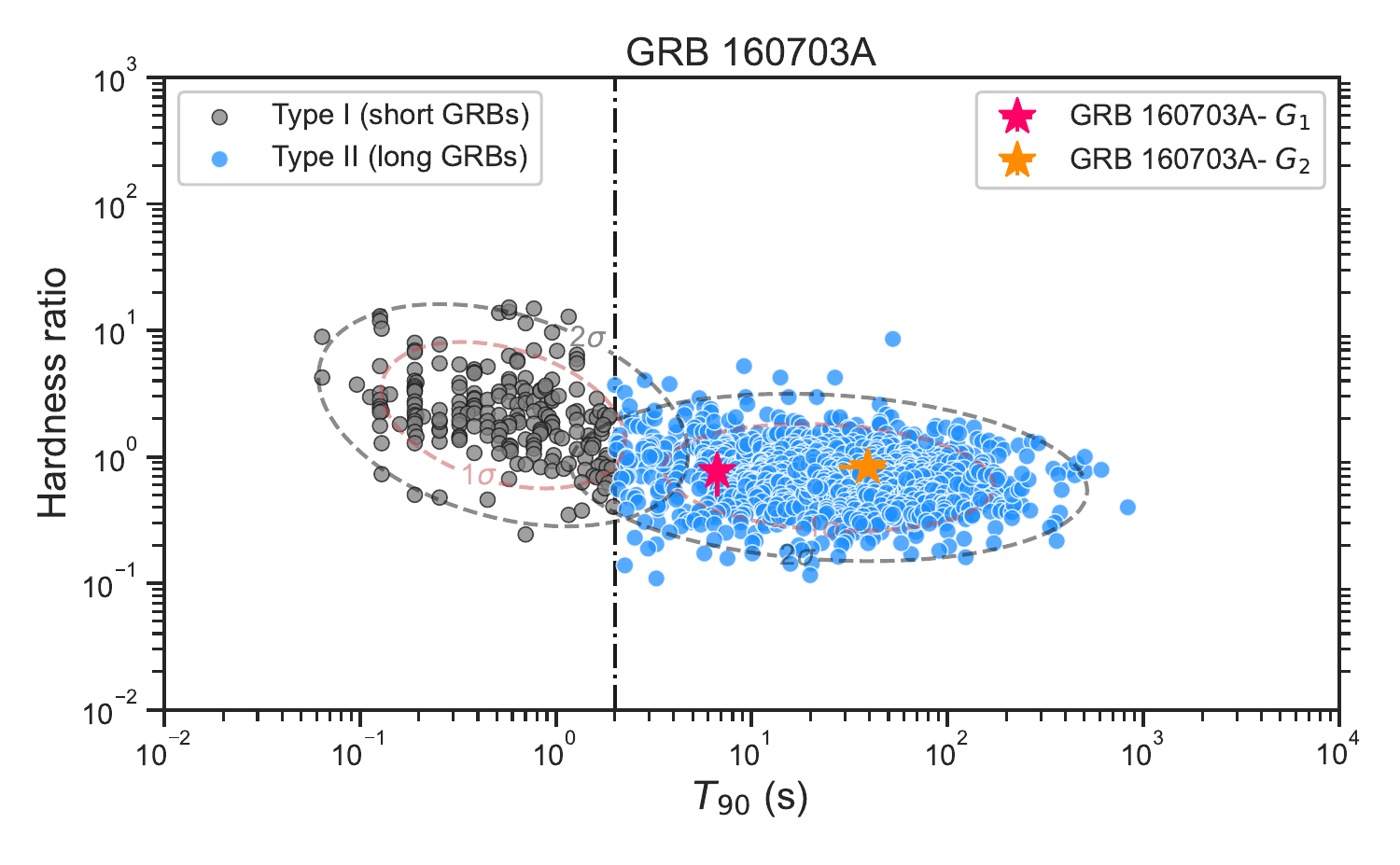}
\includegraphics[width=0.5\textwidth]{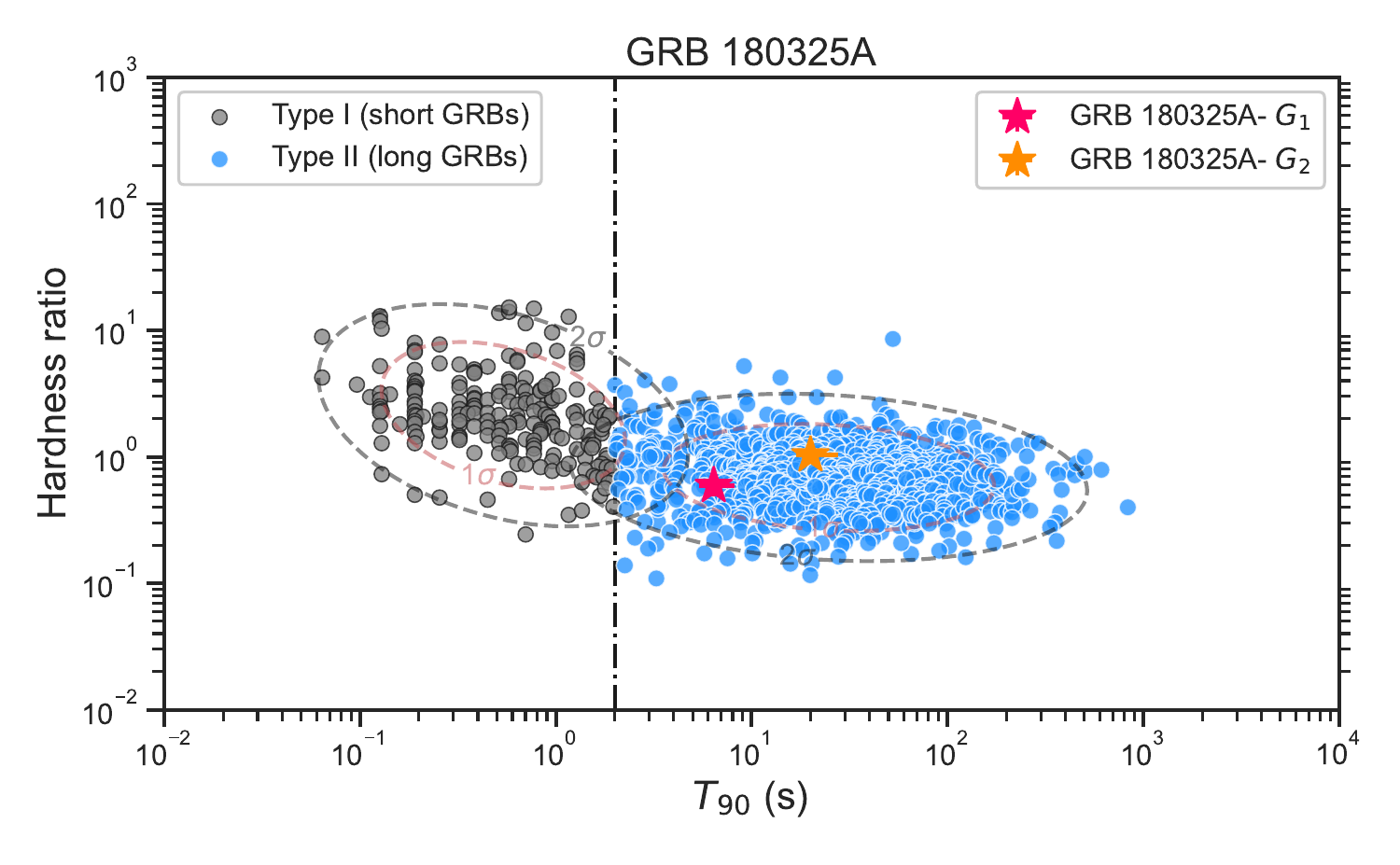}
\includegraphics[width=0.5\textwidth]{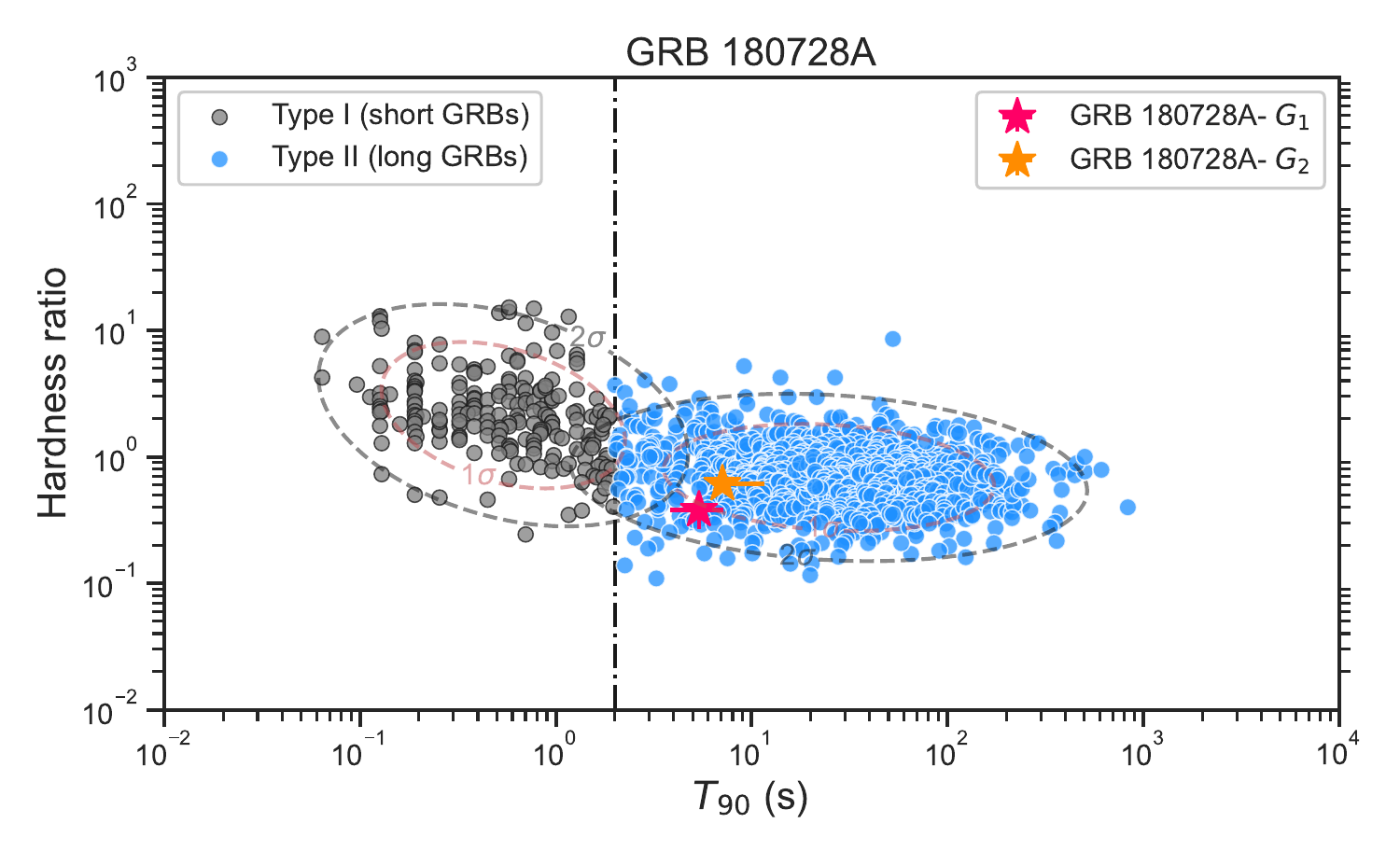}
\includegraphics[width=0.5\textwidth]{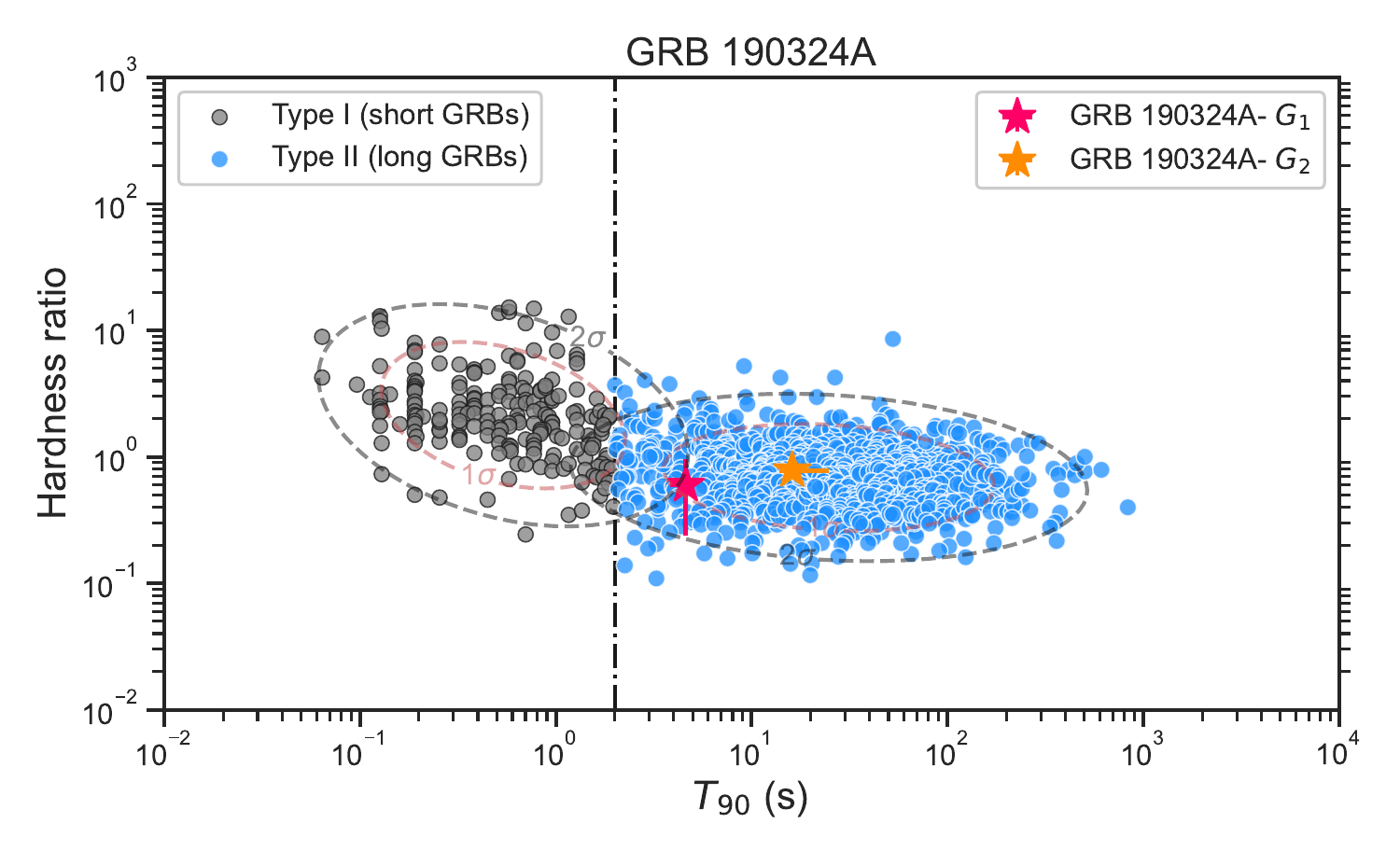}
\center{Figure \ref{fig:HR}--- Continued}
\end{figure*}

\begin{figure*}
\includegraphics[width=0.5\textwidth]{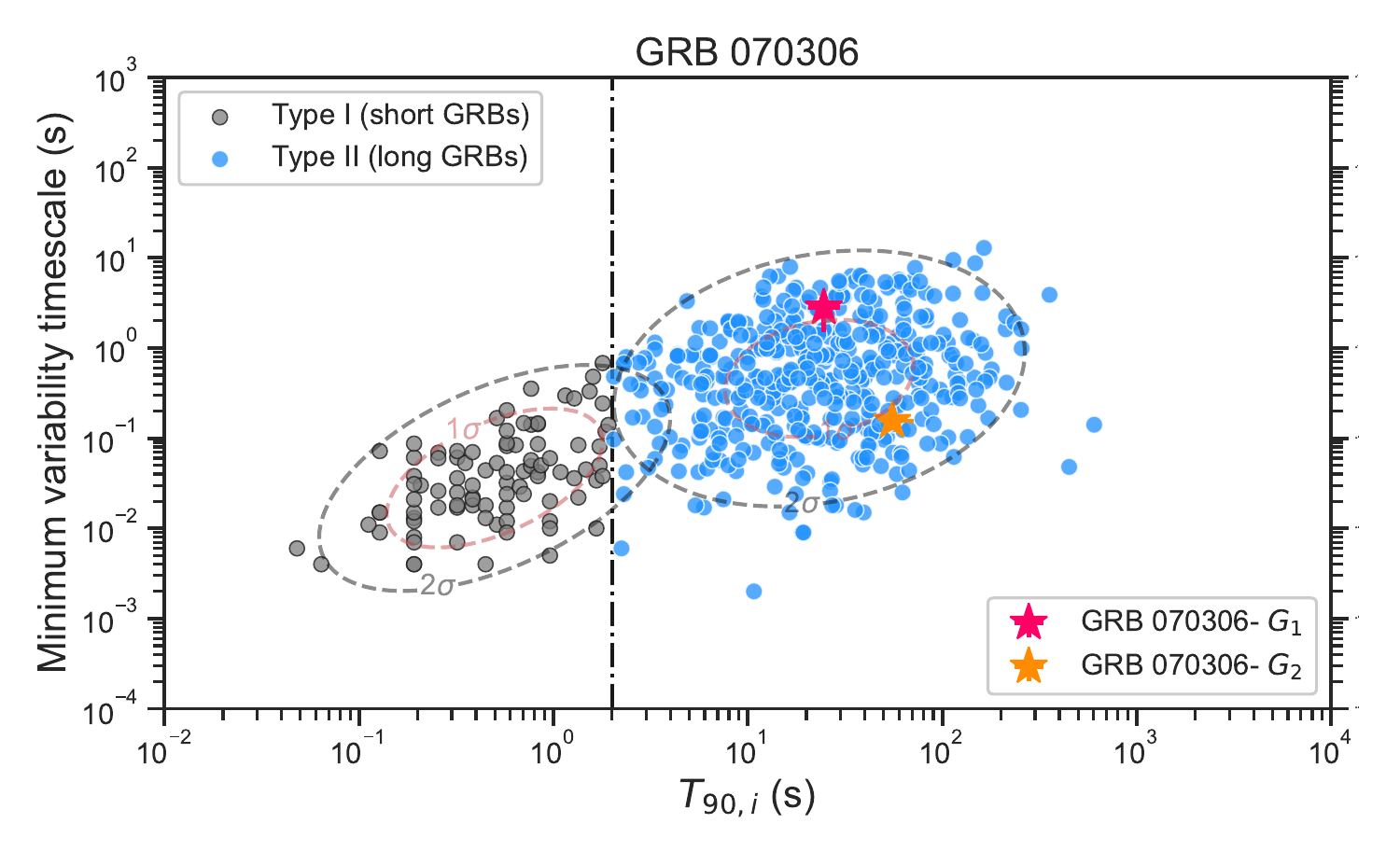}
\includegraphics[width=0.5\textwidth]{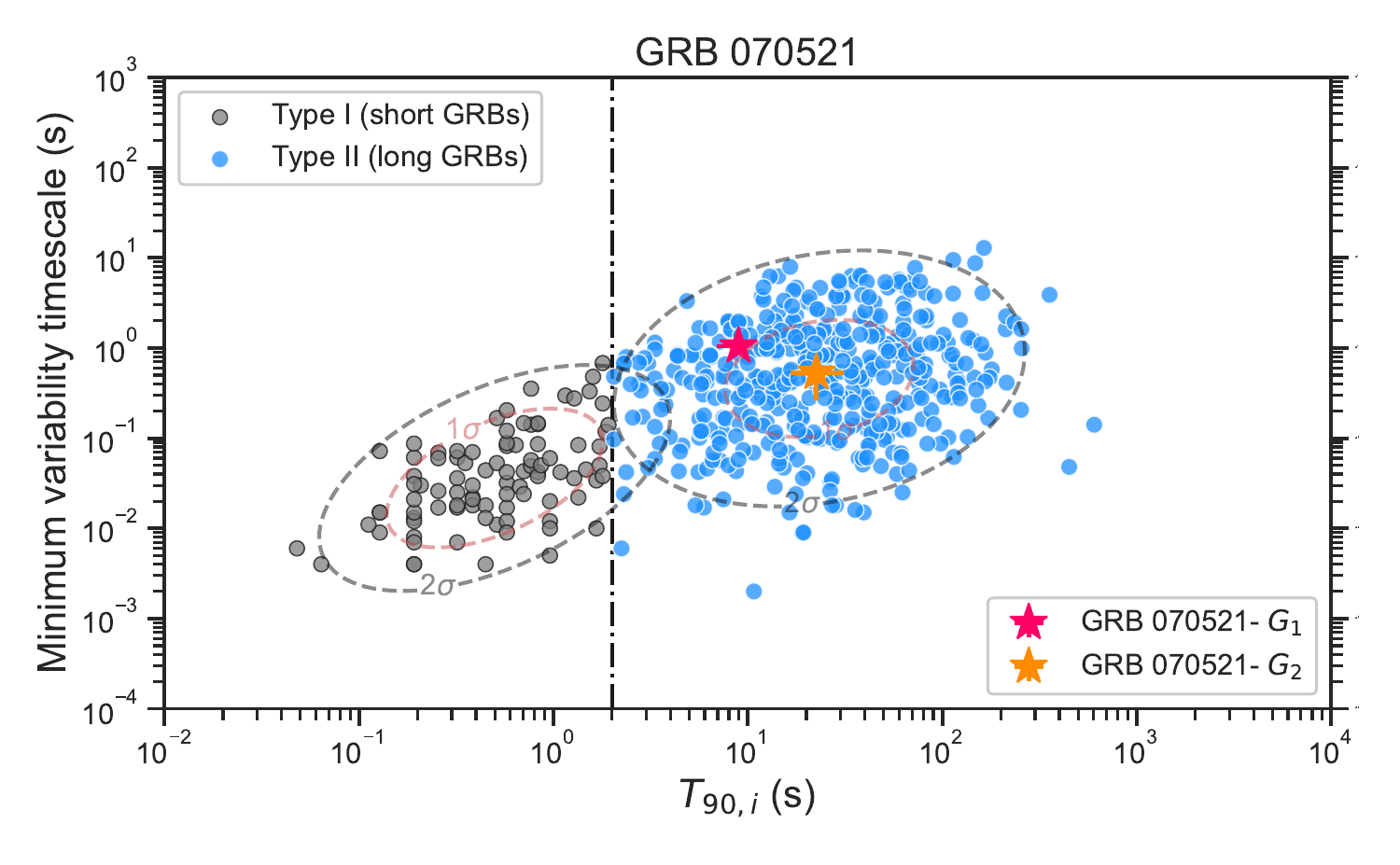}
\includegraphics[width=0.5\textwidth]{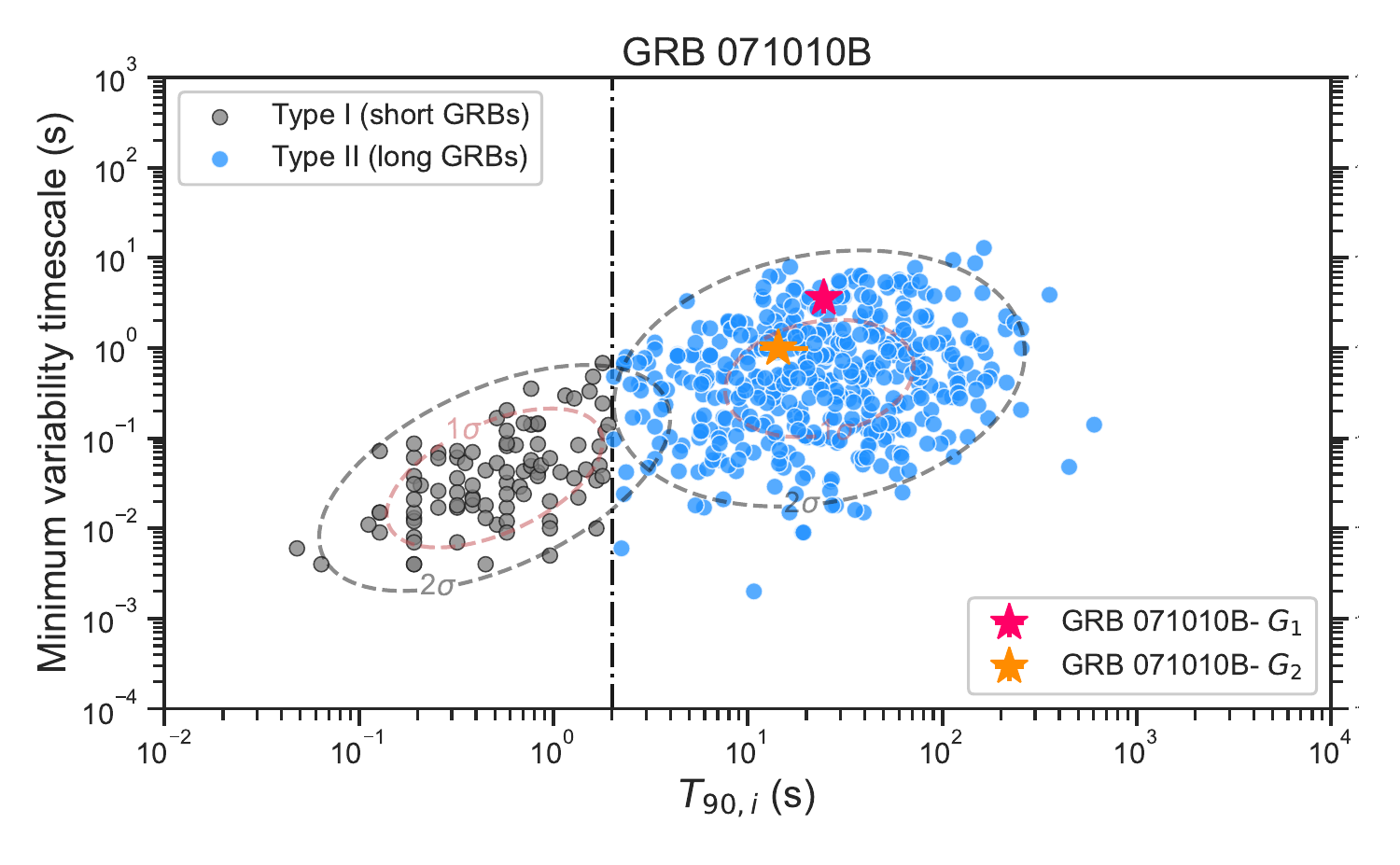}
\includegraphics[width=0.5\textwidth]{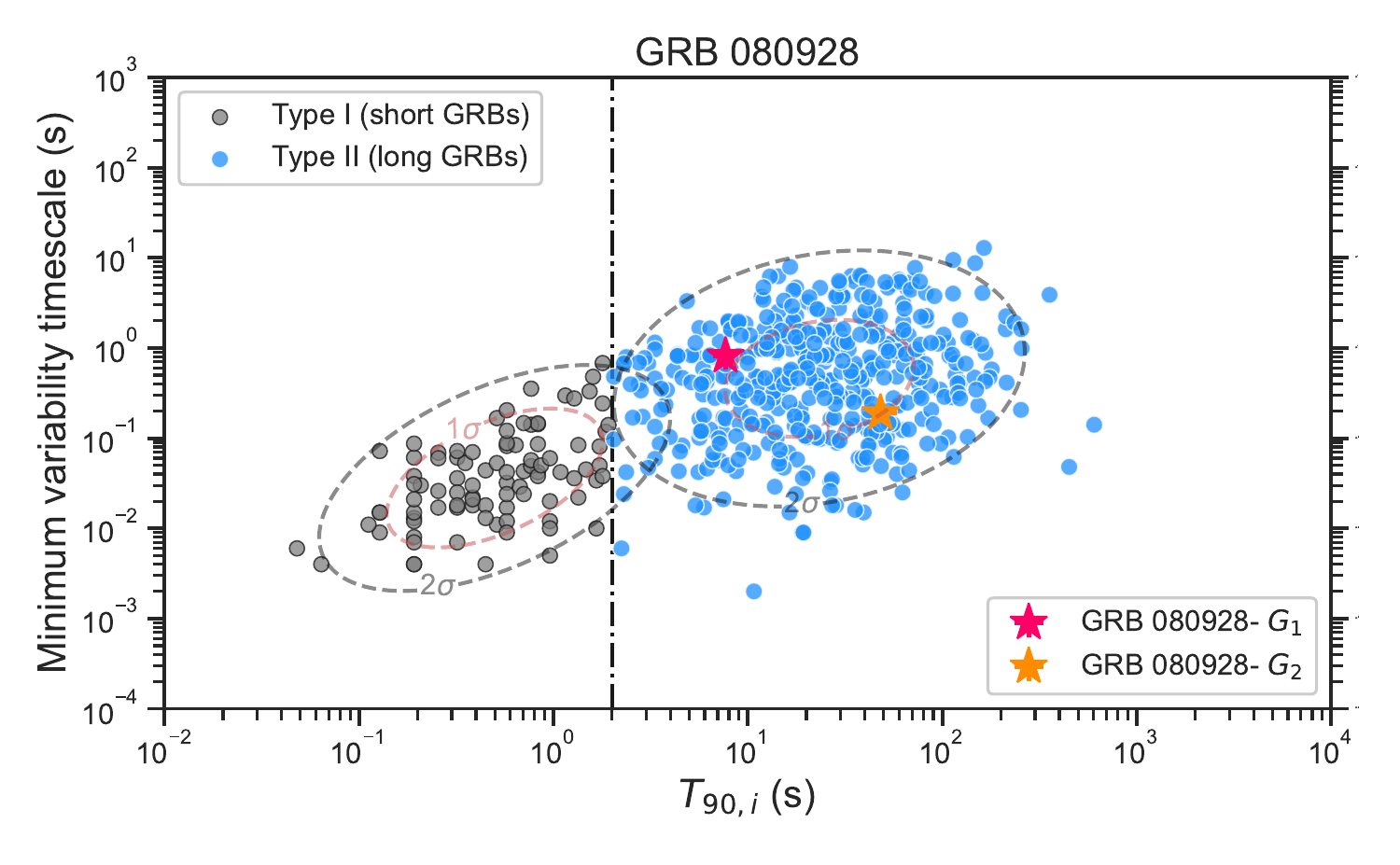}
\includegraphics[width=0.5\textwidth]{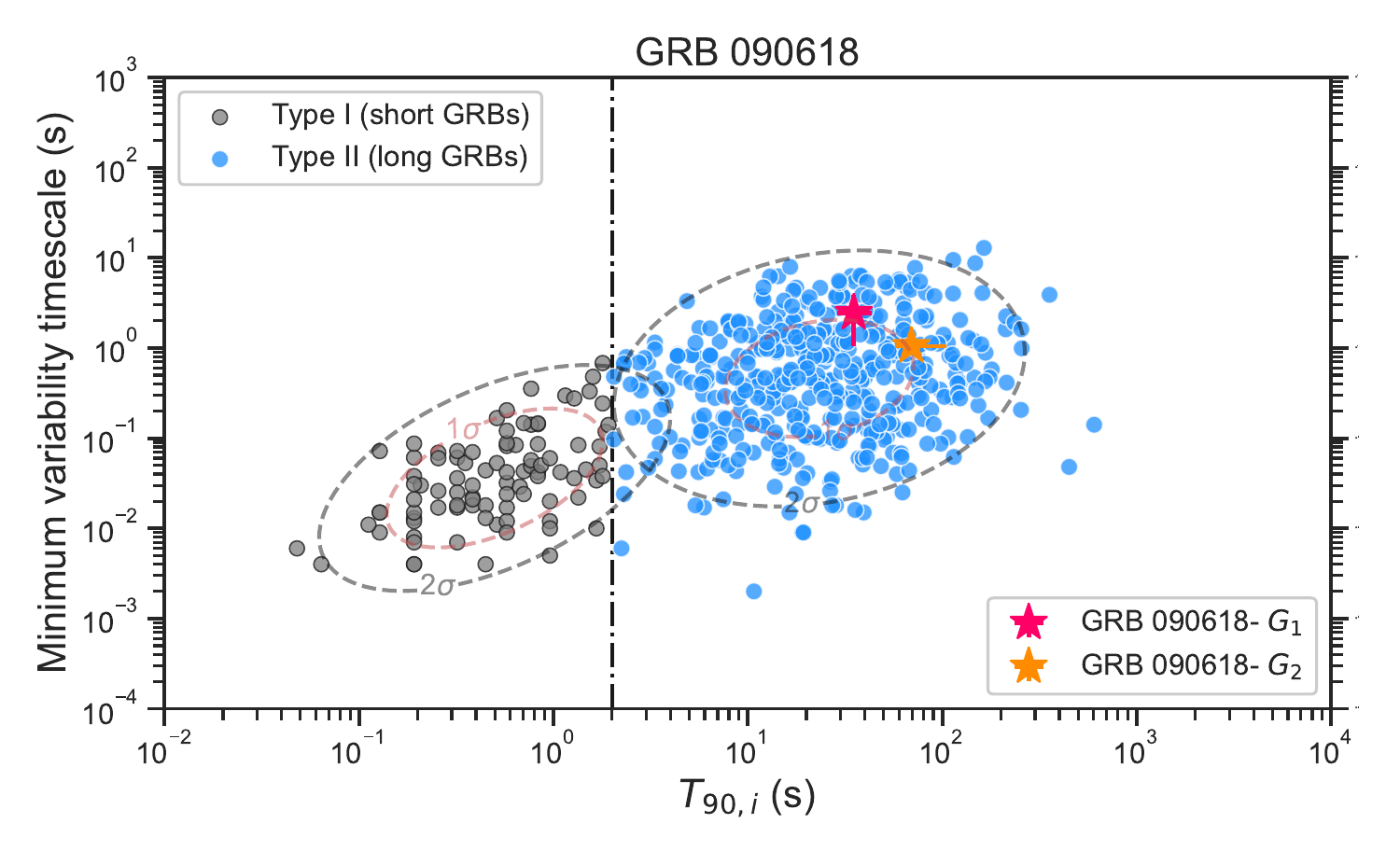}
\includegraphics[width=0.5\textwidth]{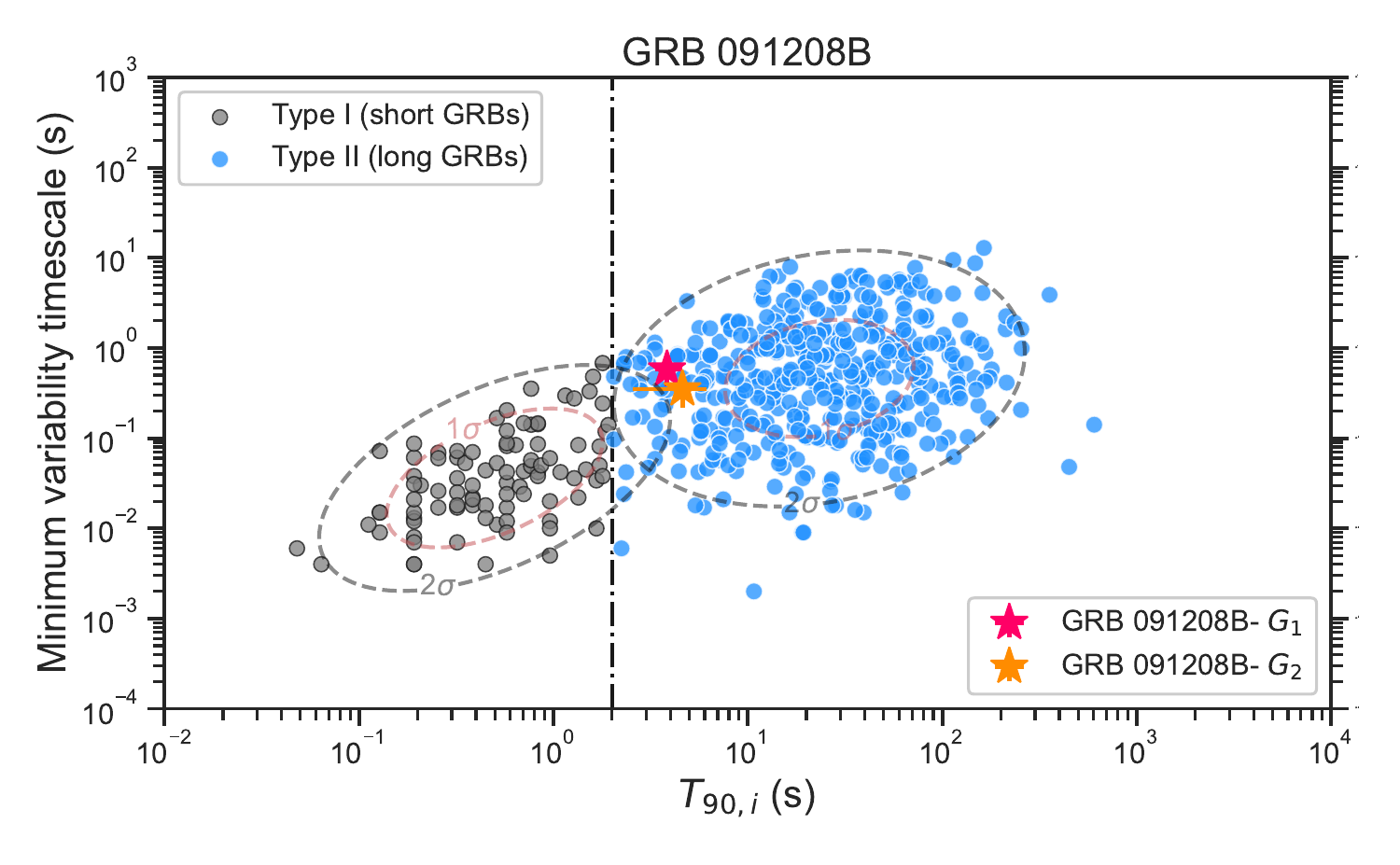}
\center{Figure \ref{fig:MVT_T90}--- Continued}
\end{figure*}
\begin{figure*}
\includegraphics[width=0.5\textwidth]{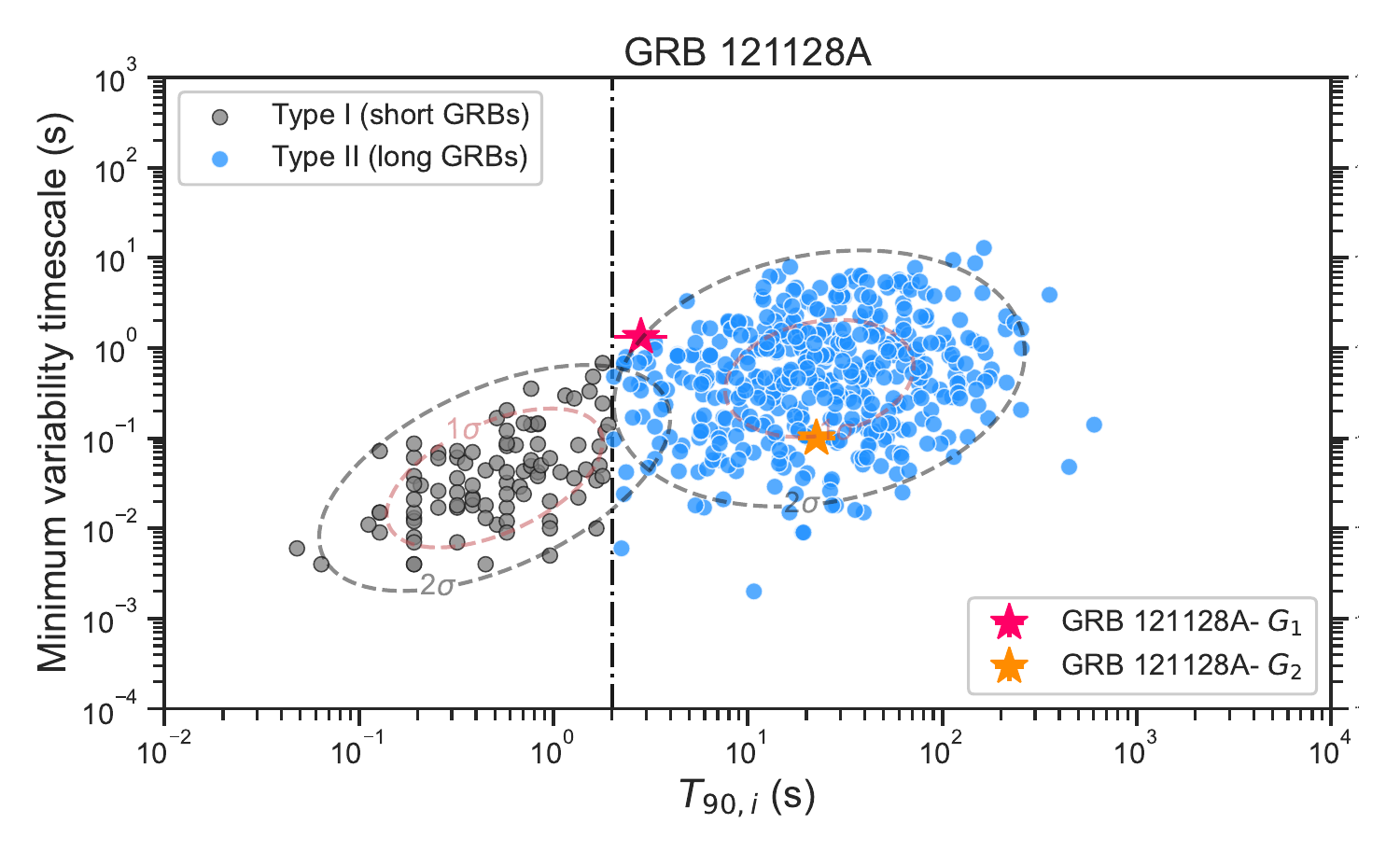}
\includegraphics[width=0.5\textwidth]{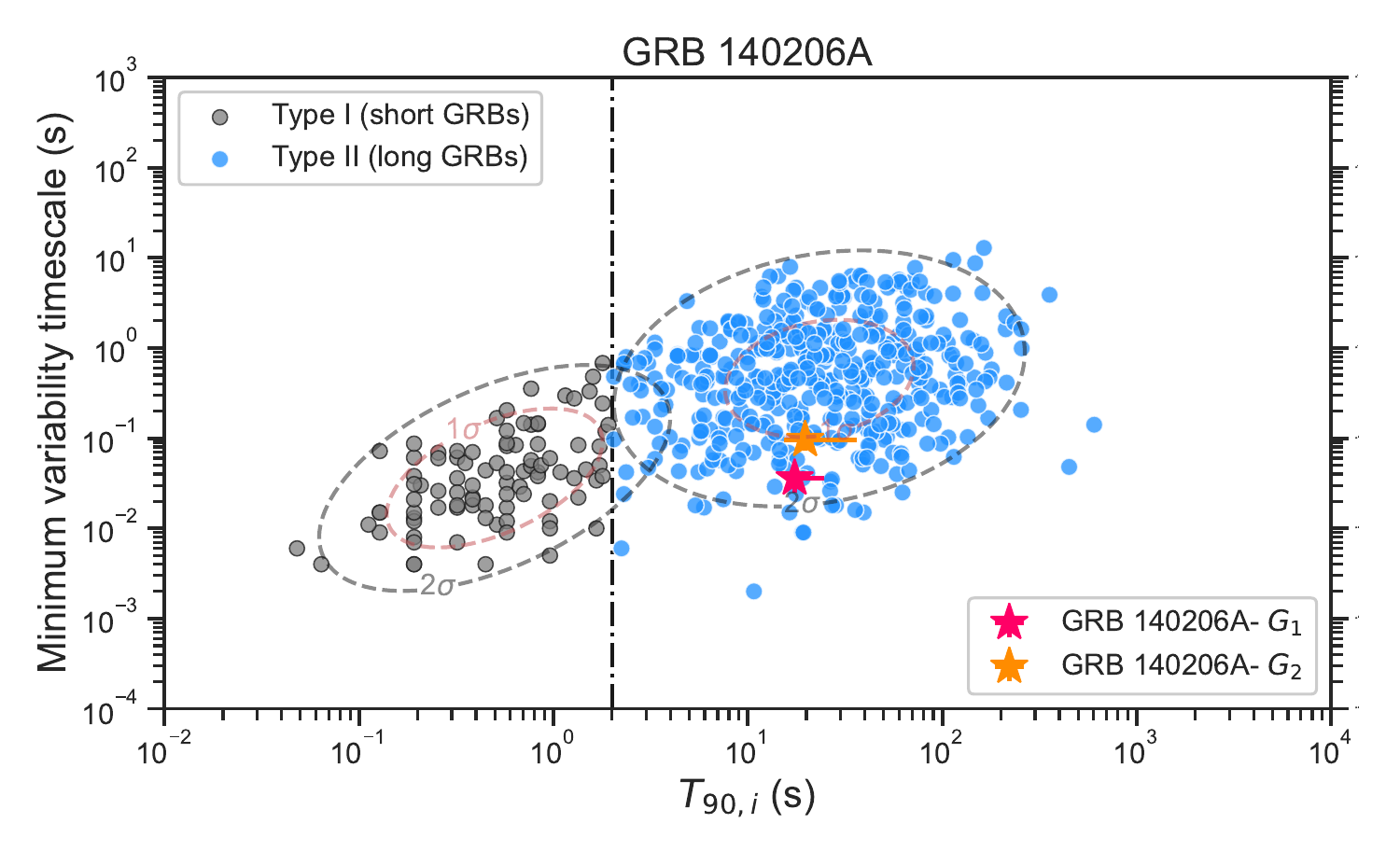}
\includegraphics[width=0.5\textwidth]{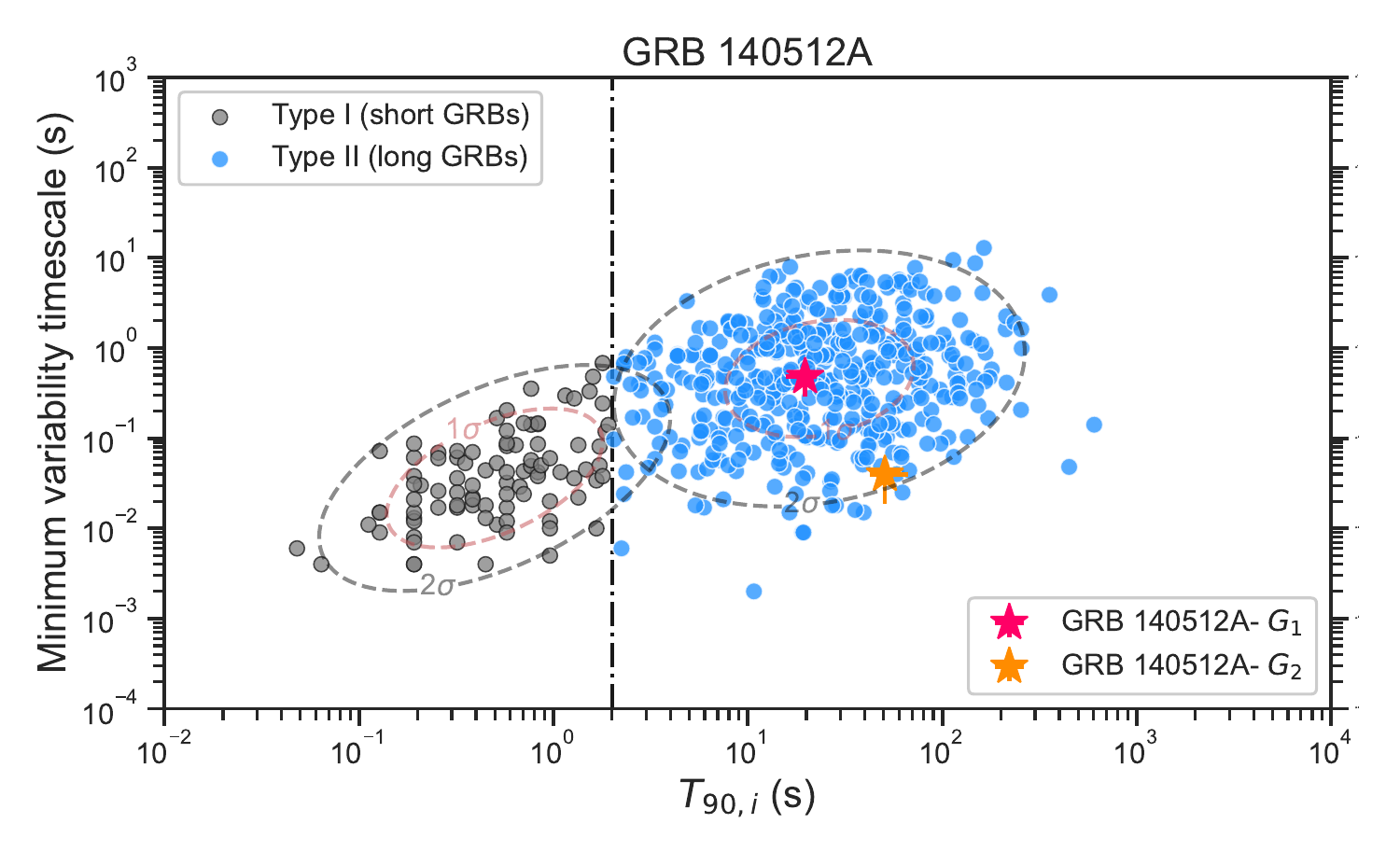}
\includegraphics[width=0.5\textwidth]{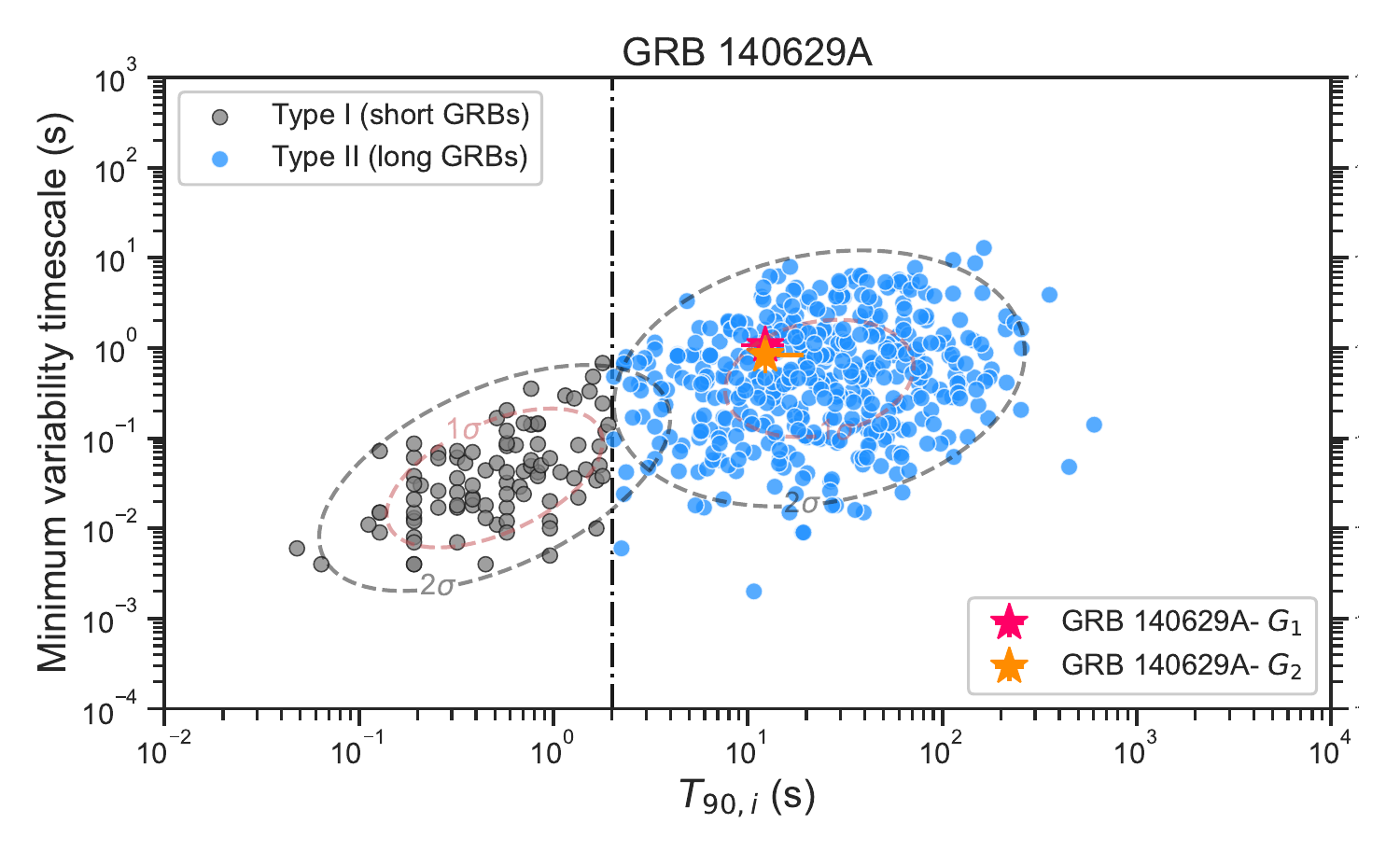}
\includegraphics[width=0.5\textwidth]{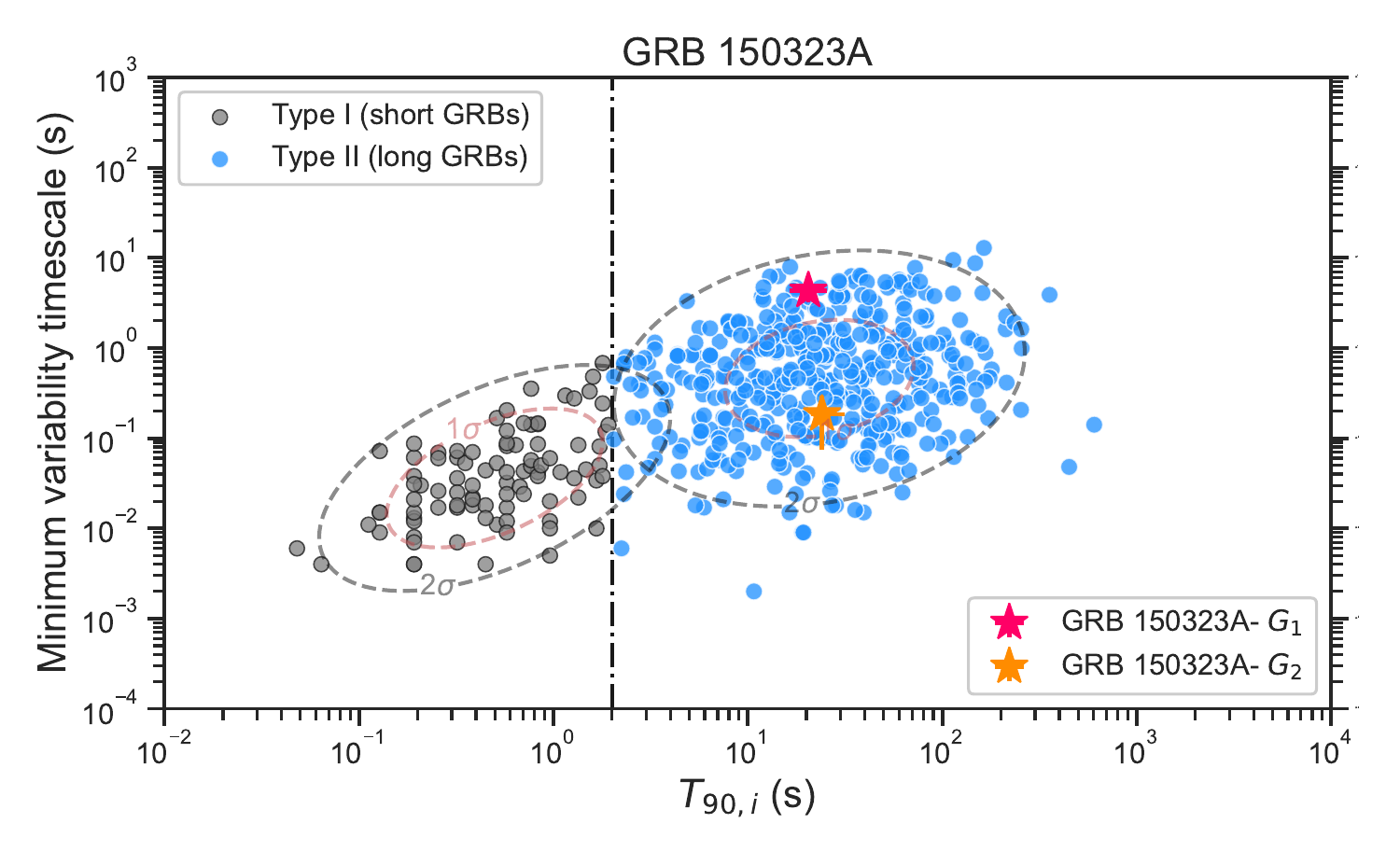}
\includegraphics[width=0.5\textwidth]{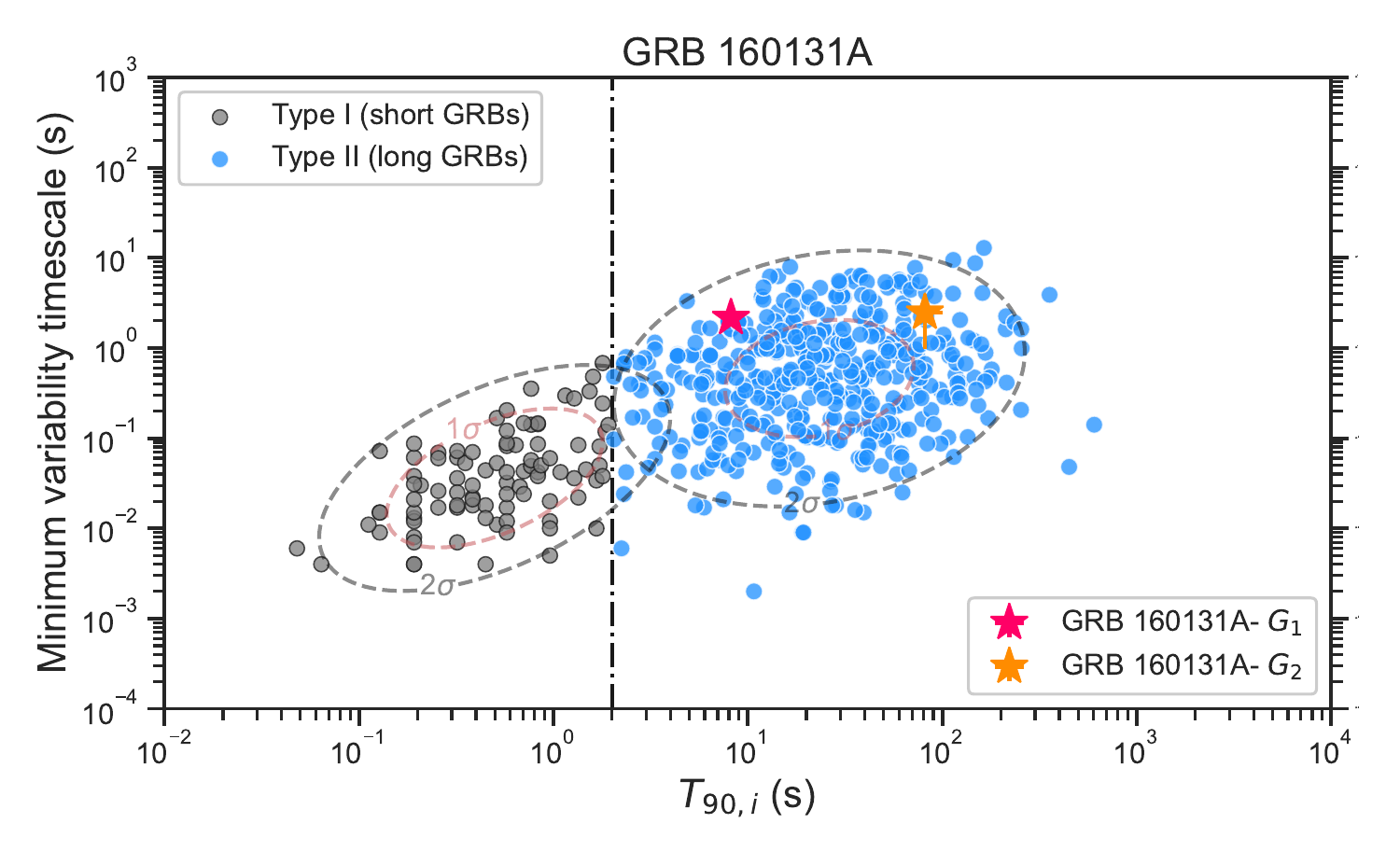}
\center{Figure \ref{fig:MVT_T90}--- Continued}
\end{figure*}
\begin{figure*}
\includegraphics[width=0.5\textwidth]{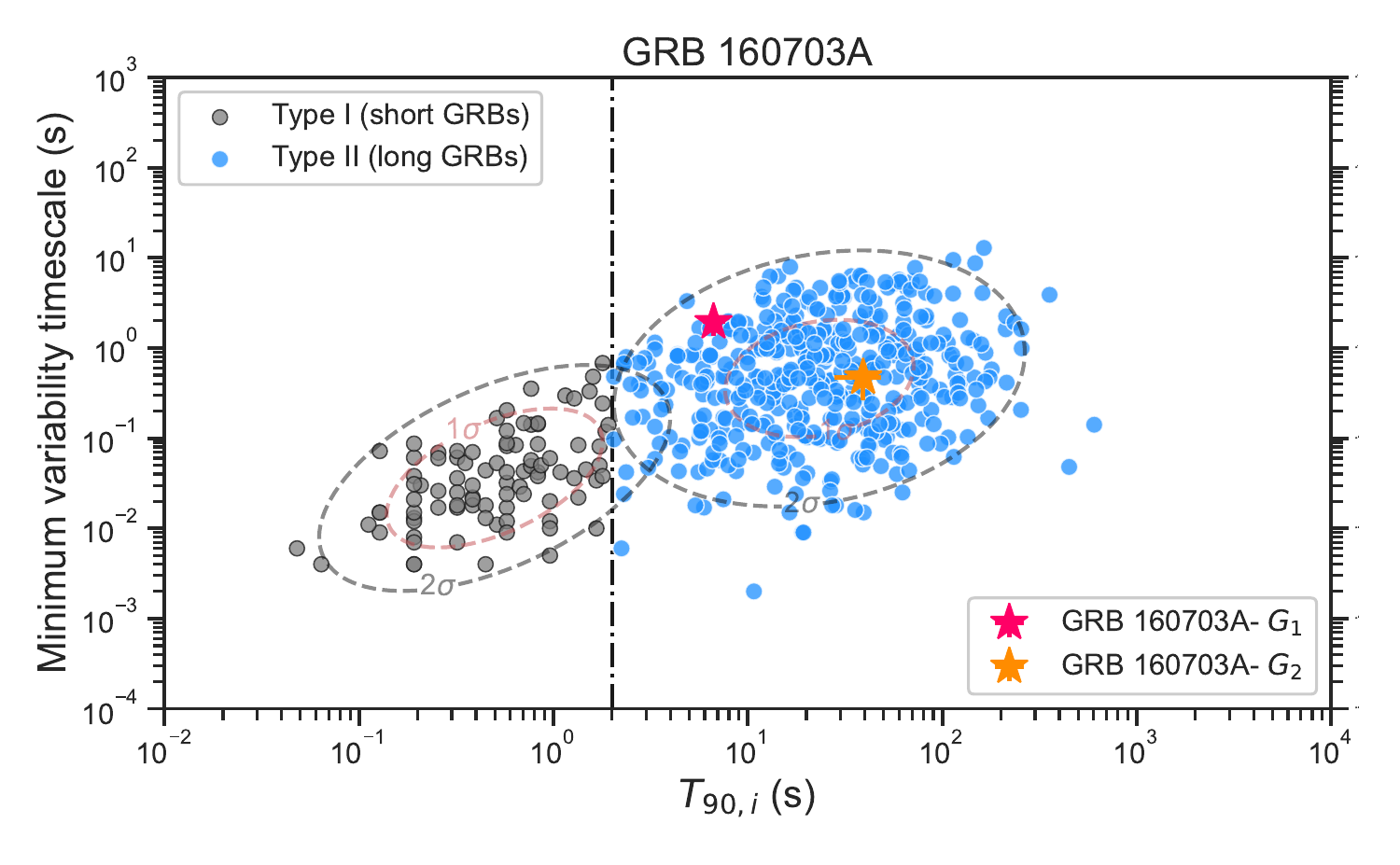}
\includegraphics[width=0.5\textwidth]{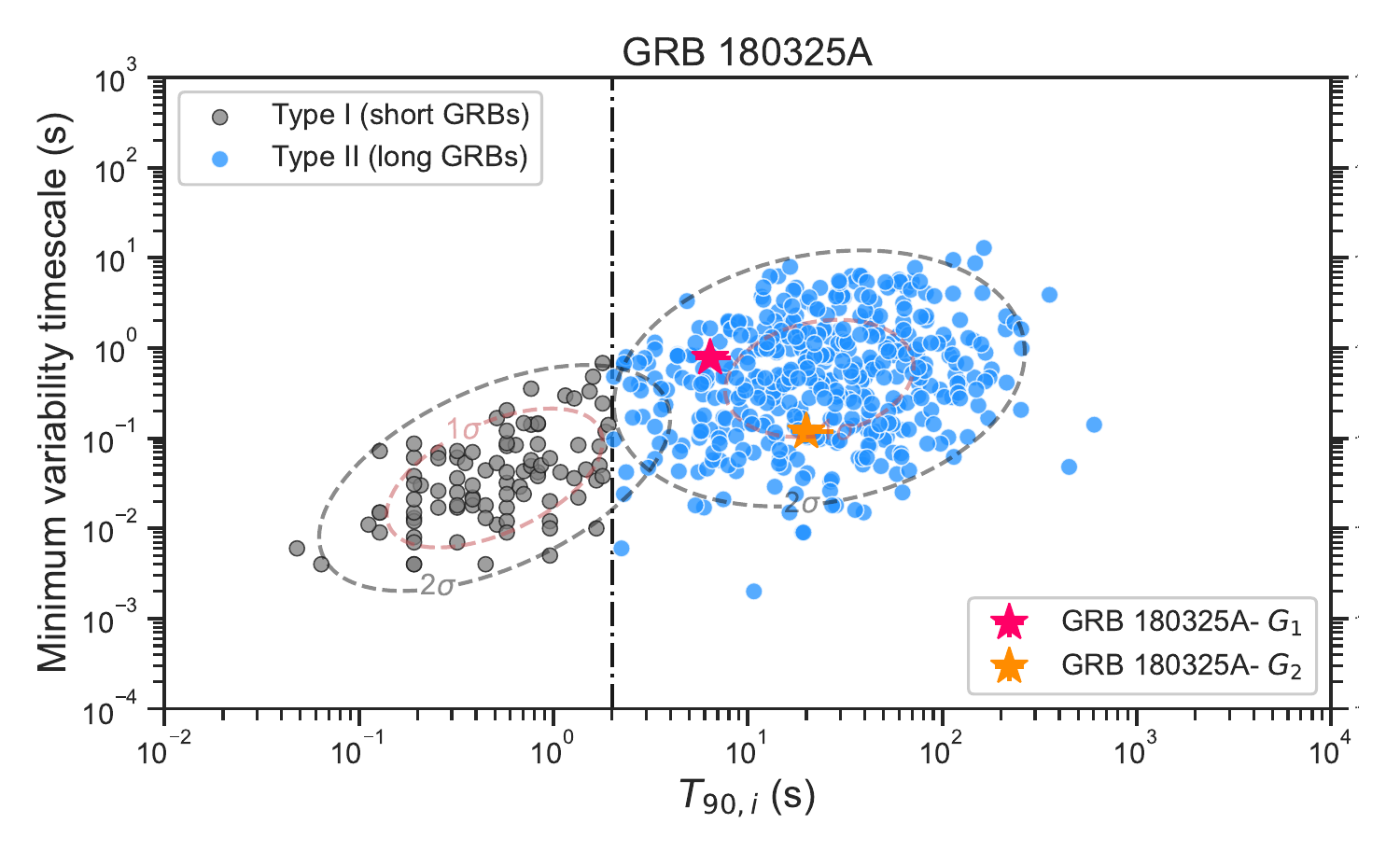}
\includegraphics[width=0.5\textwidth]{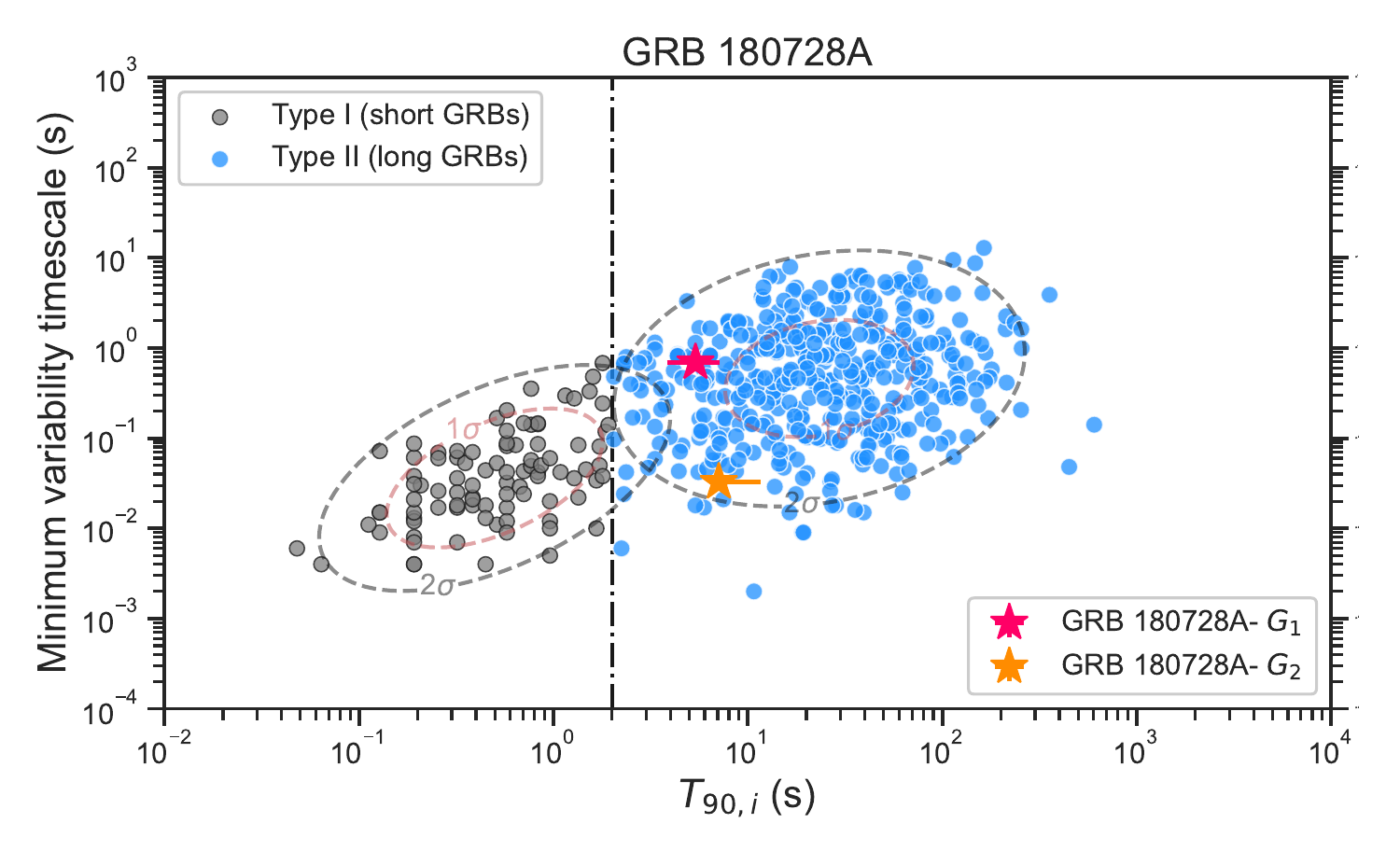}
\includegraphics[width=0.5\textwidth]{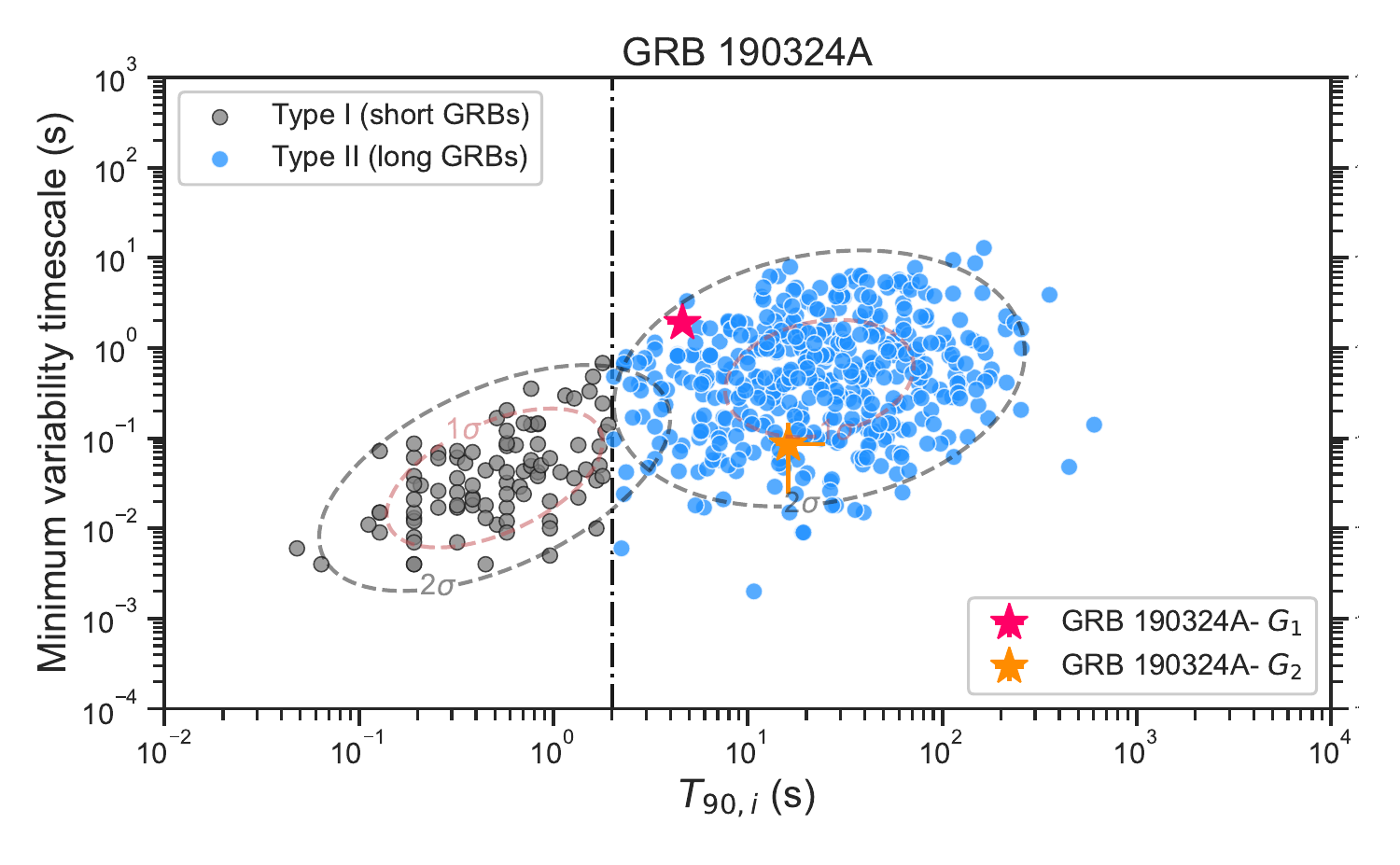}
\center{Figure \ref{fig:MVT_T90}--- Continued}
\end{figure*}

\begin{figure*}
\includegraphics[angle=0,scale=0.35]{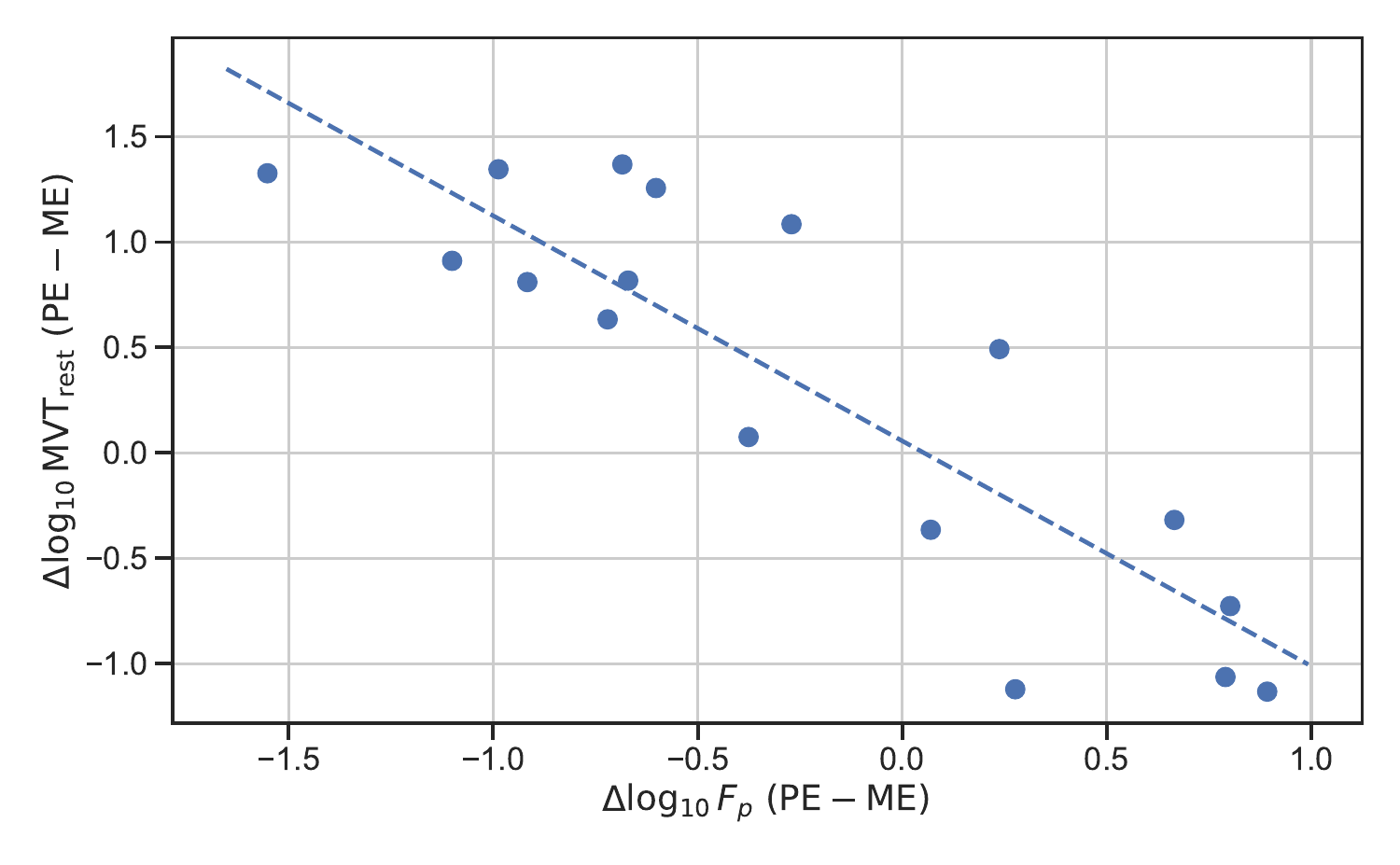}
\includegraphics[angle=0,scale=0.35]{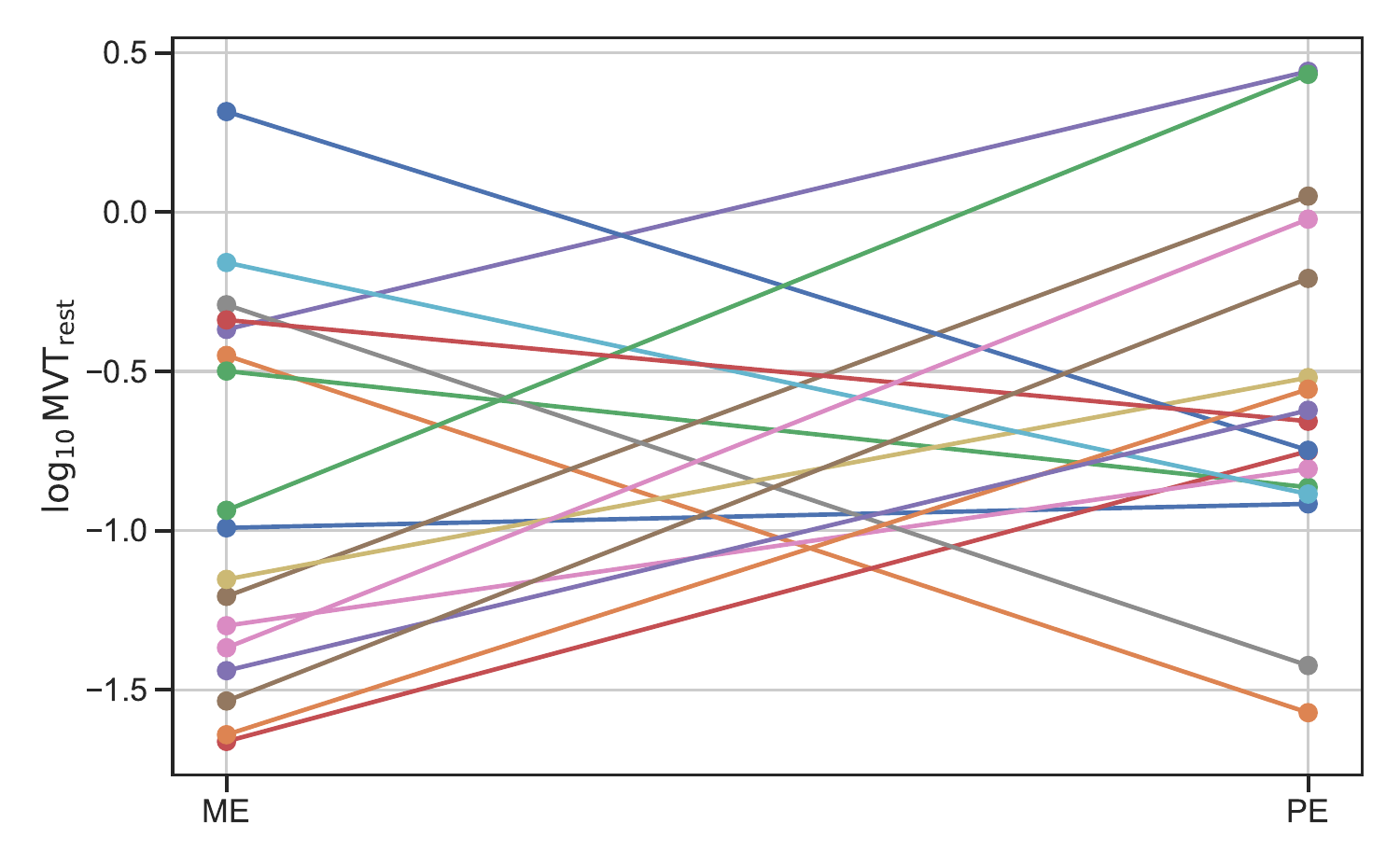}
\caption{\textbf{Left-panel:} Paired regression: dlogMVT vs dlogFp (BAT 15–150 keV). Within-burst log-difference of rest-frame minimum variability timescale, $\Delta\log_{10}\mathrm{MVT}{\rm rest}$, versus the log-difference of peak flux, $\Delta\log{10}F_p$, for the precursor (PE) relative to the main emission (ME), measured in the same BAT band (15–150 keV). The dashed line shows the OLS fit to the paired data, yielding a significantly negative slope ($\beta=-1.07^{+0.25}_{-0.28}$, $R^2=0.75$; Pearson $r=-0.869$, $p=6.0\times10^{-6}$; Spearman $\rho=-0.819$, $p=5.9\times10^{-5}$). The strong anti-correlation demonstrates that weaker (lower-SNR) precursors admit larger measurable MVT, consistent with threshold-driven variability detection. \textbf{Right-panel:} Within-GRB paired MVT (rest frame).Paired line plot of $\log_{10}\mathrm{MVT}_{\rm rest}$ from the main emission (ME) to the precursor (PE) for each burst. Most trajectories follow the global trend that weaker precursors have larger MVT (consistent with the SNR bias inferred in Figure~A), while a minority of events exhibit stronger and more rapidly variable precursors (shorter PE MVT despite higher PE flux), pointing to genuine physical differences in the onset phase (e.g., early internal-shock or magnetically dominated dissipation).}
\label{fig:MVT_SNRs_plot}
\end{figure*}

\end{document}